\documentclass[apsprl,twocolumn,superscriptaddress]{revtex4-2}

\usepackage{graphicx,subfigure}
\usepackage{changes}
\usepackage{amsmath,amssymb}
\usepackage{mathtools}
\usepackage{multirow}
\usepackage{textcomp}
\usepackage{float}
\usepackage{xcolor}
\usepackage{titletoc}
\usepackage{ulem}
\usepackage{dsfont}
\usepackage{siunitx}
\DeclareSIUnit\gauss{G}
\usepackage{bibunits}
\usepackage{import}
\usepackage{braket}
\usepackage{comment}
\usepackage{bm}
\usepackage{bbm}
\usepackage[english]{babel}
\usepackage{hyperref}
\hypersetup{colorlinks=true,linkcolor=blue,citecolor=blue,breaklinks}

\newcommand{\micro}[1]{\ensuremath{\mu\mathrm{#1}}}

\renewcommand{\micro}[1]{\ensuremath \mu\mathrm{#1}}

\renewcommand{\vec}[1]{\ensuremath{\mathbf{#1}}}

\let\oldsfdefault\sfdefault
\usepackage[scaled=0.92]{helvet}
\renewcommand{\sfdefault}{\oldsfdefault}

\definechangesauthor[name=Immanuel Bloch, color=red]{ib}

\newcommand{\ValueDiffusionConstant}{0.88(5)}
\newcommand{\ValueScalingExponentZero}{1.07(6)}
\newcommand{\ValueScalingExponentHalf}{1.4(2)}
\newcommand{\ValueScalingExponentOne}{2.2(4)}

\defaultbibliographystyle{apsrev4-2}

\begin{document}

\begin{bibunit}

\title{Emergence of fluctuating hydrodynamics in chaotic quantum systems}

\author{Julian F. Wienand}
\author{Simon Karch}
\author{Alexander Impertro}
\author{Christian Schweizer}
    \affiliation{Fakult\"{a}t f\"{u}r Physik, Ludwig-Maximilians-Universit\"{a}t, 80799 Munich, Germany}
    \affiliation{Max-Planck-Institut f\"{u}r Quantenoptik, 85748 Garching, Germany}
    \affiliation{Munich Center for Quantum Science and Technology (MCQST), 80799 Munich, Germany}
\author{Ewan~McCulloch}
    \affiliation{Department of Physics, University of Massachusetts, Amherst, MA 01003, USA}

\author{Romain Vasseur}
    \affiliation{Department of Physics, University of Massachusetts, Amherst, MA 01003, USA}

\author{Sarang Gopalakrishnan}
    \affiliation{Department of Electrical and Computer Engineering, Princeton University, Princeton, NJ 08544, USA}

\author{Monika Aidelsburger}
\author{Immanuel Bloch}
    \affiliation{Fakult\"{a}t f\"{u}r Physik, Ludwig-Maximilians-Universit\"{a}t, 80799 Munich, Germany}
    \affiliation{Max-Planck-Institut f\"{u}r Quantenoptik, 85748 Garching, Germany}
    \affiliation{Munich Center for Quantum Science and Technology (MCQST), 80799 Munich, Germany}

\date{\today}


\begin{abstract}
A fundamental principle of chaotic quantum dynamics is that local subsystems eventually approach a thermal equilibrium state. Large subsystems thermalize slower: their approach to equilibrium is limited by the hydrodynamic build-up of large-scale fluctuations. For classical out-of-equilibrium systems, the framework of macroscopic fluctuation theory (MFT) was recently developed to model the hydrodynamics of fluctuations. We perform large-scale quantum simulations that monitor the full counting statistics of particle-number fluctuations in hard-core boson ladders, contrasting systems with ballistic and chaotic dynamics. We find excellent agreement between our results and MFT predictions, which allows us to accurately extract diffusion constants from fluctuation growth. Our results suggest that large-scale fluctuations of isolated quantum systems display emergent hydrodynamic behavior, expanding the applicability of MFT to the quantum regime.

\end{abstract}
\maketitle


\textbf{Introduction}. -- Many-body quantum dynamics is intractable in general. However, in chaotic quantum systems, the expectation values of local observables evolve simply~\cite{nandkishore_many-body_2015, dalessio_quantum_2016}: Starting from a general initial state, these observables rapidly reach local equilibrium values corresponding to spatially-varying temperatures and chemical potentials~\cite{deutsch_quantum_1991, srednicki_chaos_1994, rigol_thermalization_2008}. On longer timescales, these spatial variations relax through hydrodynamic processes like diffusion~\cite{lux_hydrodynamic_2014, spohn_large_1991}. It might seem that initial states without large-scale density variations -- e.g., translation-invariant initial product states -- will rapidly relax. Even such states, however, exhibit slow  timescales: the equilibrium state has much larger density fluctuations and entanglement than the initial state, and these quantities can only build up by slow hydrodynamic processes
~\cite{kaufman_quantum_2016, rispoli_quantum_2019,von_keyserlingk_operator_2022}. The equilibration of fluctuations goes beyond standard thermalization, as it involves highly nonlocal observables. Even in the simpler setting of classical stochastic dynamics, a framework for understanding this process -- named macroscopic fluctuation theory (MFT)~\cite{bertini_macroscopic_2015} -- was only recently developed.

\begin{figure}[htb!]
\includegraphics[width=1\columnwidth]{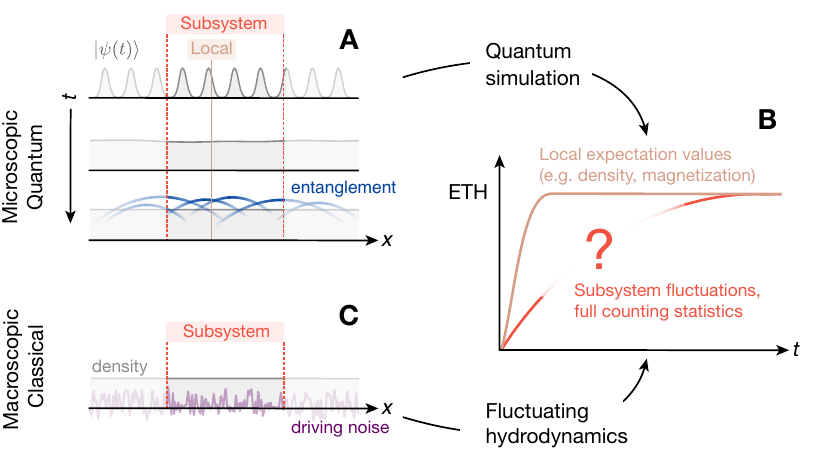}
     \caption{\textbf{Emergence of hydrodynamic fluctuations in a chaotic quantum system} (A) In an out-of-equilibrium quantum system without large-scale density variations, local expectation values (such as density) rapidly relax, while entanglement keeps spreading across the system on much longer timescales. (B) Thus, a subsystem becomes increasingly entangled with its environment, leading to fluctuations of observables in the subsystem that equilibrate on a much slower timescale than local expectation values. 
     (C) This slow hydrodynamic equilibration of fluctuations is conjectured to be classically described by macroscopic fluctuation theory (MFT) which predicts the time evolution of the statistics of a coarse-grained density $n(x,t)$ driven by statistical noise [see Eq.~(\ref{fhd})]. }
     \label{fig:sketch}
\end{figure}

The central thesis of MFT is that the dynamics of fluctuations starting from generic initial states is completely fixed by the equilibrium diffusion constant $D$. Whether MFT extends to deterministic quantum evolution remains a fundamental open question~\cite{mcculloch_full_2023}, and so does the fate of MFT in nearly integrable systems~\cite{doyon_ballistic_2022}. Experimentally, quantum simulators such as quantum gas microscopes with single-site resolution~\cite{bakr_quantum_2009,  sherson_single-atom-resolved_2010, gross_quantum_2021}, provide suitable platforms for studying this phenomenon. However, so far the dynamics of fluctuations has only been explored in relatively small quantum systems, in which slow hydrodynamic timescales are challenging to extract~\cite{kaufman_quantum_2016, rispoli_quantum_2019, lukin_probing_2019}.

In this work, we investigate the equilibration of fluctuations in large quantum systems using a $^{133}$Cs quantum gas microscope~\cite{klostermann_fast_2022, impertro_unsupervised_2022}. The atoms are arranged in a ladder geometry containing up to $100$ sites, with adjustable rung couplings and interactions set to the regime of hardcore bosons. By decoupling the chains of the ladder, we realize the integrable limit of free fermions~\cite{kinoshita_quantum_2006}, while coupling the chains places the system in the strongly chaotic regime. 

We initiate our system as a rung charge density wave (CDW) in the ladder geometry (see Fig.\,\ref{fig:experiment}A)~\cite{trotzky_probing_2012,schreiber_observation_2015}. This 
state is spatially uniform on large scales but displays strongly suppressed particle number fluctuations compared to an equilibrium state. To characterize the build-up of fluctuations, we measure the full counting statistics (FCS)~\cite{levitov_electron_1996, ranabhat_dynamics_2022, humeniuk_full_2017, honeychurch_full_2020, malossi_full_2014, calabrese_full_2020, levitov_electron_1996, stephan_full_2017, collura_full_2017, nazarov_full_2003, mcculloch_full_2023, devillard_full_2020, lovas_full_2017, gopalakrishnan_theory_2022, zheng_efficiently_2022, dall_ideal_2013} of the particle number inside subsystems of variable size and for variable evolution times after quenching on the dynamics (for previous work on FCS using quantum gas microscopes, see Refs.~\cite{wei_quantum_2022, rispoli_quantum_2019, herce_full_2022, lukin_probing_2019}). As we tune the dynamics from ballistic to diffusive, we observe that local mean densities relax increasingly fast, while the growth of nonlocal fluctuations significantly slows down. This clear separation of equilibration timescales highlights the distinction between the relaxation of local expectation values (mean density) and that of non-local quantities (fluctuations). Our work also marks the first direct observation of the crossover from ballistic to diffusive correlation growth in an isolated quantum system.

\begin{figure}[htb!]
\includegraphics[]{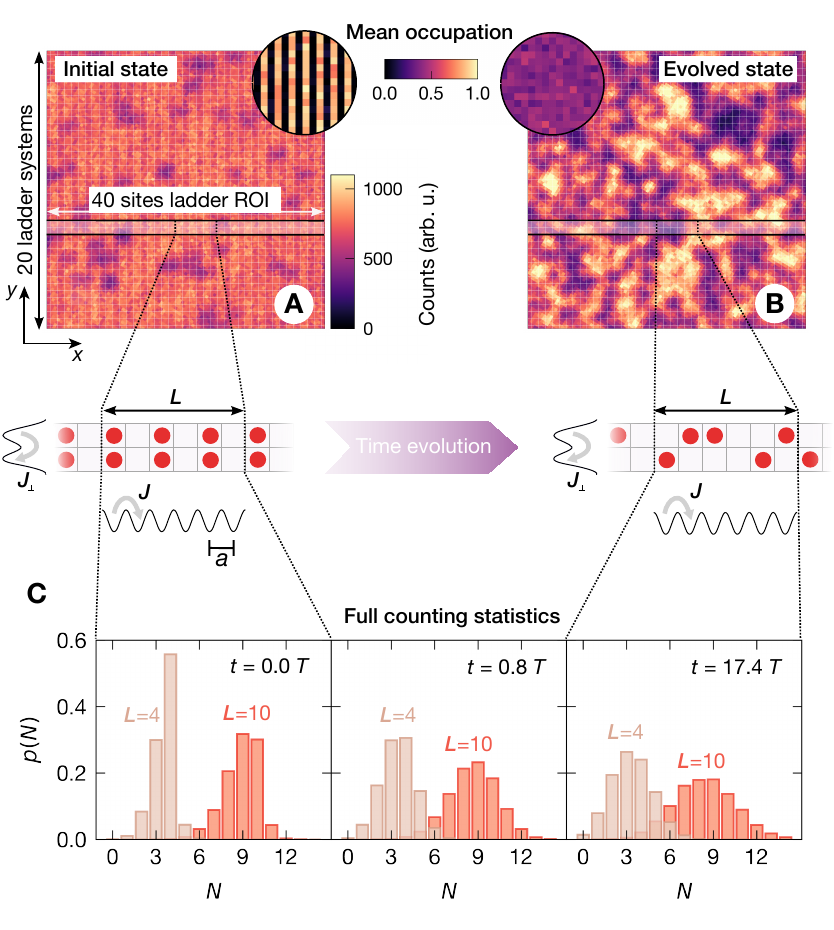}
     \caption{\textbf{Equilibration in ladder systems with tunable rung coupling.} Multiple copies of large homogeneous ladder systems are realized using an optical superlattice in the $y$-direction and a simple lattice in the $x$-direction. Each 1D chain has a total length of up to 50 sites (delimited by potential walls using a DMD), the central 40 of which are in the region of interest (ROI) used for data analysis. Adjusting $J_{\perp}/J$ allows to smoothly tune between integrable uncoupled 1D systems of hardcore bosons ($J_{\perp}/J=0$) and a fully-coupled ladder system ($J_{\perp}/J=1$) with chaotic dynamics. (A) The initial state is a CDW with a period of two lattice constants, prepared using an optical superlattice in the $x$-direction. The round inset shows the reconstructed site occupation averaged over $32$ images. (B) After quenching the system to large tunnel couplings, the CDW rapidly evolves into a state with uniform filling and slowly growing subsystem fluctuations. (C) Using single-site resolution, we obtain the FCS $p(N)$ of the total particle number $N$ in subsystems of length $L$ with $2\times L$ sites (here shown for two subsystem sizes $L=4,10$ and three different evolution times $t$ in tunnelling times $T=\hbar/J$) and use it to track the relaxation dynamics after the quench. }
     \label{fig:experiment}
\end{figure}

In the chaotic regime, we compare the fluctuation dynamics to predictions from 
MFT and find excellent quantitative agreement with experimental results. 
This agreement is a stringent test of fluctuation-dissipation theorems out of equilibrium~\cite{bertini_macroscopic_2015} and has a further practical significance: It allows to accurately determine linear-response diffusion constants from quantum simulations of far-from-equilibrium dynamics and demonstrates the capability of quantum simulators to compute quantities that are difficult to obtain using numerical methods~\cite{DAOE2020,von_keyserlingk_operator_2022}.


\begin{figure*}[htb!]
\includegraphics[bb = 0 0 6.82in 3.81in, width=1.75\columnwidth]{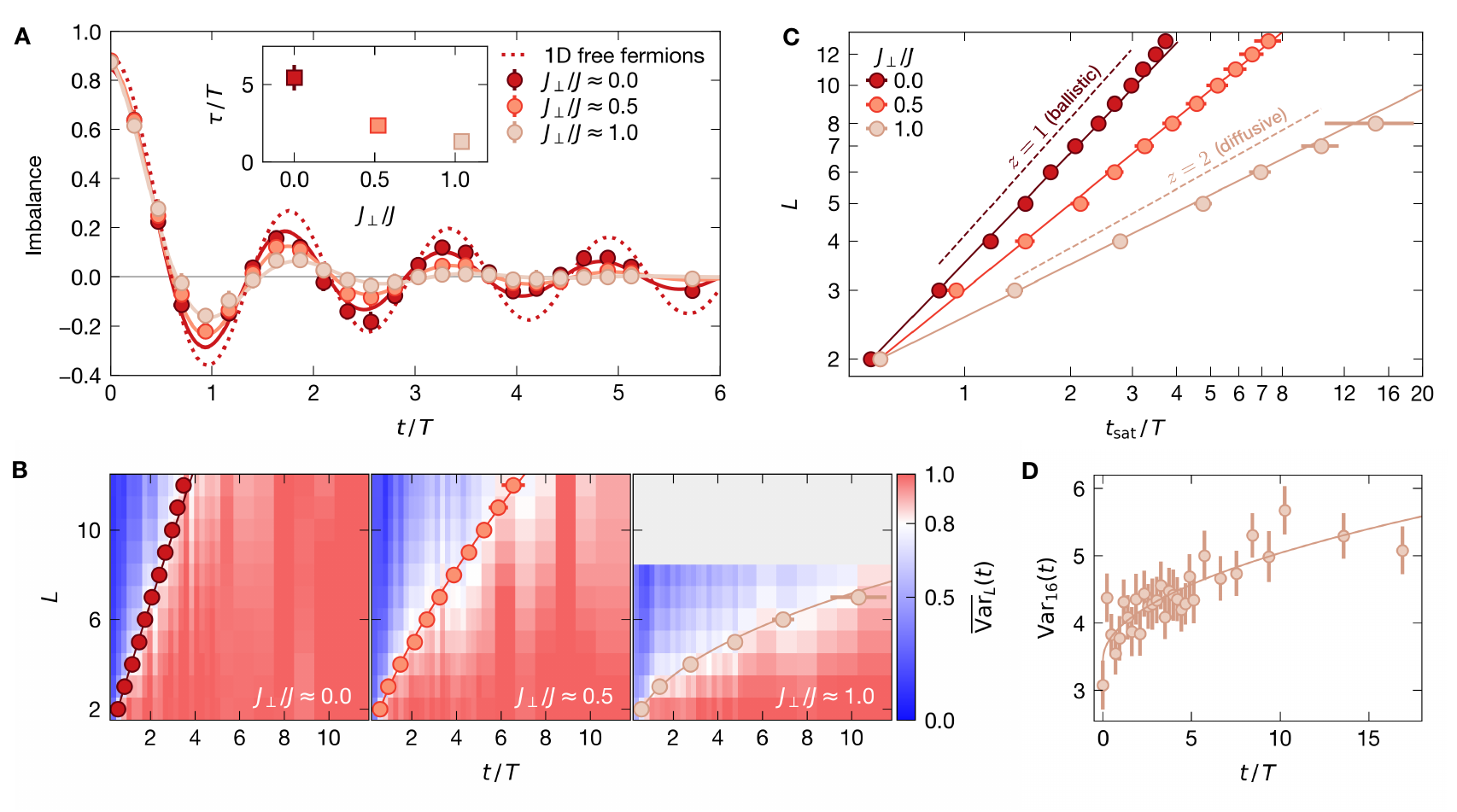}
    
     \caption{\textbf{Time evolution of the local mean density and particle number fluctuations.} (A) Imbalance as a function of evolution time, for $J_{\perp}/J  \approx 0.0, 0.5, 1.0$. Each data point is obtained from averaging over the $40 \times 40$ sites ROI in about $35$ fluorescence images. Solid curves are Bessel function fits to the experimental data (see main text for details).  The dotted curve is the theoretical expectation for the 1D chain, derived from free-fermion theory and taking into account the imperfect experimental initial state. The inset at the top depicts the fitted decay constant $\tau$ as a function of $J_{\perp}/J$. Error bars denote the standard deviation. (B) Normalized atom-number variances $\overline{\mathrm{Var}}_L(t)$ in ladder subsystems of size $2 \times L$. 
     The data points indicate the time $t_{\mathrm{sat}}$ when the variance reaches $80\%$ of its fitted saturation value, as a function of system size $L$ and for different $J_{\perp}/J$. The gray area in the right panel marks the regime of very large subsystem sizes for which we cannot reliably determine the saturation value as the fluctuation growth is too slow. The error bars indicate the standard error of the fit used for determining $t_{\mathrm{sat}}$ (see main text). (C) Threshold time $t_{\mathrm{sat}}$ as a function of system size $L$ and $J_{\perp}/J$ in log-log scale. The solid lines are linear fits to obtain the dynamical exponent $L \sim t_{\mathrm{sat}}^{1/z}$. For reference, the dashed lines indicate ideal slopes corresponding to $z=1, 2$ for ballistic and diffusive dynamics. (D) Atom number variance for a subsystem of size $L=16$ as a function of evolution time. The solid line is the MFT prediction fitted to the experimental data yielding a diffusion constant of $D=1.11(25)\, Ja^2/\hbar$. Error bars denote the standard deviation. }
     \label{fig:time_evolution}
\end{figure*}

\begin{figure*}[t]
\includegraphics[bb = 0 0 8.21in 1.94in,width=2\columnwidth]{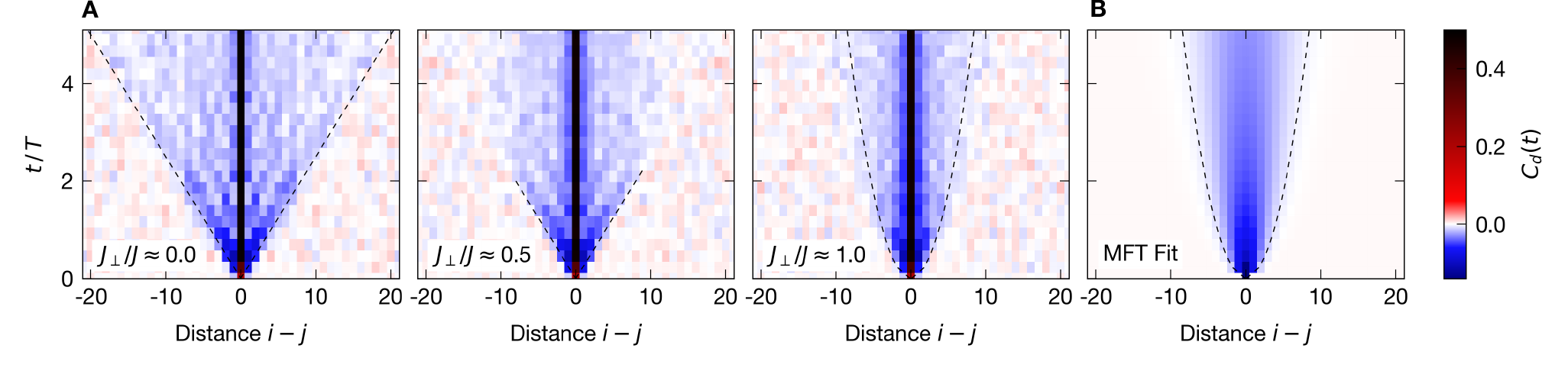}
    
     \caption{\textbf{Time evolution of two-point rung-density correlations.} Rung density-density correlations $C_{d}(t)$ showing a cone that indicates ballistic spreading for $J_{\perp}/J \approx 0.0$ and diffusive spreading for $J_{\perp}/J \approx 1.0$. The dashed lines in the left and center panel indicate the Lieb-Robinson velocity $4J a /\hbar$ in the ballistic regime~\cite{lieb_finite_1972}. The correlations at each point in time have been obtained from $\sim 35$ fluorescence images. Right panel: Gaussian fit function for the density-density correlations predicted by MFT [see Eq.~(\ref{density-density theory}) of the supplementary material~\cite{supp}], fitted for $J_{\perp}/J \approx 1.0$ and distances $1 \leq d \leq 20$, yielding a diffusion constant of $D = \ValueDiffusionConstant\,Ja^2/\hbar$. The dashed curves in (A, right panel) and (B) indicate the $2\,\sigma_c$-envelope of the Gaussian fit function, where $\sigma_c = \sqrt{4Dt/a^2}$.
     }
     \label{fig:cones_main}
\end{figure*}

\textbf{Experimental protocol. -- } For our experiments, we use a strongly-interacting quantum gas of $^{133}\mathrm{Cs}$~atoms imaged through a high-NA quantum gas microscopy setup.
After preparing a Bose-Einstein condensate in a single plane of a vertical optical lattice at the focus of the objective, the atoms are loaded into a two-dimensional (2D) superlattice potential, consisting of an optical superlattice in the $y$-direction ($\lambda_{y,\mathrm{short}}= \SI{767}{\nano\meter}$ and $\lambda_{y,\mathrm{long}}= \SI{1534}{\nano\meter}$) and a simple lattice in the $x$-direction ($\lambda_{x,\mathrm{short}}= \SI{767}{\nano\meter}$). The resulting potential is characterized by chains of double wells coupled in the $y$-direction (see Fig.~\ref{fig:experiment}) and enables us to realize the Bose-Hubbard model in ladder geometries, as expressed by the Hamiltonian:
\begin{equation}
    \begin{aligned}
        \hat H &= - J \left( \sum_{\alpha, i} \hat a_{\alpha, i}^{\dagger} \hat a_{\alpha, i+1}^{\phantom{\ast}} + \mathrm{h.c.} \right) \\ &-  J_{\perp} \left( \sum_{i} \hat a_{1,i}^{\dagger} \hat a_{2, i}^{\phantom{\ast}} + \mathrm{h.c.} \right)  + \frac{U}{2} \sum_{\alpha, i} \hat n_{\alpha, i} (\hat n_{\alpha, i}+1).
    \end{aligned}
    \label{eq:hamiltonian}
\end{equation}
Here, $\hat a_{\alpha, i}^{\phantom{\ast}}$, $\hat a_{\alpha, i}^{\dagger}$ and $\hat n_i = \hat a_{\alpha, i}^{\dagger} \hat a_{\alpha, i}^{\phantom{\ast}}$ are the bosonic annihilation, creation and particle number operators for site $i$ in leg $\alpha=1,2$ of the ladder~\cite{crepin_phase_2011, donohue_mott-superfluid_2001}. The atoms can tunnel along (perpendicular) to the ladder with strength $J$ ($J_{\perp}$). The on-site interaction energy is denoted by~$U$. All measurements were performed in the hard-core regime with $U/J > 6.5$, such that the probability for two atoms to occupy the same lattice site is $\lesssim 3\%$ for all values of $J_{\perp}/J$ probed in this work.  In the experiment, 20 identical uncoupled ladders are present (see Fig.~\ref{fig:experiment}A), each of which contains up to $2 \times 50$ sites. The harmonic confinement of the vertical lattice is compensated by a tailored light profile shaped using a digital mirror device (DMD), realizing a homogeneous box potential with hard walls.

Our experiments begin by preparing a period-two CDW using an additional optical superlattice potential in the $x$-direction ($\lambda_{x,\mathrm{short}}= \SI{767}{\nano\meter}$ and $\lambda_{x,\mathrm{long}}= \SI{1534}{\nano\meter}$). 
This initial CDW is close to a product state where every other site along the ladder is occupied by one atom, as shown in  Fig.~\ref{fig:experiment}A~\cite{trotzky_probing_2012,schreiber_observation_2015, luschen_observation_2017,rubio-abadal_many-body_2019, kohlert_observation_2019}. Typically, we achieve a filling of $84(8)\%$ in the occupied and $4(3)\%$ in the unoccupied rows.
First, all dynamics is frozen, i.e. $J_{\perp} \approx J \approx 0$. We then quench on the tunnel coupling to $J/h=\SI{96 \pm 3}{\hertz}$ and $J_{\perp}/J=\SI{0.0029(3)}{}, \SI{0.55(2)}{} , \SI{1.04(3)}{}$, setting the coupling between the legs of the ladder and letting the system evolve for a controllable evolution time~$t$ (see supplementary material for more details on the experimental setup and sequence~\cite{supp}).


\textbf{Local mean density decay. -- } 
Since the initial (perfect) CDW state lacks large-scale density gradients, hydrodynamics predicts that local expectation values -- the simplest observables -- should rapidly relax. For this initial state, a natural expectation value is the average imbalance $\mathcal{I} = (\langle \hat n_{\mathrm{even}} \rangle - \langle \hat n_{\mathrm{odd}} \rangle)/( \langle \hat n_{\mathrm{even}} \rangle + \langle \hat n_{\mathrm{odd}} \rangle)$ of all ladder systems in the region of interest (ROI) of $40 \times 40$ sites. It compares the average filling of even ($\langle\hat n_{\mathrm{even}} \rangle = \langle \hat n_{2i} \rangle_i$) and odd sites ($\langle \hat n_{\mathrm{odd}} \rangle = \langle \hat n_{2i+1} \rangle_i$) ~\cite{trotzky_probing_2012,schreiber_observation_2015}. 
As shown in Fig.~\ref{fig:time_evolution}A, the imbalance decays to zero on timescales comparable to the tunnelling strength $J$ for all $J_\perp/J$. 
We extract the decay constant by fitting an exponentially decaying Bessel function $\mathcal{I}(t)\,=A \,\mathcal{J}_{0}(4\,t/T)\; e^{-t/\tau}$~\cite{cramer_exploring_2008}.

To motivate this fit, we note that in the one-dimensional (1D) case ($J_\perp/J = 0$) the dynamics can be mapped onto that of free fermions through a Jordan-Wigner transformation.
Free-fermion theory predicts a purely Bessel-type decay with $\tau \to \infty$ and $A=1$ (dashed curve). However, due to the imperfect filling of the initial state with $\mathcal{I}(0) \sim 0.9$, the fitted amplitude is reduced to $A=0.9$. Further, the finite decay constant measured even for $J_{\perp}/J=0$ can most likely be attributed to residual disorder~\cite{supp}. Notably, the decay of the oscillation is enhanced for larger $J_{\perp}/J$. This is due to the integrability-breaking interactions between adjacent chains, which dephase the oscillations that occur in the free-fermion limit and entail a faster decay of the imbalance.


\textbf{Number fluctuations. -- } While the density reaches equilibrium rapidly, subsystem number fluctuations show a strikingly different behaviour. We quantify fluctuations through the variance of the particle number inside a ladder region of length $L$ with $2 \times L$ sites, i.e., $\mathrm{Var}_L \equiv \mathrm{Var}(\sum_i^L  \hat N_i)$, where $\hat N_i = \hat n_{1,i} + \hat n_{2,i}$ is the total particle number in the $i$-th rung of the ladder. This quantity is computed from density-density correlators (see below), in order to mitigate systematic reconstruction errors (see~\cite{supp}).

In the perfect CDW state, the variance is zero, but as the subsystem interacts (and becomes entangled) with its surroundings, this variance eventually builds up to its thermal equilibrium value, which is expected to be close to the infinite temperature value $ 2\, L\,\overline{n}\,(1-\overline{n})$ for an infinite disorder-free system in the hard-core regime ($\overline{n}$ is the average filling). Since the equilibrium fluctuations must build up due to the subsystem becoming entangled with its environment, they should emerge on timescales that are ballistic in the integrable limit and diffusive in the chaotic limit. Thus, while chaotic dynamics lead to a faster equilibration of local mean quantities (see Fig.~\ref{fig:time_evolution}A), they are expected to slow down the relaxation of non-local quantities, including atom number fluctuations. Fig.~\ref{fig:time_evolution}B depicts the time evolution of the normalized variance $\overline{\mathrm{Var}}_L(t) = [\mathrm{Var}_L(t) - \mathrm{Var}_L(0)]/[\mathrm{Var}_L(\infty) - \mathrm{Var}_L(0)]$ for various system sizes $L$ and $J_{\perp}/J\approx0.0, 0.5, 1.0$. Here, $\mathrm{Var}_L(\infty)$ is the saturation value for $t \to \infty$, which is extracted from a fitting procedure that uses an empirical function (see the supplementary material for details~\cite{supp}). To quantify the rate of growth of fluctuations, we define the saturation time $t_{\mathrm{sat}}$ at which the variance reaches $80\%$ of its long-time saturation value. We find that larger subsystems exhibit a variance growth that is substantially slower in the ladder ($J_{\perp}/J\approx1$) compared to the 1D case ($J_{\perp}/J\approx0$).

This is visualized by the data points in Fig.~\ref{fig:time_evolution}B which show $t_{\mathrm{sat}}$ as a function of subsystem size $L$. In Fig.~\ref{fig:time_evolution}C, the same data is shown with log-log scaling, revealing that the saturation time $t_{\mathrm{sat}}$ scales polynomially with subsystem size $L$, i.e., $L \sim t_{\mathrm{sat}}^{1/z}$ with scaling exponent $z$.
Using a linear fit to extract the slope, we find $z=\ValueScalingExponentZero, \ValueScalingExponentHalf, \ValueScalingExponentOne$ for $J_{\perp}/J \approx 0.0, 0.5, 1.0$, respectively. This is consistent with the expectation that the dynamics in the decoupled chains ($J_{\perp}/J = 0.0$) is ballistic ($z=1$) and the dynamics in the system with fully coupled legs ($J_{\perp}/J = 1.0$) is diffusive ($z=2$). For $J_\perp/J = 0.5$ we find an intermediate value, which, as we discuss below, can be interpreted as a crossover from ballistic to diffusive dynamics.

We observe that large scale fluctuations (number fluctuations in subsystems that are a sizeable fraction of the total system) fail to fully thermalize~\cite{supp}.


\textbf{Hydrodynamics of fluctuations. --} 
We have presented evidence that fluctuations starting from a non-equilibrium initial state continue to obey simple hydrodynamic behavior. This motivates the conjecture (analogous to classical MFT) that the growth dynamics of fluctuations is in fact controlled by a single parameter, the diffusion constant $D(n)$. It depends on the particle density $n$ and can be determined from the equilibrium structure factor (i.e., through linear response). We note that this conjecture is much stronger than, for instance, the eigenstate thermalization hypothesis~\cite{deutsch_quantum_1991, srednicki_chaos_1994}, which is only concerned with whether local expectation values eventually relax to their thermal equilibrium values.

Applying MFT implies that the dynamics of large-scale particle fluctuations in closed many-body systems is governed by the stochastic equations of fluctuating hydrodynamics~\cite{bertini_macroscopic_2015}, 
\begin{equation}\label{fhd}
\partial_t n + \partial_x j = 0, \quad j = -D(n)\,\partial_x n + \sqrt{2\,D(n)\,\chi(n)}\,\xi,
\end{equation}
where $j$ is the particle current, $\chi(n) = n\,(1-n)$ is the static susceptibility and $\xi(x,t)$ is a Gaussian white noise process with unit variance. In this picture, the noise is playing the role of fast (non-hydrodynamic) modes, which act as a source of thermal fluctuations for slow hydrodynamic modes, in this case, the particle density. From the perspective of unitary dynamics, large scale particle fluctuations in a closed quantum system build up through the growth of entanglement (unitary dynamics leads to a complex superposition of states with different particle-number configurations), whereas, in the effective hydrodynamic picture, these fluctuations build up as a result of the classical noisy dynamics. While the former is exceedingly hard to calculate, the latter provides an approximate classical description that can be studied numerically far more readily than the full many-body quantum dynamics, and in some cases, can even be solved exactly~\cite{mallick_exact_2022}.

One test of this effective description is to study the particle transfer between a subsystem and the rest of the system which, in an initial state with sharp particle number, corresponds to studying fluctuations in the subsystem’s total atom number. The non-linear fluctuating hydrodynamic equation~\eqref{fhd} is hard to solve directly~\cite{mallick_exact_2022}. Instead, one can study the macroscopic properties of any microscopic model, classical or quantum, that shares Eq.~\eqref{fhd} as its effective hydrodynamic theory. We take this approach and compute the number fluctuations and density correlations for a simplified diffusive model (see supplementary material~\cite{supp}). We predict,
\begin{equation}\label{eq:particle-transfer-variance}
\text{Var}_L^{\text{CDW}}(t) \approx  \sqrt{\frac{2Dt}{\pi a^2 }}, \quad t \ll (L\,a)^2/D,
\end{equation}
where $a$ is the lattice spacing. This expression describes the growth of the particle number variance for a subsystem of size $L$, starting from a perfect CDW state, before its eventual equilibration at times $\sim (La)^2/D$, 
and enables the determination of the diffusion constant from the particle number fluctuations in a state with far-from-equilibrium fluctuations. By comparing the diffusion constant, obtained far-from-equilibrium in our quantum simulation, with numerically obtained equilibrium values, we are able to test classical fluctuating hydrodynamics based on MFT in a quantum setting.

In Fig.~\ref{fig:time_evolution}D we show the atom number variance in subsystems of size $L=16$ as a function of evolution time. Due to the large size of the subsystems, saturation will occur at very long times and not affect the fluctuation growth in the range of measured evolution times.
By fitting the data using a modified version of Eq.~(\ref{eq:particle-transfer-variance}) which takes into account experimental imperfections, we obtain a diffusion constant of $D=1.11(25)\, Ja^2/\hbar$. This value is in agreement with numerical estimates for the equilibrium infinite-temperature diffusion constant, reported to be $D=0.95 \, Ja^2/\hbar$~\cite{steinigeweg_2014} and $D \approx 0.97 \,Ja^2/\hbar$~\cite{DAOE2020}.


\textbf{Density-density correlations. --}
Using our quantum gas microscope, we measure the rung density correlation function $C_{d}=\langle \hat N_i \hat N_j \rangle  - \langle \hat N_i \rangle \langle \hat N_j \rangle$ over distances up to $|d|=21$, where $d=i-j$ denotes the real space distance between two sites with index $i$ and $j$.
The particle number variance studied previously combines these correlations into a single quantity. By studying the spatial profile of density correlations over a large range of distances, we can improve our estimate for the diffusion constant, as shown below. Additionally, we can use the spatially-resolved correlations to further shed light on the dynamics that govern the thermalization process as we crossover from integrable ($J_\perp/J \approx 0$) to chaotic dynamics ($J_\perp/J > 0$).

Fig.~\ref{fig:cones_main} shows $C_{d}$ as a function of distance and evolution time and reveals a cone illustrating how the correlations emerge after the quench and how they grow spatially during equilibration~\cite{cheneau_light-cone-like_2012,takasu_energy_2020, tajik_experimental_2022}.  The slope of the cone boundary indicates the maximum speed of correlation spreading. For the 1D integrable system ($J_{\perp}/J \approx 0.0$) the boundary of the cone is linear with a slope consistent with $4J a /\hbar$ (see dashed line, $a = \SI{383.5}{\nano\meter}$),  suggesting ballistic spreading at the Lieb-Robinson velocity predicted by free-fermion theory 
\cite{lieb_finite_1972, cheneau_light-cone-like_2012}. For $J_{\perp}/J>0$, the onset of the cone at short times is characterized by the same slope. However, at later times, the cone's expansion slows down according to a square-root law, indicating that the correlations eventually spread in a diffusive fashion.

For $J_{\perp}/J \approx 1.0$ we can obtain the diffusion constant by fitting the correlation cone in Fig.~\ref{fig:cones_main}~(right panel), which we expect to spread as a Gaussian with width $\sqrt{4Dt/a^2}$ for diffusive systems~\cite{supp},
yielding $D = \ValueDiffusionConstant\,Ja^2/\hbar$. This value is in good agreement with both the value obtained from charge transfer fluctuations in large subsystems [see Eq.~(\ref{eq:particle-transfer-variance})] as well as with the estimates from recent theoretical equilibrium linear response predictions~\cite{steinigeweg_2014,DAOE2020}. By extracting the diffusion constant from the correlation cone, we build a bridge between equal-time correlation functions following a quench from a far-from-equilibrium initial state and two-time correlation functions about an equilibrium initial state. The former manifest themselves as number fluctuations, growing from the initial CDW state, while the latter define linear-response coefficients, including the diffusion constant, which uniquely defines the entire macroscopic time evolution of charge fluctuations.

 
\textbf{Discussion}. -- Most studies of thermalization in isolated quantum systems have focused on local mean values, such as the average density or the imbalance.
By studying the equilibration dynamics of non-local quantities, such as particle number fluctuations and density-density correlators in very large systems, we have shown that the dynamics of a chaotic quantum system can be effectively described by a single quantity -- the diffusion constant. In doing so we have provided strong evidence that MFT provides an excellent description of the large-scale density fluctuations in quantum many-body systems, and that the macroscopic dynamics do not depend on the quantum nature of the microscopic processes. Whether MFT can be extended to include quantum coherent phenomena such as entanglement spreading remains an important question for future work~\cite{Bernard_QMFT}. Another interesting question concerns the relationship between the equilibration of FCS and the emergence of quantum state designs under Hamiltonian dynamics~\cite{mark2022benchmarking, Choi2023}.

\begin{figure}[htb!]
\includegraphics[width=1\columnwidth]{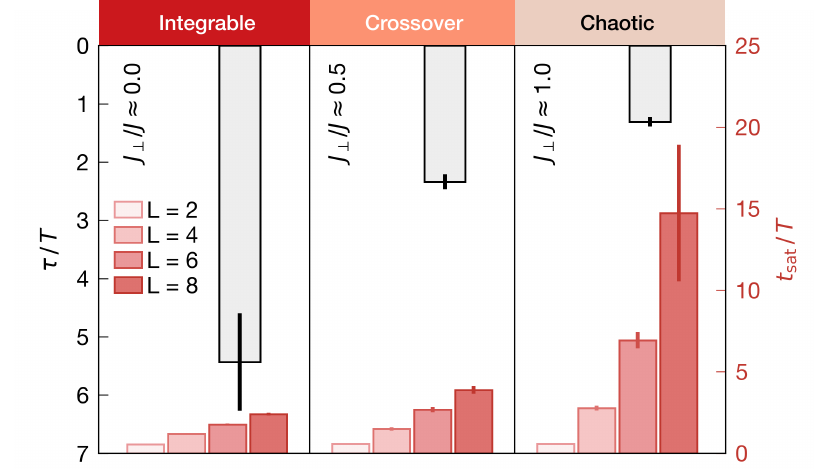}
    
     \caption{\textbf{Separation of equilibration timescales.} Imbalance decay constant $\tau$ (see Fig.~\ref{fig:time_evolution}A) and the variance growth threshold time $t_{\mathrm{sat}}$ for various system sizes (see Fig.~\ref{fig:time_evolution}B,C) as a function of $J_{\perp}/J$, showing opposite behavior in the limits of $J_{\perp}/J\approx0.0$ and $1.0$. Whereas local expectation values relax faster in the diffusive limit, subsystem fluctuation growth slows down,  creating a separation of equilibration timescales between different moments of the particle number distribution.}
     \label{fig:scaling}
\end{figure}

Our results demonstrate that, even when expectation values relax rapidly, experimentally accessible observables such as large-scale fluctuations can exhibit slow hydrodynamic timescales. This separation of equilibration timescales becomes increasingly pronounced as the coupling between the ladder is increased, making the system chaotic (Fig.~\ref{fig:scaling}). Thus, while a state might look locally like it has thermalized (because the initial CDW has decayed), it actually has not and the dynamics continue to play out in the fluctuations rather than in the mean. For instance, the growth of fluctuations was recently argued to yield a long timescale in the relaxation of the momentum distribution of a Tonks-Girardeau gas after a quench~\cite{le_direct_2022}.

By leveraging fluctuating hydrodynamics, we have explicitly related a quantity that is challenging to measure or compute -- the equilibrium diffusion constant -- to one that is easily measurable, the growth of fluctuations after a quench from the CDW state.  Measuring transport through the growth of fluctuations has several practical advantages over doing so through a quench from a domain-wall initial state~\cite{choi_exploring_2016, wei_quantum_2022}: the CDW state rapidly achieves a uniform density throughout, and the growth of fluctuations probes the diffusion constant at this uniform density, whereas in a domain-wall setup the diffusion constant is generally spatially inhomogeneous and keeps changing as the density profile evolves. Quenches from a CDW allow to directly probe equilibrium transport quantities while working far from a linear-response limit and thus retaining high signal-to-noise ratio.

This opens the door for exploring transport in a variety of many-body systems that lie at or beyond the edge of current computational capabilities, including those with finite interactions beyond the hard-core approximation and those described by ballistic MFT~\cite{doyon_ballistic_2022, castro-alvaredo_emergent_2016}. Our large system size of $\sim 2500$ sites might particularly benefit detailed studies of prethermalization~\cite{rubio-abadal_many-body_2019, gring_relaxation_2012, ueda_quantum_2020} and MBL~\cite{leonard_probing_2023,bordia_probing_2017,rispoli_quantum_2019} in one and two dimensions. The ability to prepare a large variety of initial states could facilitate the quantum simulation of Hilbert space fragmentation and many-body scars~\cite{moudgalya_quantum_2022, turner_quantum_2018, scherg_observing_2021, kohlert_exploring_2023, su_observation_2023} under the microscope, shedding light on systems with exotic thermalization properties.

While our results only systematically address the first two moments of physical observables, a natural question is whether each higher moment of these observables relaxes on a separate timescale. In addition to the duration of these relaxation processes, it is not well understood under what conditions these processes can reach infinite-temperature saturation values in finite systems and to what extent thermalization fails. These are interesting questions for future theoretical and numerical work. 

\vspace{2em}
\begin{center}
\textbf{ACKNOWLEDGEMENTS}
\end{center}
\vspace{0.5em}

The authors would like to thank Jacopo De Nardis, David Huse, Vedika Khemani, Alan Morningstar and Bharath Hebbe Madhusudhana for helpful discussions. The work by the Munich team received funding from the Deutsche Forschungsgemeinschaft (DFG, German Research Foundation) via Research Unit FOR 2414 under project number 277974659 and under Germany’s Excellence Strategy – EXC-2111 – 390814868 and from the German Federal Ministry of Education and Research via the funding program quantum technologies – from basic research to market (contract number 13N15895 FermiQP). J.F.W. acknowledges support from the German Academic Scholarship Foundation and the Marianne-Plehn-Program. S.K. receives funding from the International Max Planck Research School (IMPRS) for Quantum Science and Technology. A.I. was supported by the Bavarian excellence network ENB via the International Ph.D. Programme of Excellence Exploring Quantum Matter (ExQM). C.S. has received funding from the European Union’s Framework Programme for Research and Innovation Horizon 2020 (2014-2020) under the Marie Sk{\l}odowska-Curie Grant Agreement No.~754388 (LMUResearchFellows) and from LMUexcellent,
funded by the BMBF and the Free State of Bavaria under the
Excellence Strategy of the German Federal Government and the L\"{a}nder. Further, we acknowledge support from NSF DMR-2103938 (S.G., E.M.), DMR-2104141 (E.M., R.V.) and the Alfred P. Sloan Foundation through Sloan Research Fellowships (R.V.) .


\vspace{2em}
\begin{center}
\textbf{REFERENCES}
\end{center}
\vspace{0.5em}

\putbib[manuscript]
\end{bibunit}


\clearpage

\begin{bibunit}
\setcounter{equation}{0}
\setcounter{figure}{0}
\setcounter{table}{0}
\renewcommand{\theequation}{S\arabic{equation}}
\renewcommand{\theHequation}{S\arabic{equation}}
\renewcommand{\thefigure}{S\arabic{figure}}
\renewcommand{\theHfigure}{S\arabic{figure}}
\renewcommand{\thetable}{S\arabic{table}}
\renewcommand{\theHtable}{S\arabic{table}}
\setcounter{page}{1}

\title{Emergence of fluctuating hydrodynamics in chaotic quantum systems}

\author{Julian F. Wienand}
\author{Simon Karch}
\author{Alexander Impertro}
\author{Christian Schweizer}
    \affiliation{Fakult\"{a}t f\"{u}r Physik, Ludwig-Maximilians-Universit\"{a}t, 80799 Munich, Germany}
    \affiliation{Max-Planck-Institut f\"{u}r Quantenoptik, 85748 Garching, Germany}
    \affiliation{Munich Center for Quantum Science and Technology (MCQST), 80799 Munich, Germany}
\author{Ewan~McCulloch}
    \affiliation{Department of Physics, University of Massachusetts, Amherst, MA 01003, USA}

\author{Romain Vasseur}
    \affiliation{Department of Physics, University of Massachusetts, Amherst, MA 01003, USA}

\author{Sarang Gopalakrishnan}
    \affiliation{Department of Electrical and Computer Engineering, Princeton University, Princeton, NJ 08544, USA}

\author{Monika Aidelsburger}
\author{Immanuel Bloch}
    \affiliation{Fakult\"{a}t f\"{u}r Physik, Ludwig-Maximilians-Universit\"{a}t, 80799 Munich, Germany}
    \affiliation{Max-Planck-Institut f\"{u}r Quantenoptik, 85748 Garching, Germany}
    \affiliation{Munich Center for Quantum Science and Technology (MCQST), 80799 Munich, Germany}

\title{Supplementary material for: \\ Emergence of fluctuating hydrodynamics in chaotic quantum system} 

\maketitle

\tableofcontents


\section{Experimental details} \label{sec:supp_exp_details}
\subsection{Setup} \label{sec:supp_setup}

\begin{figure*}[t]
\includegraphics[bb = 0 0 7.05in 4.53in, width=0.9\textwidth]{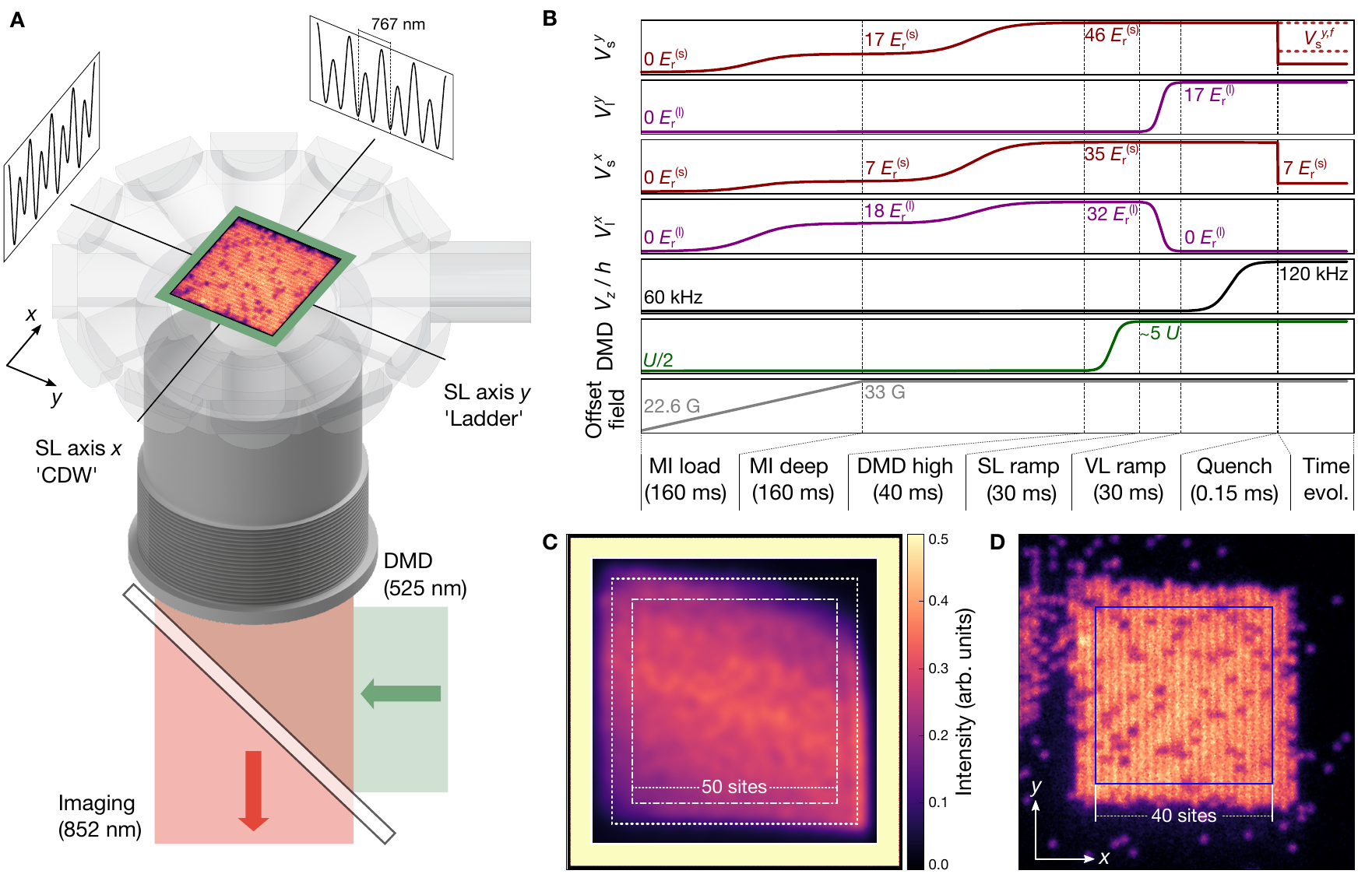}
     \caption{\textbf{Experimental setup and sequence.} (A) Sketch of the experimental setup illustrating high-resolution imaging, superlattice configuration and DMD. Drawing not to scale. (B) Experimental sequence. The depths of the short and long horizontal lattices $V_i^j$ are given in terms of their respective recoil energies. $V_s^{y,f}$ is the final depth of the short lattice in $y$-direction that is chosen to realize a specific $J_\perp$. (C) Flattening mask as obtained in the iterative potential compensation algorithm. The mask is projected onto the atomic plane with the DMD to compensate the external harmonic confinement. The dashed square denotes the $60\times60$ site region used for the compensation algorithm, and the dash-dotted square is the $50\times50$ region used in the experiment. (D) For data evaluation, we use a $40\times40$ sites ROI inside the $50\times50$ sites flat box potential. Shown is a single fluorescence image of the CDW initial state. }
     \label{fig:sm_setup_and_sequence}
\end{figure*}

The main experimental setup was described in Refs.~\cite{klostermann_fast_2022, impertro_unsupervised_2022}. The experiment takes place in a glass cell with a large degree of optical access (see Fig.~\ref{fig:sm_setup_and_sequence}A). The atoms are vertically confined by loading them into a single plane of a shallow-angle vertical lattice (wavelength $\lambda=\SI{1064}{\nano\meter}$, spacing $d=\SI{8}{\micro\meter}$). In the horizontal plane, there is a square superlattice, consisting of two sets of retro-reflected bichromatic optical lattices. Each superlattice consists of a beam with wavelength $\lambda_{x,\mathrm{ short}}=  \lambda_{y, \mathrm{short}} = \SI{767}{\nano\meter}$ (\textit{short lattice}) and wavelength $\lambda_{x, \mathrm{long}}=  \lambda_{y, \mathrm{long}} = \SI{1534}{\nano\meter}$ (\textit{long lattice}). The long lattice is frequency-locked to the short lattice, allowing us to freely vary the superlattice phase at the location of the atoms by introducing a frequency offset~\cite{folling_direct_2007}. The lock is realized by frequency-doubling the light from the $\SI{1534}{\nano\meter}$ laser and beating it with the light from the short lattice laser. The beat signal is stabilized to a variable reference frequency via a phase-locked loop (Vescent D2-135), using the frequency-modulation piezo of the long lattice laser as the actuator.

The atomic configuration in the 2D lattice is read out using fluorescence imaging~\cite{impertro_unsupervised_2022}. To this end, we freeze the configuration by quenching the horizontal lattices and the vertical lattice up to a depth of around $\SI{400}{\micro\kelvin}$. We then turn on optical molasses cooling on the D2 line ($\lambda=\SI{852}{\nano\meter}$), which causes the atoms to scatter fluorescence photons and at the same time cools them in their respective lattice site. The scattered fluorescence photons are collected using a high-resolution objective ($\mathrm{NA}=0.8$) and focused onto a sCMOS camera (Teledyne Photometrics Kinetix) with a magnification of around 40.
In addition to the imaging, we use the objective to project arbitrary repulsive potentials onto the atoms by imaging the surface of a digital micromirror device (DMD, Texas Instruments DLP6500, interface by bbs Bild- und Lichtsysteme GmbH). The DMD is illuminated by light coming from a multi-mode diode laser which emits at a wavelength of $\lambda=\SI{525}{nm}$ and has a spectral width of several nanometers (Wavespectrum WSLX525-005-400M-H), resulting in a low coherence to reduce any interference speckle pattern.

\subsection{Initial state preparation and experimental sequence}
\label{sec:supp_sequence}

The experimental sequence is depicted in Fig.~\ref{fig:sm_setup_and_sequence}B. We prepare the initial state by loading a rectangular Mott insulator in a repulsive box potential. To this end, we start with a 2D cloud of $^{133}\mathrm{Cs}$ atoms confined in a single plane of the vertical lattice. The horizontal confinement is provided by projecting a repulsive box using the DMD. Before adding the horizontal lattices, the phase of the superlattice in $x$-direction, which we use to prepare the initial state, is chosen to realize an array of strongly tilted double-wells at the position of the atoms (see Fig.~\ref{fig:sm_setup_and_sequence}A) \cite{trotzky_probing_2012}. To prepare the rectangular Mott insulator, we ramp up the superlattice in the $x$-direction as well as the short lattice in the $y$-direction using a two-step sigmoidal ramp deep into the Mott insulating regime with a total duration of $\SI{320}{ms}$. Additionally, we ramp the offset field from $\SI{22.6}{G}$ to $\SI{33}{G}$ in order to increase the scattering length from $a=280a_0$ to $a=680a_0$, where $a_0$ is the Bohr radius. To avoid doublon generation, we choose the height of the box to be around $U/2$, causing surplus atoms to spill over the central region when crossing the transition. This allows us to realize rectangular Mott insulating states with a filling of typically $84(8)\%$ in the even and $4(3)\%$ in the odd rows. At this point, the density distribution is frozen out, realizing the initial state that resembles a 1D charge density wave (CDW, see Fig.~\ref{fig:sm_setup_and_sequence}D).

In preparation for the quench experiment, we first increase the height of the DMD walls to $\sim 5\, U$ in order to prevent atoms from tunnelling into or out of the box region. Further, in order to prepare long ladder systems, we completely remove the long lattice along the $x$-direction and simultaneously ramp up the long lattice in the $y$-direction to $17\,E_\mathrm{r}^\mathrm{(l)}$, where $E_\mathrm{r}^\mathrm{(l,s)} = h^2/(2m\lambda_\mathrm{(l,s)}^2)$ is the recoil energy in the short and long horizontal lattice, respectively. The depth of the long lattice is chosen sufficiently large so that tunneling between neighboring ladders is suppressed. Finally, the vertical lattice is increased to a depth of $V_z/h=\SI{120}{kHz}$ to further increase the interaction energy during the time evolution. This moves the system deeper into the hard-core limit and ensures that the doublon fraction stays negligibly small ($\lesssim 3\%$ of doubly-occupied sites) for all values of $J_{\perp}/J$.

The time evolution is turned on by quenching the short lattice along the $x$-direction down to $7\,E_\mathrm{r}^\mathrm{(s)}$ over $\SI{150}{\micro\second}$, corresponding to a tunnel coupling of $J/h =\SI{96(3)}{Hz}$. When studying the equilibration in ladder systems, we quench the short lattice along the $y$-direction at the same time to realize a specific value of $J_{\perp}$. The tunnel couplings $J$ and $J_{\perp}$ have been calibrated using parametric heating resonances and tunnelling oscillations in isolated double wells, respectively.

\subsection{Potential flattening} \label{sec:supp_flattening}

To observe spatially homogeneous dynamics, we correct for external confinement within the repulsive box potential using an iterative potential compensation algorithm. To this end, we prepare a cold superfluid in the box, as its density distribution is particularly sensitive to potential corrugations. The first compensation mask is then generated by dividing the mean occupation averaged over the entire box region by the measured spatial distribution of atoms averaged over ten repetitions. Using a PI-controlled feedback loop, this procedure is repeated over typically ten iterations, where each step gradually improves the compensation mask from the previous step. A typical compensation mask is shown in Fig.~\ref{fig:sm_setup_and_sequence}C, suggesting that the external confinement has a dominant contribution that is oriented diagonally with respect to the lattice axes. We can attribute this profile to the in-plane confinement of the shallow-angle vertical lattice. To avoid edge effects occurring close to the border of the compensation region, we perform the compensation algorithm using a box region that is significantly larger compared to the $50 \times 50$ sites box that we use in the experiment. 

\subsection{Reconstruction of the lattice occupation}

Using high-resolution fluorescence imaging on the D2 line, we achieve a resolution of about $\SI{850}{nm}$. To resolve atoms in the short lattice with spacing $a = \SI{383.5}{nm}$, which is less than half the imaging resolution, we employ a reconstruction algorithm based on a convolutional neural network that is trained directly with experimental data~\cite{impertro_unsupervised_2022}. We estimate that this allows us to determine the occupation with a fidelity of at least $96\%$~\cite{impertro_unsupervised_2022}.


\section{Data analysis} \label{sec:supp_analysis}

\subsection{Post-selection on atom number} \label{sec:supp_postsel}

We use a $40 \times 40$ sites large region of interest (ROI) in the flat box potential region containing $50 \times 50$ sites. The ROI is chosen such that there is a buffer of around five sites between the ROI and the edges of the box region (see Fig.~\ref{fig:sm_setup_and_sequence}D). A small percentage of fluorescence images contain few or no atoms due to errors in the sequence execution. We remove these shots based on the average filling $n$ in the ROI, using $\bar{n}=0.36$ as a threshold value. After this procedure, the average filling in the ROI is $\bar{n}=0.43(2)$.

\begin{figure}[t]
\includegraphics[bb = 0 0 2.85in 1.77in, width=1\columnwidth]{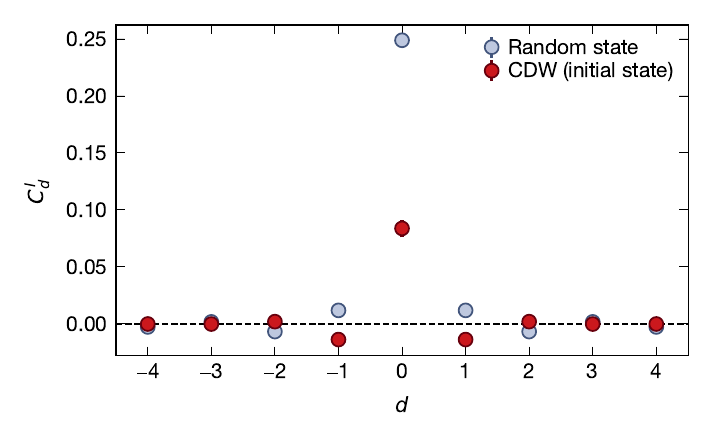}
    
     \caption{\textbf{Calibration measurements for detecting systematic errors in the single-site reconstruction.} Spatial correlation profile $C_{d}^{I}$ of a charge density wave with high imbalance $\sim 0.9$ and a random state with near-zero imbalance. The profiles have been obtained from about $40$ and $70$ fluorescence images, respectively. Error bars represent the standard deviation. }
     \label{fig:corr_calib}
\end{figure}

\subsection{Density-density correlations}
\label{sec:supp_densitydensitycorrl}

The (two-point) density-density correlators are defined as
\begin{equation}
    \begin{aligned}
        C_{d}^{\alpha,\beta} = \left< \hat{n}_{\alpha,i} \hat{n}_{\beta,j} \right> - \left< \hat{n}_{\alpha,i} \right> \left< \hat{n}_{\beta,j} \right>
    \end{aligned}
\end{equation}
where $\alpha, \beta = 1, 2$ denote the legs of the ladder and $i, j$ are indices of sites at distance $d = i-j$ in the respective legs of the ladder. The average $\langle \ldots \rangle = \langle \langle \ldots \rangle_k \rangle_d$ is first taken over all ladders (index $k$) in all fluorescence images and then over all combinations of indices leading to the same distance $d=i-j$.

In our ladder geometry, it is possible to compute three different kinds of density-density correlators.

\begin{itemize}
    \item If $\alpha=\beta$, we calculate correlations along the 1D chains
    \begin{equation}
        C_{d}^{I} = \frac{1}{2} \left(C_{d}^{1,1} + C_{d}^{2,2}\right).
    \end{equation}

    \item Secondly, we can compute the correlations between the legs of the ladder along the direction of the chains:
    \begin{equation}
    C_{d}^{II} = \frac{1}{2} \left( C_{d}^{2,1} + C_{d}^{1,2} \right).
    \label{eq:cd2}
    \end{equation}
    
    \item Lastly, one can consider the ladder as a quasi one-dimensional system and compute the correlations of the total atom number in the rungs of the ladder:

    \begin{equation}
        \begin{aligned}
        C_{d}^{III} &=  2\left(C_{d}^{I} + C_{d}^{II}\right) \\ &= \langle \hat{N}_{i} \hat{N}_{j} \rangle  - \langle \hat{N}_{i} \rangle \langle \hat{N}_{j} \rangle,
        \end{aligned}
    \end{equation}
    
    where $\hat{N}_i = \hat{n}_{1, i} + \hat{n}_{2, i}$. In the main text, we make use of this correlator as it provides us  with the highest signal-to-noise ratio when studying the details of the post-quench dynamics. Further, it is a natural choice when investigating the transition from a (quasi-one-dimensional) ladder system to a (truly one-dimensional) system of decoupled chains.
\end{itemize}

\begin{figure*}[hbt!]
\includegraphics[bb = 0 0 7.55in 2.51in, width=2\columnwidth]{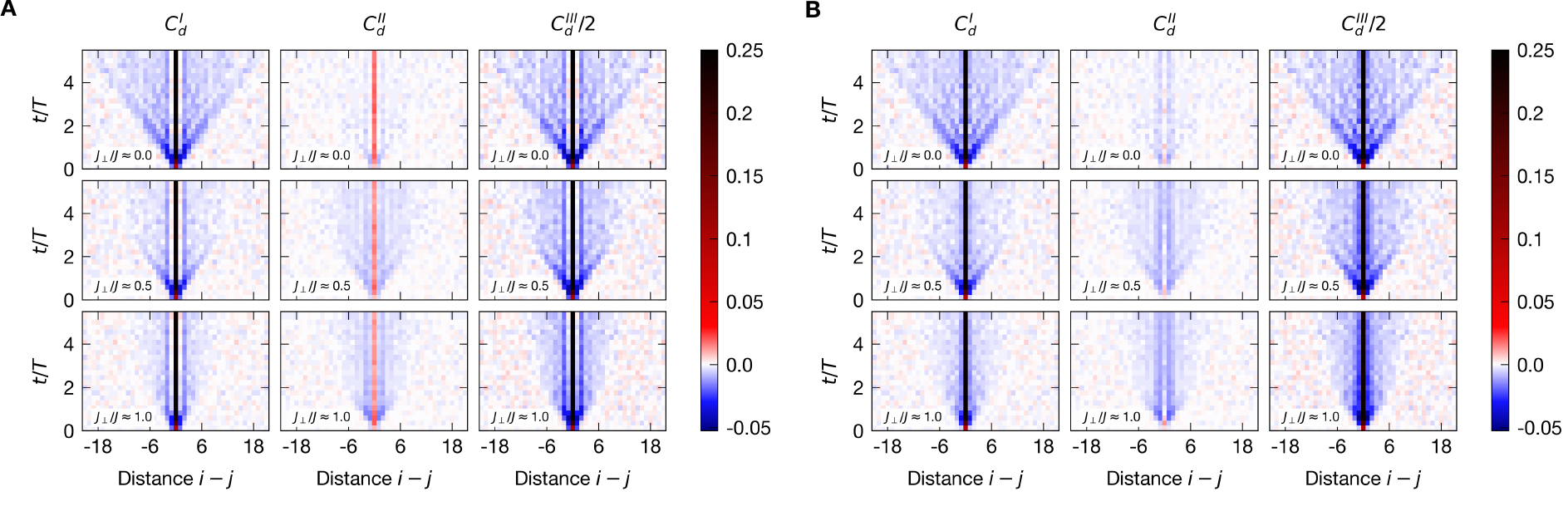}
    
     \caption{\textbf{Impact of the reconstruction correction on the density-density correlations.} Density-density correlations along the legs of the ladder ($C_{d}^{I}$, left column), between the legs of the ladder ($C_{d}^{II}$, center column) and correlations along the ladder of atom numbers summed in the rungs ($C_{d}^{III}$, right column) for $J_{\perp}/J \approx 0.0,0.5,1.0$ (top, center and bottom row, respectively) as a function of distance $d=i-j$ and evolution time $t$. (A) Before the correction described in Section~\ref{sec:corr_corr}.  (B) After the correction described in Section~\ref{sec:corr_corr}. }
     \label{fig:full_light_cones}
\end{figure*}

\subsection{Correcting systematic errors in the single-site reconstruction}
\label{sec:corr_corr}

Due to the finite fidelity of our reconstruction algorithm there may be correlated errors that impact the measured short-distance correlations, presumably on the order of $1-\mathcal{F}$, where $\mathcal{F}$ is the fidelity of the reconstruction process \cite{impertro_unsupervised_2022, cheuk_quantum-gas_2015}.
This is because the errors of the reconstruction are not entirely random but are more likely to happen for certain occupation patterns. For instance, the presence of an atom in a particular site might enhance the probability that the reconstruction algorithm will detect an atom on the adjacent site even though there is no atom.

In order to investigate whether these artificial correlations occur, we perform two calibration measurements for two different states with different imbalances that can be assumed to lack any density-density correlations for $|d|>0$. The two states are:
\begin{itemize}
  \item the CDW in the initial state before the quench with filling $\bar{n}_{\mathrm{initial}}=0.44(2)$ and imbalance $\mathcal{I}_{\mathrm{initial}} = 0.91(7)$, prepared by adiabatically ramping up a tilted superlattice, and
  \item a random state with filling $\bar{n}_{\mathrm{random}}=0.48(2)$ and imbalance $\mathcal{I}_{\mathrm{random}}=0.00(13)$, prepared by taking an $n=1$ Mott insulator in the short lattices and removing about half of the atoms using microwave addressing and a blow-out pulse.
\end{itemize}

Fig.~\ref{fig:corr_calib} shows the density-density correlations $C_{d}^{I}$ measured in these two states as a function of distance $d = i-j$. Since we do not expect to find any non-zero density-density correlations for $|d|>0$, the visible deviations from zero, particularly for $|d|=1,\,2$, have to be attributed to an artefact of the reconstruction process with sign and strength depending on the imbalance. In order to correct the data for this artefact, we compute an imbalance-dependent correction by interpolating between the spatial correlation profiles of the two calibration measurements. We assume that this interpolation is linear, neglecting any (small) non-linear effects.

For imbalance $\mathcal{I}_{\text{initial}}$ of the initial state, we measure a correlation profile $\vec{C}_{\text{initial}}$, while for random state with imbalance $\mathcal{I}_{\text{random}}$, we measure a correlation profile $\vec{C}_{\text{random}}$ (each component corresponds to a distance $d=i-j$).

We obtain the corrected correlations
\begin{equation}
\vec{C}_{\text{corrected}} = \vec{C}_{\text{measured}} - \vec{C}_{\text{correction}}(\mathcal{I})
\label{eq:acorr1}
\end{equation}
by subtracting the imbalance-dependent correction $\vec{C}_{\text{correction}}(\mathcal{I})$ from the measured correlations $\vec{C}_{\text{measured}}$. Assuming that the correction depends linearly on the imbalance $\mathcal{I}$, we can construct the correction from the calibration measurements as follows:
\begin{equation}
    \begin{aligned}
        \vec{C}_{\text{correction}}(\mathcal{I}) &= \vec{C}_{\text{random}} + \frac{\mathcal{I}}{\mathcal{I}_{\text{initial}}} \left( \vec{C}_{\text{initial}} - \vec{C}_{\text{random}} \right).
    \end{aligned}
\label{eq:acorr2}
\end{equation}

Similarly, we can find a correction for $C_{d}^{II}$ and $C_{d}^{III}$. Note that the correction is only applied for $|d|=1,2$, i.e., for distances that are likely to be affected by reconstruction-caused density-density correlations due to the point spread function (PSF) of an atom leaking into other sites up to two sites away. This is supported by the spatial profile in Fig.~\ref{fig:corr_calib}, which shows the artificial correlations dropping quickly to zero within distance two. For $C_{d}^{II}$ we additionally correct $d=0$ as $C_{d=0}^{II}$ is the density-density correlation between two adjacent sites in different legs of the ladder, corresponding to a real-space distance of one. The correlation cones before and after applying the correction are shown in Fig.~\ref{fig:full_light_cones}. Notably, Fig.~\ref{fig:full_light_cones}B (center column) shows how the $C_{d}^{II}$ correlations grow when $J_{\perp}/J > 0$ is increased and the two legs of the ladders start interacting with each other. In contrast, for $J_{\perp} = 0$ we measure approximately $C_{d}^{II} = 0$.

\begin{figure*}[hbt!]
\includegraphics[bb = 0 0 6.3in 3.0in, width=2\columnwidth]{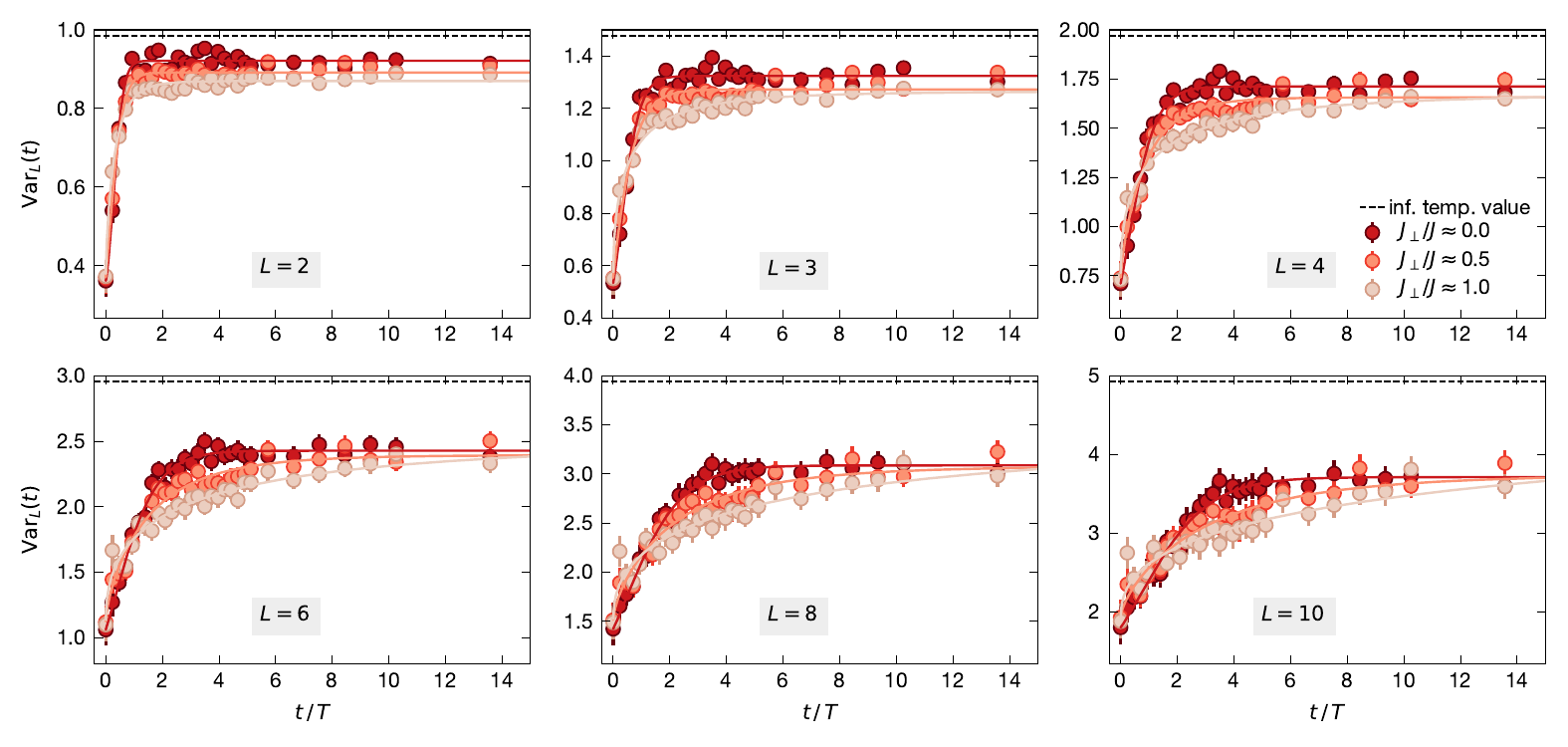}
     \caption{\textbf{Variance growth for different subsystem sizes $L$ and $J_{\perp}/J\approx 0.0, 0.5, 1.0$.} Solid curves are the fits using the empirical fit function in Eq.~(\ref{eq:empirical}). 
     }
     \label{fig:full_variances}
\end{figure*}

\subsection{Computing variances from density-density correlations}
\label{sec:supp_variances_msd}

The atom number variance in a ladder subsystem of size $2 \times L$ is closely linked to the correlators $C_{i,j}^{\alpha, \beta}$ evaluated at distances $|d| < L$. By expanding the definition of the atom number variance in a ladder subsystem of size $2 \times L$ we find that:
\begin{equation}
    \begin{aligned}
        \mathrm{Var}_{L}(t) &= \left< (\hat{n}_{1,1} + \hat{n}_{2,1} + \cdots + \hat{n}_{1,L} + \hat{n}_{2,L})^2 \right> \\ &- \left< \hat{n}_{1,1} + \hat{n}_{2,1} + \cdots + \hat{n}_{1,L} + \hat{n}_{2,L} \right>^2 \\ &= \sum_{i,j = 1}^L \left( C_{i,j}^{1,1} + C_{i,j}^{1,2} + C_{i,j}^{2,1} + C_{i,j}^{2,2}\right).
    \end{aligned}
\end{equation}
The sums in this expression can be further expanded showing directly the connection to the correlation cone. For instance:
\begin{equation}
    \begin{aligned}
        \sum_{i,j = 1}^L C_{i,j}^{1,1} &= L\,C_{0}^{1,1} + 2\,(L-1)\,C_{1}^{1,1} \\ &+ 2\,(L-2)\,C_{2}^{1,1} + \cdots + 2\,C_{L-1}^{1,1}.
    \end{aligned}
\end{equation}

In the limit of decoupled 1D chains ($J_{\perp}/J=0$), $C_{i,j}^{1,2} = C_{i,j}^{2,1} = 0$ and the variance of the $2 \times L$ subsystem is just a sum of variances of the two $1 \times L$ subsystems.

While we cannot easily correct the real-space densities $\hat{n}_{\alpha,i}$ for the reconstruction artefact, we can correct the correlations (see Section~\ref{sec:corr_corr}) and therefore the atom number variances.

\subsection{Determination of the diffusion constant}
\label{sec:computing_diffusion_const}

For the fully coupled ladder ($J_{\perp}/J = 1$) -- a chaotic system -- we find that the dynamics is diffusive. In the diffusive limit,  for initial states $\rho$ that are diagonal in the number basis, the density-density correlator $C_{d}^{\alpha,\beta}$ takes the following form within MFT (see Section~\ref{supp_sec:hydrodynamics_deriv} for a detailed derivation):
\begin{equation}\label{density-density theory}
    C_{d}^{\alpha,\beta} = (\langle \text{Var}(\hat{n}_{\alpha, j}) \rangle_{\alpha, j} - \chi(\overline{n}))\frac{a\,e^{-\frac{d^2 a^2}{8Dt}}}{2\sqrt{8 \pi D t}} + \chi(\overline{n})\delta_{\alpha,\beta}\delta_{d,0}, 
\end{equation}
where $\overline{n}$ is the average occupation (filling) in the initial state, $\chi(n)=n(1-n)$ is the susceptibility, and $ \langle \text{Var}(\hat{n}_{\alpha, j}) \rangle_{\alpha, j}$ is the spatial average of the number variance $\text{Var}(\hat{n}_{\alpha, j})$ in the initial state. In this expression, $D$ denotes the equilibrium diffusion constant, $d$ is the distance along the ladder between lattice site $i$ in leg $\alpha$ and lattice site $j$ in leg $\beta$ in units of the lattice spacing $a$. The correlations as a function of distance and time form a cone of a Gaussian shape with width increasing over time as $\sigma_c(t) = \sqrt{4 D t / a^2}$.

In order to obtain the diffusion constant from our experimental data, we use a simplified form of the expression in Eq.~(\ref{density-density theory}) which includes an offset parameter $c$,

\begin{equation}
    C_{d}^{III} = \sum_{\alpha, \beta = 1,2} C_{d}^{\alpha,\beta} = A\frac{a\,e^{-\frac{d^2 a^2}{8Dt}}}{2\sqrt{8\pi D t}} + c
    \label{supp_eq:diffusion_fit_function}
\end{equation}
and fit the measured correlations $C_{d}^{{III}}$ for $1~\le~d~\le~20$ and all time data points with free parameters $A$, $D$ and $c$. The result of the fit, which yields $D = \ValueDiffusionConstant\,Ja^2/\hbar$, is shown in Fig. \ref{fig:cones_main}.

\subsection{Fitting the number variance time evolution}
\label{sec:supp_fitting_variance}

In order to characterize the variance growth, we fit all variance time evolution curves using the following empirical fit function:

\begin{align}
\mathrm{Var}(t, \mathrm{Var}_{\infty}, k, d) &= (\mathrm{Var}_{\infty} -\mathrm{Var}_0) \left( \frac{2}{1 + e^{-k t^d}}-1 \right) \nonumber \\ &+\mathrm{Var}_{0},
\label{eq:empirical}
\end{align}
where $\mathrm{Var}_{0} = \mathrm{Var}(t=0)$ and $\mathrm{Var}_{\infty} = \mathrm{Var}(t \to \infty)$. The free fit parameters are $\mathrm{Var}_{\infty}$, $d$ and $k$, while $\mathrm{Var}_{0}$ is fixed to the variance measured at $t=0$. Fig.~\ref{fig:full_variances} shows the atom number variance as a function of evolution time $t$ and subsystem size $L$ for $J_{\perp}/J\approx 0.0,0.5,1$. The solid lines are the fits using Eq.~(\ref{eq:empirical}(. For the thermalization of an infinite temperature state, the observed saturation values should equal $2L\bar{n}(1-\bar{n}) = L/2$ for an infinite system and perfect initial state with $\bar{n}=1/2$. However, due to residual errors after correcting systematic reconstruction errors (see Section~\ref{sec:corr_corr}), the presence of disorder (see Section~\ref{sec:supp_disorder}), finite size effects (see Section~\ref{sec:supp_finitesize}) and reduced filling $\bar{n}<1/2$, we expect the atom number variance to saturate at lower levels.

In order to quantify the timescale of the variance growth, we determine the intersection of the fit result with a threshold value of $80\%$ of the fitted saturation value (i.e. $0\%$ corresponds to $\mathrm{Var}_0$ and $100\%$ is equivalent to $\mathrm{Var}_{\infty}$), which defines the saturation time $t_{\mathrm{sat}}$. The uncertainty of $t_{\mathrm{sat}}$ is obtained by evaluating the inverse of Eq.~(\ref{eq:empirical}) at $t_{\mathrm{sat}}$, taking into account the uncertainties of all fitted variables. The saturation time $t_{\mathrm{sat}}$ and its scaling with increasing system size is plotted in Figs. \ref{fig:time_evolution}B,~C of the main text.


\begin{figure}[t!]
\includegraphics[bb = 0 0 2.95in 1.97in, width=1\columnwidth]{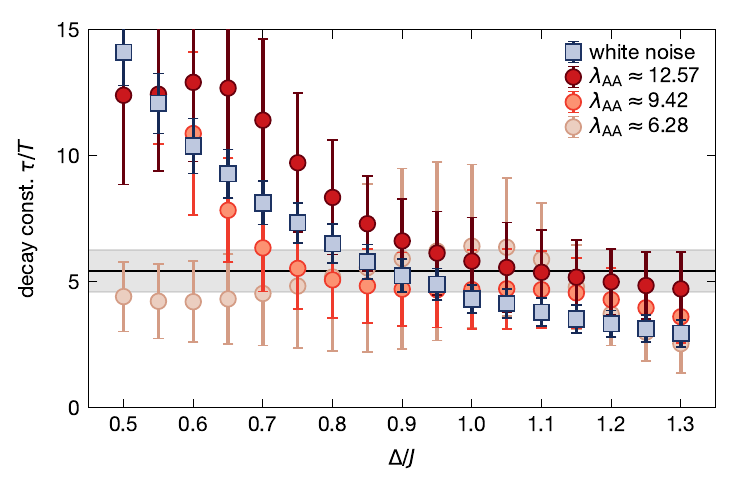}
    
     \caption{\textbf{Estimation of the disorder strength.} Result of an exponentially damped Bessel fit to the simulated imbalance oscillations for white noise disorder and Aubry-André disorder of three different periods $\lambda_{\mathrm{AA}}$ (in units of lattice sites) and different disorder amplitudes. The error bars are the fit error bars. The black line with the shaded region shows the experimentally measured damping constant $\tau$ and its uncertainty.}
     \label{fig:disorder_imbalance_fit}
\end{figure}

\section{Benchmarking of experimental imperfections}
\label{sec:supp_benchmarking}

\subsection{Free-fermion approach}
\label{sec:supp_freefermion_approach}

We have performed numerical simulations of the one dimensional chains to understand the role of experimental imperfections, such as disorder, quality of the initial state and finite system size effects. Hard-core bosons in one dimension can be mapped onto spinless fermions using the Jordan-Wigner transformation: It states that the bosonic creation and annihilation operators $\hat{a}^{\dagger}_i$ and $\hat{a}_i$ that fulfill the commutation relation $\left[ \hat{a}_i, \hat{a}_j^{\dagger} \right] = \delta_{i,j}$ can be replaced by corrresponding fermionic operators $\hat{c}^{\dagger}_i$ and $\hat{c}_i$ that fulfil $\left[ \hat{c}_i, \hat{c}_j^{\dagger} \right]_+ = \delta_{i,j}$. The resulting Hamiltonian reads
\begin{equation}
    \begin{aligned}
        \hat{H}_{\mathrm{ff}} = - J \left( \sum_{i} \hat{c}_{i}^{\dagger} \hat{c}_{i+1} + \mathrm{h.c.} \right) + \sum_{i} V_i \hat{n}_{i},
    \label{eq:free_fermion_hamiltonian}
    \end{aligned}
\end{equation}
where $V_i$ is the potential energy at site $i$. 
Note that the interaction term has dropped out.

For a perfectly homogeneous (i.e. disorder-free) infinite one-dimensional system and a defect-free initial state, the imbalance and the connected density-density correlator will evolve in time according to \cite{smith_disorder-free_2019}:

\begin{equation}
    \begin{aligned}
        \mathcal{I}(t) = \mathcal{J}_0\left( 4tJ/\hbar \right),
    \label{eq:perfect_imbalance}
    \end{aligned}
\end{equation}

\begin{equation}
    \begin{aligned}
        \left< \hat{n}_j(t) \hat{n}_k(t)\right>-\left< \hat{n}_j(t) \right> \left< \hat{n}_k(t)\right> = \frac{1}{4} \delta_{j,k} - \frac{1}{4}\mathcal{J}_{j-k}(4tJ/\hbar)^2
    \label{eq:perfect_two_point_correlator}
    \end{aligned}
\end{equation}
where $\delta_{j,k}$ is the Kronecker delta and $\mathcal{J}_{m}$ is the $m$th-order Bessel function of the first kind.

For non-zero on-site potential terms (such as disorder potential), an imperfect initial state and finite system size, we do not have such analytic predictions. Following the calculation outlined in \cite{smith_disorder-free_2019}, we diagonalize the free-fermion Hamiltonian $\hat{H}_{\mathrm{ff}}$ using a Fourier transform and obtain expressions for $\left< \hat{n}_j(t) \right>$ and $\left< \hat{n}_j(t) \hat{n}_k(t)\right>$ in terms of the time evolution operator $\hat{U}(t) = e^{-i \hat{H}_{\mathrm{ff}} t/\hbar}$. This operator is then computed numerically and gives us access to on-site occupations, two-point density-density correlators and therefore also atom number variances (see SM Section~\ref{sec:supp_densitydensitycorrl}).

\begin{figure}[t!]
\includegraphics[bb = 0 0 2.95in 1.97in, width=1\columnwidth]{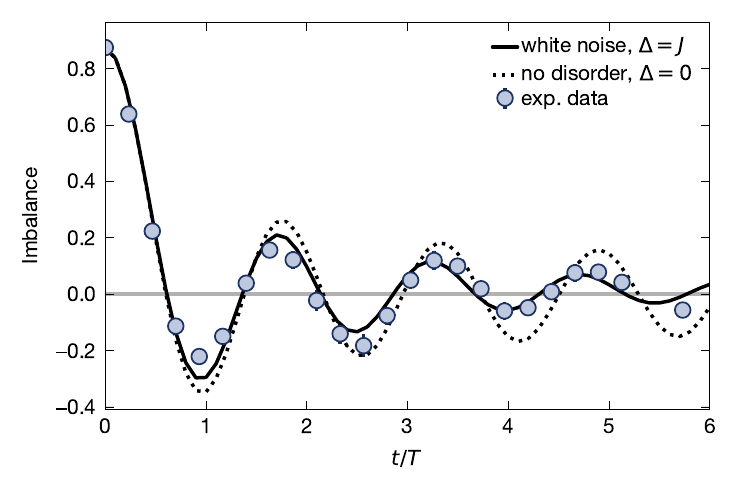}
    
     \caption{\textbf{Effect of disorder on the imbalance.} Free-fermion simulations of imbalance oscillation for systems with no disorder (dotted line) and white-noise disorder of amplitude $\Delta = J$ (solid line) for which we found reasonable agreement with the experimental data.}
     \label{fig:best_imbalance_fit}
\end{figure}

\subsection{Disorder and initial state quality}
\label{sec:supp_disorder}

As outlined in Section~\ref{sec:supp_flattening}, we remove the large-scale harmonic confinement using a potential compensation algorithm. Thus, we assume that remaining inhomogeneities are caused by disorder in the form of random potential corrugations. To investigate the impact of disorder on our experimental results, estimate the disorder strength and see whether it can explain our experimental observations, we compare our measurements for $J_{\perp}/J \approx 0.0$ with results obtained from one-dimensional free-fermion simulations (see Section~\ref{sec:supp_freefermion_approach}). We consider both white noise and Aubry-André-type disorder, exploring the impact of random (uncorrelated) and deterministic disorder potentials, respectively. For the white noise disorder model, $V_i$ in Eq.~(\ref{eq:free_fermion_hamiltonian}) is randomly sampled from the interval $\left[ -\Delta, \Delta \right]$. For Aubry-André disorder, $V_i = \Delta \sin\left(2 \pi i/\lambda_{\mathrm{AA}} + \phi \right)$, where $\lambda_{\mathrm{AA}}$ is the period of the periodic wave potential in units of lattice sites and $\phi$ is its phase that is randomly sampled from the interval $\left[ 0, 2\pi\right]$. In both cases $\Delta$ denotes the disorder amplitude. While it is unclear precisely what type of disorder governs our system, comparing the impact of the two disorder models allows us to obtain an estimate for $\Delta$, as shown below.

The free-fermion simulation is performed on a one-dimensional lattice of length $N = 50$ sites and with open boundary conditions. Just as in the experiment, we use a ROI of 40 sites for evaluation, leaving a buffer zone of $5$ sites between the edges of the ROI and the edge of the overall system. We compute the time evolution of 600 initial states that are randomly sampled using the experimentally measured fillings of the odd and even sites (see SM Section~\ref{sec:supp_setup}).

\begin{figure}[t!]
\includegraphics[bb = 0 0 3.0in 1.53in, width=1\columnwidth]{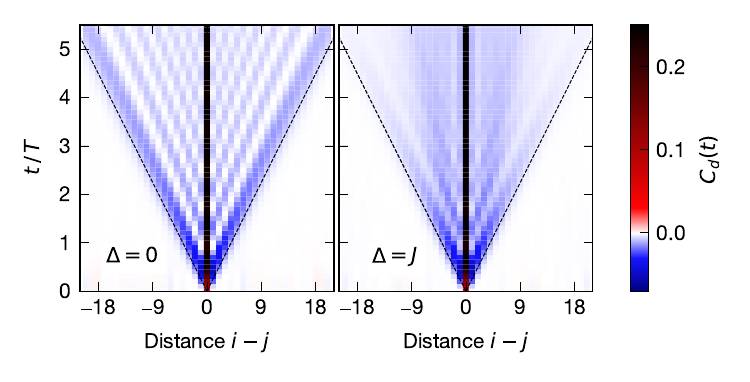}
    
     \caption{\textbf{Effect of disorder on correlation spreading.} Ballistic correlation cones calculated using the free-fermion simulation for an imperfect initial state and cases with no disorder ($\Delta = 0$) and white noise disorder $\Delta = J$. Disorder reduces the visibility of the outer arm of the cone at larger distances and leads to smearing of the correlations, hiding the internal structure of the correlation cone. The dashed line indicates the Lieb-Robinson velocity $4J a /\hbar$.}
     \label{fig:cones_benchmarking}
\end{figure}

In order to estimate the disorder strength in our experiment, we fit the decay constant $\tau$ of the simulated imbalance oscillations and compare it to the experimentally measured one (see Fig.~\ref{fig:time_evolution}A).
Fig.~\ref{fig:disorder_imbalance_fit} shows the fitted decay constants as a function of disorder strength for both white noise disorder and Aubry-André disorder of selected wavelengths. We find that the measured imbalance decay constant is consistent with disorder strength $\Delta \sim J$ with some dependence on the disorder model. Assuming white noise disorder of strength $\Delta = J$, we find good agreement with the experimental imbalance oscillations, as highlighted in   Fig.~\ref{fig:best_imbalance_fit}. Fig.~\ref{fig:cones_benchmarking} shows how the presence of disorder reduces the contrast of the correlation cone, hiding its inner structure, and reducing the visibility of the outer arm. The spreading velocity, as indicated by the outer edge, however remains unaffected.

\begin{figure}[t!]
\includegraphics[bb = 0 0 2.95in 1.97in, width=1\columnwidth]{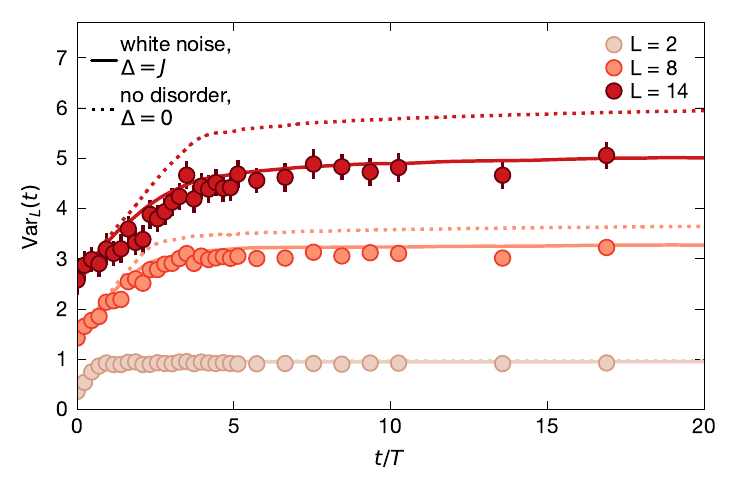}
    
     \caption{\textbf{Effect of disorder on the variance growth.} Experimentally measured atom number variances in subsystems of size $L = 2, 8$ and $14$ and respective free-fermion predictions that have been calculated from the correlation cones shown in Fig.~\ref{fig:cones_benchmarking}. The solid lines show the white noise $\Delta = J$ prediction, while the dotted lines show the zero disorder prediction. Disorder affects the dynamics for $t \gtrapprox 2T$, effectively reducing the variance growth and saturation value. Variance at $t=0$ is non-zero due to the initial state quality.} 
     \label{fig:variance_benchmarking}
\end{figure}

Fig.~\ref{fig:variance_benchmarking} shows that white noise disorder of amplitude $\Delta \approx J$, independently benchmarked using the imbalance oscillations, leads to a very good agreement with the experimentally measured atom number variances.

Using the free-fermion simulation and the approach outlined in \cite{karamlou_quantum_2022}, we can estimate the Anderson localization length corresponding to our white noise disorder strength. For a disorder amplitude of $\Delta \approx J$, we obtain the localization length of approximately $18$ lattice sites, which is larger than the subsystem sizes that we use for studying fluctuations in the ballistic case ($J_{\perp}/J = 0$).

\subsection{Impact of finite system size effects}
\label{sec:supp_finitesize}

In this section we investigate the impact of finite system size effects, in particular the impact on the saturation value of the atom number variance, using both simulations and experimental data.

\subsubsection{Experimental data}

\begin{figure}[t!]
\includegraphics[bb = 0 0 2.95in 3.94in, width=1\columnwidth]{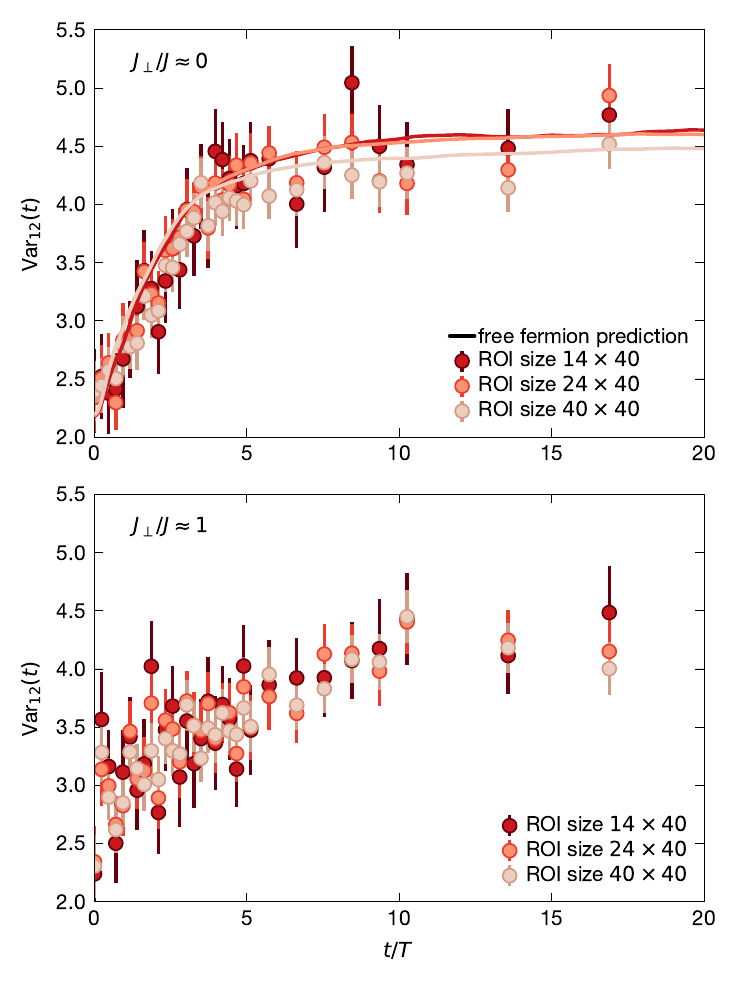}
    
     \caption{\textbf{Edge effects for $J_{\perp}/J \approx 0.0, 1.0$.} Variance growth for subsystem size $L = 12$ evaluated for three different ROI sizes: $12 \times 40$, $20 \times 40$ and $40 \times 40$ sites (full ROI, see Fig.~\ref{fig:experiment}). Solid lines show the free-fermion prediction for $J_{\perp}/J \approx 0$ with white noise disorder $\Delta = J$ evaluated for the same ROI sizes and overall system size 50.}
     \label{fig:edge_effects0}
\end{figure}

In order to ensure that the choice of our $40\times40$ sites ROI inside the full $50\times50$ system does not introduce significant finite size effects, we compare the data evaluation in central ROIs of three different sizes. Fig.~\ref{fig:edge_effects0} shows the time evolution of the atom number variance in a subsystem of size $L = 12$ for the three different ROIs,  $14 \times 40$, $24 \times 40$ and $40 \times 40$ sites for $J_{\perp}/J \approx 0.0$ and $1.0$. In both cases, the three curves for different ROI sizes agree within error bars. Thus, we conclude that the reduction of the variance saturation value compared to the infinite-temperature expectation is not a consequence of of our large ROI, which we have chosen for the sake of better statistics.

\subsubsection{Simulations of decoupled 1D chains}

\begin{figure}[t!]
\includegraphics[bb = 0 0 2.95in 1.97in, width=1\columnwidth]{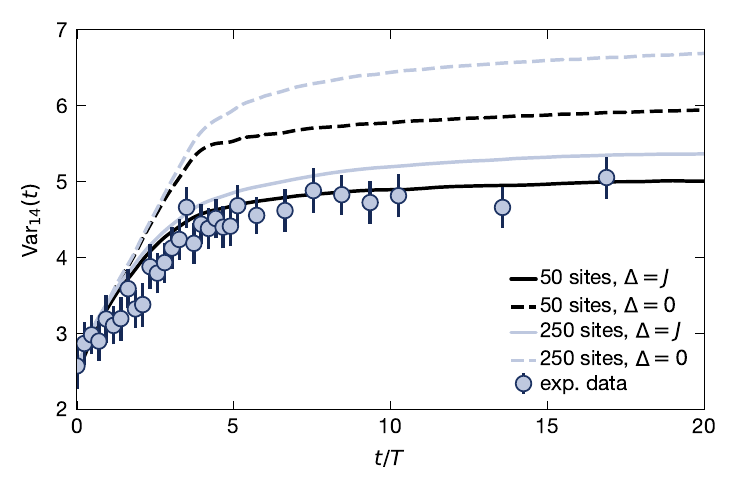}
    
     \caption{\textbf{Combined effect of finite system size and disorder for $J_{\perp}/J=0$.} The lines show the results of a free-fermion simulation for system sizes 50 and 250 evaluated in a central ROI of 40 sites, both for no disorder and for white noise disorder of amplitude $\Delta = J$. Note that in the large system of 250 sites (that is not affected by finite system size effects for the times shown) the variance grows almost up to the theory prediction $\approx L/2$. Both finite system size effects and disorder lead to reduction in the variance growth and smaller saturation value.}
     \label{fig:edge_effects_simulation}
\end{figure}

We perform a free-fermion simulation for system sizes 50 and 250 for cases with no disorder and disorder of amplitude $\Delta = J$, evaluated in a central ROI of 40 sites. Fig.~\ref{fig:edge_effects_simulation} shows the time evolution of the atom number variance in a subsystem of size $L=14$. Information about the edges propagates towards the center ballistically with Lieb-Robinson velocity $4Ja/\hbar$ and ROI of size 40 inside a system of size 250 is therefore not affected by finite system size effects for the times shown in Fig.~\ref{fig:edge_effects_simulation} ($t/T<20$). We can see that in a clean system, that is not affected by finite size effects, the variance saturation value is reaching the theoretical prediction $\approx L/2$.

For system size of 50 sites (realized in the experiment) we observe a clear reduction of the saturation value due to finite size effects, both in the clean system and in the presence of disorder with an amplitude comparable to our experiment ($\Delta \sim J$). The combined effect of disorder and finite system size then leads to good agreement with the experimentally-measured atom number variances. In contrast to the subsystem variances, the local imbalance is not significantly affected by finite size effects.

\subsubsection{Simulations of fully coupled ladders}

\begin{figure}[t!]
\includegraphics[bb = 0 0 2.95in 1.60in, width=1\columnwidth]{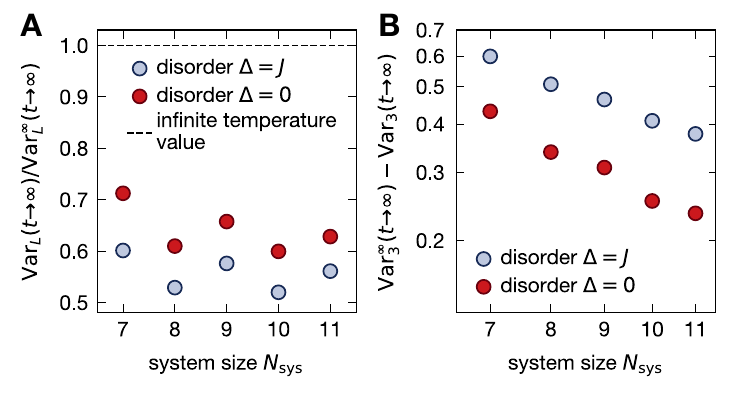}
    
     \caption{\textbf{Fully coupled ladder small scale numerics.} The saturation value of fluctuations in simulations of small systems ($7 \leq N_\mathrm{sys} \leq 11$). $\mathrm{Var}^\infty_{L}(t)$ denotes the infinite temperature prediction. (A) The (normalized) saturation value of the half system number variance for different total system sizes, showing a robust suppression of fluctuations as system size is increased. (B) The deviation of saturation value for the number variance from the infinite temperature value for a subsystem of size $L=3$ at different total system sizes $7 \leq N_\mathrm{sys} \leq 11$.}
     \label{fig:finite-size-ladder}
\end{figure}

To investigate the effects of disorder and finite system size in the fully coupled ladder ($J_{\perp}/J = 1$), we numerically study this system (using MPS-TEBD simulations using the open-source package TeNPy \cite{hauschild_efficient_2018}) for different system sizes. The rapidly entangling evolution of the fully coupled ladder limits us to small system sizes, with the number of rungs, $N_{\text{sys}}$, at most $N_{\text{sys}}=11$ (i.e. $22$ sites). In both the clean and disordered system, we observe that large scale fluctuations (fluctuations of observables of a large fraction of the system) fail to reach the infinite temperature value. In Fig.~\ref{fig:finite-size-ladder}A we show the number variance for subsystem size equal to half the total system size $L=\left \lfloor {N_{\text{sys}}/2}\right \rfloor$. This shows the half-system number variance remaining strongly suppressed from the thermal value as system size is increased, even as expectation values thermalize.

If the opposite limit is taken, $N_{\text{sys}}\to\infty$ with fixed subsystem size, we find evidence that the fluctuations do thermalize, as expected. Fig.~\ref{fig:finite-size-ladder}B shows the deviation in saturation values from the infinite temperature value ($L/2$) for subsystem size $L = 3$ for different total system size. We see, in both the clean and disordered case, a power law approach to the infinite temperature value as system size increases, with the disordered case having a more pronounced suppression.

In the experiment, we observe the saturation of fluctuations in subsystems up to $1/5$ of the total system size, and find a $\sim 20\%$ suppression of the number variance for these subsystem sizes (see Fig.~\ref{fig:full_variances}). This is comparable to the reduction seen in the small scale numerics for similarly sized subsystems (in proportion to the total system). For example, taking a subsystem of $3$ rungs in a total system of $11$ rungs, we find a reduction of $\sim 25\% $. We thus note that the finite size effects can be relevant even in the case of fully coupled ladder with diffusively spreading correlations.

\subsection{Experimentally measured inhomogeneities}

\begin{figure}[t!]
\includegraphics[width=1\columnwidth]{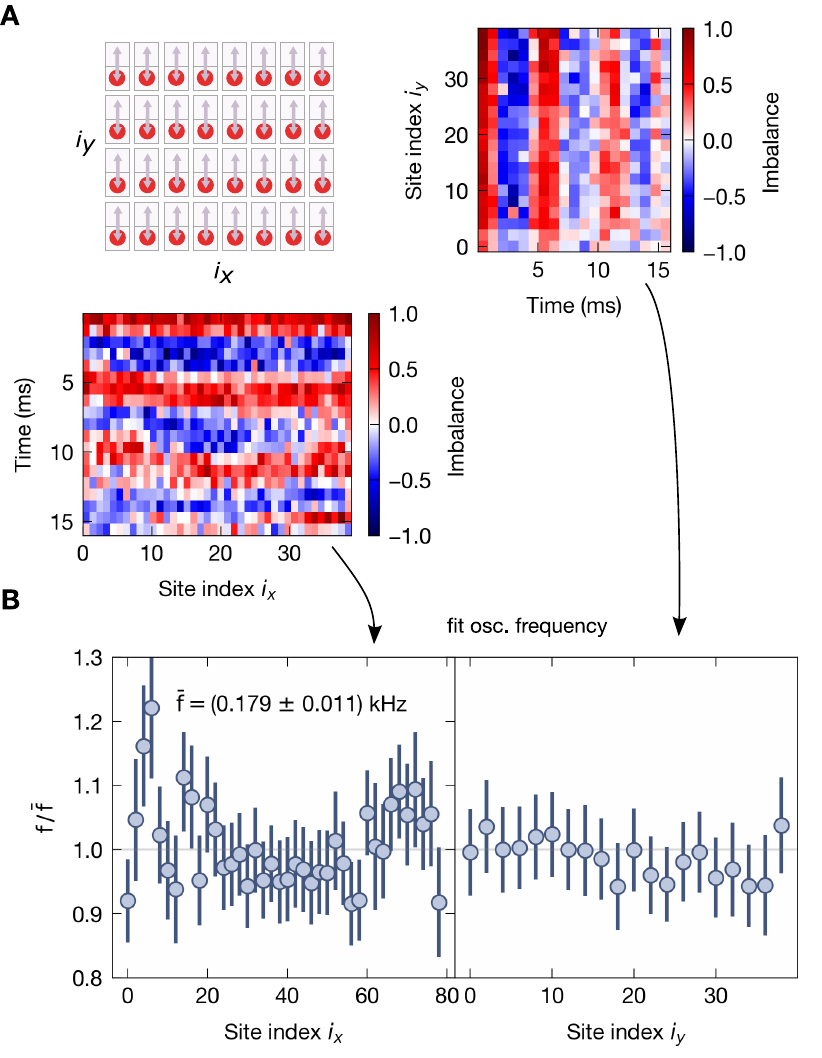}
    
     \caption{\textbf{Spatially resolved double well oscillations.} (A) In order to quantify the homogeneity of the superlattice potential that we use to realize the ladder systems, we observe oscillations in isolated double wells. The local time evolution of the imbalance is averaged along either the $x$- or the $y$-direction (i.e. along or perpendicular to the ladder, respectively, see \ref{fig:experiment}A for definition of $x$ and $y$) and (B)  fitted using a sine function to obtain the double well oscillation frequency as a function of position.}
     \label{fig:tunnosc}
\end{figure}

To characterize inhomogeneities in our superlattice potential that we use to generate the ladder systems, we perform double well oscillations by adiabatically preparing a CDW with each atom being localized in an isolated double well. Then, we quench on the tunnelling strength between the two sites of each double well and observe the atom oscillating between the two sites. Due to inhomogeneities inside our ROI, some double wells will be more tilted than others, locally varying the double well oscillation frequency. By measuring the spatial variation of this frequency in our ROI, we get a spatial profile of double well tilts in the superlattice potential. Fig.~{\ref{fig:tunnosc}A} shows the imbalance as a function of time, averaged along columns and rows of our square superlattice. The imbalance can be seen oscillating at an average frequency of $\bar{f}=\SI{179(11)}{\hertz}$ with variations on the scale of $10\%$ indicating double well tilts up to $\SI{30}{\hertz}$.

\subsection{Atom loss}
\label{supp_sec:atom_loss}

\begin{figure}[t!]
\includegraphics[bb = 0 0 3.13in 2.03in, width=1\columnwidth]{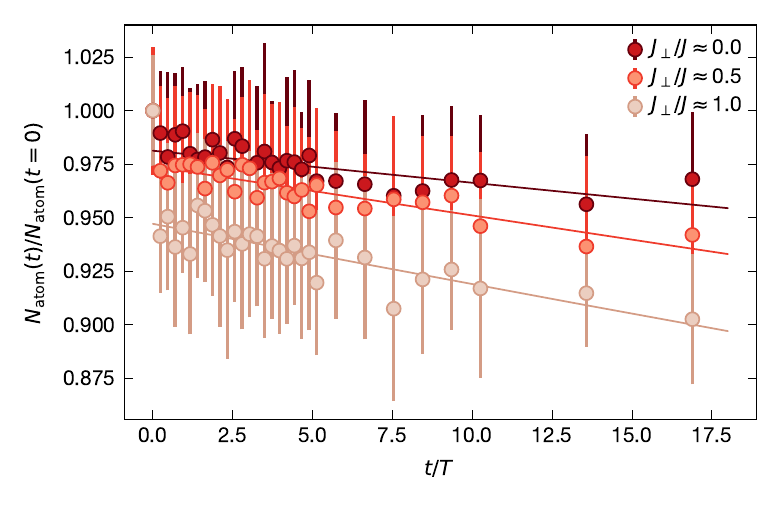}
    
     \caption{\textbf{Atom loss during the time evolution after the quench.} Fraction of remaining atoms (after parity projection) as a function of time, normalized to the initial state value at $t=0$. The initial drop in the atom number for the case of fully coupled ladder ($J_{\perp}/J\approx1.0$) can be attributed to doublon generation.}
     \label{fig:atomloss}
\end{figure}

During the time evolution we experience atom loss due to heating of the atoms in the optical lattice and due to the finite lifetime in the vacuum. Further, owing to the finite interaction energy $U$, there is a small doublon fraction $\lesssim 3\,\%$ which gets parity projected during fluorescence imaging and reduces the number of visible atoms. The latter effect is enhanced in the ladder case ($J_{\perp}/J \approx 1.0$), where $U/J$ is smaller (but still in the hard-core limit). As shown in Fig.~\ref{fig:atomloss}, the total atom loss stays below $10\,\%$ during the entire time evolution and does not have a significant impact on our results. 

\section{Fluctuating hydrodynamics of diffusive quantum systems}
\label{supp_sec:hydrodynamics}
In this section, we discuss how classical fluctuating hydrodynamics emerges in chaotic diffusive quantum systems through the lens of operator spreading, following Ref.~\cite{von_keyserlingk_operator_2022}. In particular, we derive hydrodynamic formulas for the equal-time correlations and the atom number transfer variance after a quench from a fully polarized charge density wave state. We also adapt this to imperfect initial states.

We assume that the dynamics of particle density is decoupled from energy density and that the system has a fixed diffusion constant, which we will take to be the diffusion constant of fully coupled ladder at half filling, $D=D(1/2)$. This assumption is necessary as we do not have knowledge of the full density (or temperature) dependence of $D(n)$, but reasonable, as: (1) the initial CDW state rapidly equilibrates to a constant density on average ($\overline{n}=1/2$), and (2) the diffusion constant is both analytic and weakly dependent on density and temperature around the half-filled infinite temperature state \cite{steinigeweg_2014}. Alternatively, MFT allows us to study classical stochastic models with fixed diffusion constant, such as a symmetric exclusion process. Such a stochastic model will share the same fluctuating hydrodynamics, as was shown for unitary circuit models with a conserved density \cite{mcculloch_full_2023}. Indeed, the following calculation can be adapted for classical stochastic processes, yielding the same result for the growth of fluctuations (Eq.~(\ref{eq:diffusion_fit_var})).

\subsection{Equal-time density-density correlations} \label{supp_sec:hydrodynamics_deriv}

In this section, we provide a theory for density-density correlations in diffusive systems at half-filling shown in Eq.~(\ref{density-density theory}), which includes the CDW initial state and the infinite temperature state. We will find it helpful to use a short-hand notation $\xi = (\alpha, i)$ to label both the leg $\alpha$ and rung $i$ of the ladder as well as defining charge operators $\hat{z}_{\xi}=2 \hat{n}_{\xi} - \mathds{1}$, which measure the deviation in the occupancy from half-filling on a given site. For a fully coupled ladder, charge will spread diffusively, quickly equilibrating between the legs, while longer wavelength hydrodynamic modes will describe the dynamics of charge along the extended direction (along the legs of the ladder) -- charge will spread with a width $\sqrt{2Dt/a^2}$. For concreteness, we assume a Gaussian kernel for the spreading of charge,
\begin{equation}
    \text{Tr}(\hat{z}_\xi(t) \hat{z}_\eta)/\text{Tr}(\mathds{1}) = K_{\xi,\eta}(t) \equiv \frac{1}{2}\frac{a\,e^{-\frac{(i-j)^2 a^2}{4Dt}}}{\sqrt{4 \pi D t}},
\end{equation}
where $a$ is the lattice spacing, and where the factor of $1/2$ is due to the ladder geometry (charge can be on either site of a rung). Instead of working directly with the MFT equation, we use the equivalence between MFT and the Heisenberg dynamics of charge densities in conserving circuits~\cite{von_keyserlingk_operator_2022, mcculloch_full_2023}. Under this dynamics, 
the charges can also develop overlap with products of many local charge densities, as well as non-diagonal (and therefore non-hydrodynamic) operators \cite{von_keyserlingk_operator_2022}. These higher-order terms can be regarded either through the perspective of Heisenberg operator growth, or as consequences of integrating over the noise. The Heisenberg-evolved density $\hat z_\xi(t)$ admits the operator expansion
\begin{align}\label{z-spreading}
    \hat{z}_\xi(t) = &\sum_{\xi'}K_{\xi,\xi'}(t)\hat{z}_{\xi'}\nonumber\\
    &+ \sum_{n\geq 2}\sum_{\xi_1< \cdots< \xi_n}B^{(n)}_{\xi,(\xi_1,\cdots \xi_n)}(t)\hat{z}_{\xi_1}\cdots \hat{z}_{\xi_n} + \cdots.
\end{align}
The coefficients $B^{(n)}(t)$ decay quicker than $K(t)$ (the coefficients $B^{(n)}(t)$ decay as $t^{-3/2}$ in $U(1)$ random unitary circuits \cite{von_keyserlingk_operator_2022}), and where the ellipsis denotes non-diagonal terms. This describes the diffusive dynamics of a single charge operator. As the density-density two-point function involves a product of charges, we must also write a similar expression for the Heisenberg evolution of a pair of charge operators $\hat{z}_{\xi}(t)\hat{z}_{\eta}(t)$, $\xi\neq \eta$. Showing only the slowest hydrodynamic mode, $zz\to zz$, in which the two charges undergo single file diffusion (and do not generate or annihilate further charges), we have
\begin{equation}
    \hat{z}_\xi(t) \hat{z}_\eta(t) = \sum_{\xi',\eta'}K_{(\xi,\eta),(\xi',\eta')}(t) \hat{z}_{\xi'} \hat{z}_{\eta'} + \cdots.
\end{equation}
The kernel $K_{(\xi,\eta),(\xi',\eta')}(t)$ must vanish at $\xi'=\eta'$ as much like a single charge, products of charge operators can never develop overlap with the identity operator. For diagonal (in number basis) $\rho$, all non-diagonal terms in the operator expansion vanish. By neglecting slower hydrodynamic contributions, which involve transmuting one $z$-string to another by the creation or annihilation of $z$'s, we are assuming a linearized hydrodynamic description. Under this assumption, the connected equal-time charge-charge correlator (with $\xi\neq\eta$) is given by
\begin{align}
    \left< \hat{z}_{\xi} \hat{z}_{\eta} \right>^C_{\rho(t)} &\equiv \left< \hat{z}_{\xi} \hat{z}_{\eta} \right>_{\rho(t)} - \left< \hat{z}_{\xi}  \right>_{\rho(t)}\left< \hat{z}_{\eta} \right>_{\rho(t)} \nonumber \\
    &= \sum_{\xi,\eta} \left( K_{(\xi,\eta),(\xi',\eta')}(t) \left<\hat{z}_{\xi'} \hat{z}_{\eta'}\right>_{\rho}\right.\nonumber\\
    &\quad \left.- K_{\xi,\xi'}(t)K_{\eta,\eta'}(t) \left<\hat{z}_{\xi'}\right>_{\rho_0}\left<\hat{z}_{\eta'}\right>_{\rho_0}\right).
\end{align}
The $zz\to zz$ kernel decouples at late times (the contact interactions of random walkers is RG irrelevant \cite{mcculloch_full_2023}, leading to a faster decay $\delta K \sim t^{-2}$),
\begin{align}
    K_{(\xi,\eta),(\xi',\eta')}(t) = &\ \frac{1}{2}\left(K_{\xi,\xi'}(t)K_{\eta,\eta'}(t)+ K_{\xi,\eta'}(t)K_{\eta,\xi'}(t)\right)\nonumber\\
    & + \delta K_{(\xi,\eta),(\xi',\eta')}(t).
\end{align}
Using this, and the fact that $\left< \hat{z}_{\xi'} \hat{z}_{\eta'} \right>_{\rho} = \left< \hat{z}_{\xi'} \right>_{\rho}\left< \hat{z}_{\eta'} \right>_{\rho}$ for $\xi'\neq \eta'$, we write
\begin{align}
    \left< \hat{z}_{\xi} \hat{z}_{\eta} \right>^C_{\rho(t)} = & \sum_{\xi'} K_{\xi,\xi'}(t)K_{\eta,\xi'}(t) (1-\left<\hat{z}_{\xi'}\right>_{\rho}^2)\nonumber\\
    &+ \sum_{\xi',\eta'} \delta K_{(\xi,\eta),(\xi',\eta')}(t) \left<\hat{z}_{\xi'} \hat{z}_{\eta'}\right>_{\rho}.\label{abc}
\end{align}
The correction $\delta K$ to the $zz\to zz$ kernel results from scattering events of two `hardcore' random walkers. For a scattering of $z$'s with initial positions $\xi$ and
$\eta$ and final positions $\xi'$ and $\eta'$ to occur, the $z$'s must diffuse until next to one another, scatter, and then diffuse to their final positions. Such a process is 
exponentially suppressed in the (scaled) separations $(i-j)/\sqrt{Dt/a^2}$, $(i'-j')/\sqrt{Dt/a^2}$, and the (scaled) center-of-mass
separation $(\frac{i+j}{2}-\frac{i'+j'}{2})/\sqrt{Dt/a^2}$. For states with density variations on length scales much smaller than $\sqrt{Dt/a^2}$, $\delta K$ and $K$ vary much more slowly
than $\left<\hat{z}_{\xi'}\right>_{\rho}$ and
$\left<\hat{z}_{\xi'}\right>^2_{\rho}$. We are therefore able to replace these by their spatial averages, 
$\left<\hat{z}_{\xi'}\right>_{\rho}\to 2\overline{n}-1$ and $\left<\hat{z}_{\xi'}\right>^2_{\rho}\to 1-4\langle\text{Var}(\hat{n}_{\xi})\rangle_\xi$. Putting this into Eq.~(\ref{abc}) yields
\begin{align}
    \left< \hat{z}_{\xi} \hat{z}_{\eta} \right>^C_{\rho(t)} &= \sum_{\xi'} 4\langle\text{Var}(\hat{n}_{\xi})\rangle_\xi K_{\xi,\xi'}(t),K_{\eta,\eta'}(t)\nonumber\\
    + \sum_{\xi', \eta'} &\delta K_{(\xi,\eta),(\xi',\eta')}(t) (\delta_{\xi',\eta'} + \delta_{\xi'\neq \eta'}(2\overline{n}-1)^2).
\end{align}
Together with the sum rules for $K_{(\xi,\eta),(\xi',\eta')}(t)$ and $K_{\xi,\xi'}(t)$, the condition that $K_{(\xi,\eta),(\xi',\xi')}(t)=0$ yields $\delta K_{(\xi,\eta),(\xi',\xi')}(t) = - K_{\xi,\xi'}(t)K_{\eta,\xi'}(t)$. Altogether, we find
\begin{equation}
    \left< \hat{z}_{\xi} \hat{z}_{\eta} \right>^C_{\rho(t)} = 4(\langle\text{Var}(\hat{n}_{\xi})\rangle_\xi - \chi(\overline{n})) K_{\xi,\eta}(2t) + 4\chi(\overline{n})\delta_{\xi,\eta},
\end{equation}
where we fixed the value of the correlator at $\xi=\eta$ using similar arguments to those presented in the above for $\xi\neq\eta$. Converting back to number operators $\hat{n}_{\alpha,i}$, the equal-time density-density correlator is given by Eq.~(\ref{density-density theory}) as required.

\subsection{Counting statistics of number transfer} \label{sec:number_transfer}

The number of atoms $\mathcal{N}$ transferred in and out of an ROI labeled $S$ is a stochastic variable, its variance can be related to the two-time and equal-time density-density correlators (see SM Section \ref{sec:supp_densitydensitycorrl}) as follows,
\begin{align}
    \text{Var}(\mathcal{N}(t)) =
    &\sum_{(\alpha, i), (\beta, j)\in S}\left[ C^{\alpha,\beta}_{i,j}(t) - 2\langle \hat{n}_{\alpha, i}(t) \hat{n}_{\beta, j}(0) \rangle\right.\nonumber\\
    &\quad \left. + 2\langle \hat{n}_{\alpha, i}(t)\rangle\langle \hat{n}_{\beta, j}(0) \rangle + C^{\alpha,\beta}_{i,j}(0)\right].
\end{align}
For an initial state with a sharp number of atoms in the ROI, the connected two-time functions are zero, enabling us to reconstruct the FCS from the equal-time correlators alone. Using Eq.~(\ref{density-density theory}) for the density-density correlator, we find the following prediction for the atom transfer variance across the central bond of the ladder for a fully polarized charge density wave state (in which all sites on even bonds are fully occupied and all sites on odd bonds are empty),
\begin{equation}
\text{Var}(\mathcal{N}(t))_{\text{CDW}} \approx \sqrt{\frac{Dt}{2\pi\,a^2}}.
\end{equation}
We have checked that this formula also holds for classical stochastic systems like the simple symmetric exclusion process~\cite{derrida_current_2009,mallick_exact_2022}.
This is easily adapted for finite subsystems, which have two boundaries through which atoms can transfer. For subsystem of length $L$, the variance of atom number transfer grows at twice the rate of subsystem with a single boundary until fluctuations equilibrate to the infinite temperature half-filling value ($\text{Var}(\mathcal{N}_L)\to L/2$) at times $t \sim (La)^2/D$,
\begin{equation}
    \text{Var}(\mathcal{N}_L(t))_{\text{CDW}} \approx \sqrt{\frac{2Dt}{\pi\,a^2}},\quad t \ll (La)^2/D.
\end{equation}
In the absence of two-time correlation data, and with initial state imperfections (leading to a loss of sharpness in atom number), we must modify this and instead look at the growth of variance of subsystem atom number, $\text{Var}_L(t)-\text{Var}_L(0)$. Using Eq.~(\ref{density-density theory}) for the equal-time correlators, we find
\begin{equation}
\label{eq:diffusion_fit_var}
\text{Var}_L(t)-\text{Var}_L(0) \approx 4(\chi - \langle\text{Var}(\hat{n}_{\alpha,i})\rangle_{\alpha,i})\sqrt{\frac{2Dt}{\pi\,a^2}},
\end{equation}
for times $t\ll (La)^2/D$, and where $\chi= \overline{n}\,(1-\overline{n}) = 1/4$ is the susceptibility at half filling ($\overline{n}=0.5$), and $\langle\text{Var}(\hat{n}_{\alpha, i})\rangle_{\alpha, i}$ is the number variance on a single site in the initial state, averaged over sites. We use Eq.~(\ref{eq:diffusion_fit_var}) to fit the experimental data in Fig.~\ref{fig:time_evolution}D with free fit parameters $D$ and $\text{Var}_L(0)$.

\vspace{2em}
\begin{center}
\textbf{SUPPLEMENTARY REFERENCES}
\end{center}
\vspace{0.5em}

\putbib[manuscript]
\end{bibunit}


\begin{thebibliography}{67}%
\makeatletter
\providecommand \@ifxundefined [1]{%
 \@ifx{#1\undefined}
}%
\providecommand \@ifnum [1]{%
 \ifnum #1\expandafter \@firstoftwo
 \else \expandafter \@secondoftwo
 \fi
}%
\providecommand \@ifx [1]{%
 \ifx #1\expandafter \@firstoftwo
 \else \expandafter \@secondoftwo
 \fi
}%
\providecommand \natexlab [1]{#1}%
\providecommand \enquote  [1]{``#1''}%
\providecommand \bibnamefont  [1]{#1}%
\providecommand \bibfnamefont [1]{#1}%
\providecommand \citenamefont [1]{#1}%
\providecommand \href@noop [0]{\@secondoftwo}%
\providecommand \href [0]{\begingroup \@sanitize@url \@href}%
\providecommand \@href[1]{\@@startlink{#1}\@@href}%
\providecommand \@@href[1]{\endgroup#1\@@endlink}%
\providecommand \@sanitize@url [0]{\catcode `\\12\catcode `\$12\catcode
  `\&12\catcode `\#12\catcode `\^12\catcode `\_12\catcode `\%12\relax}%
\providecommand \@@startlink[1]{}%
\providecommand \@@endlink[0]{}%
\providecommand \url  [0]{\begingroup\@sanitize@url \@url }%
\providecommand \@url [1]{\endgroup\@href {#1}{\urlprefix }}%
\providecommand \urlprefix  [0]{URL }%
\providecommand \Eprint [0]{\href }%
\providecommand \doibase [0]{https://doi.org/}%
\providecommand \selectlanguage [0]{\@gobble}%
\providecommand \bibinfo  [0]{\@secondoftwo}%
\providecommand \bibfield  [0]{\@secondoftwo}%
\providecommand \translation [1]{[#1]}%
\providecommand \BibitemOpen [0]{}%
\providecommand \bibitemStop [0]{}%
\providecommand \bibitemNoStop [0]{.\EOS\space}%
\providecommand \EOS [0]{\spacefactor3000\relax}%
\providecommand \BibitemShut  [1]{\csname bibitem#1\endcsname}%
\let\auto@bib@innerbib\@empty
\bibitem [{\citenamefont {Nandkishore}\ and\ \citenamefont
  {Huse}(2015)}]{nandkishore_many-body_2015}%
  \BibitemOpen
  \bibfield  {author} {\bibinfo {author} {\bibfnamefont {R.}~\bibnamefont
  {Nandkishore}}\ and\ \bibinfo {author} {\bibfnamefont {D.~A.}\ \bibnamefont
  {Huse}},\ }\href {https://doi.org/10.1146/annurev-conmatphys-031214-014726}
  {\bibfield  {journal} {\bibinfo  {journal} {Annu. Rev. Condens. Matter
  Phys.}\ }\textbf {\bibinfo {volume} {6}},\ \bibinfo {pages} {15} (\bibinfo
  {year} {2015})}\BibitemShut {NoStop}%
\bibitem [{\citenamefont {D'Alessio}\ \emph {et~al.}(2016)\citenamefont
  {D'Alessio}, \citenamefont {Kafri}, \citenamefont {Polkovnikov},\ and\
  \citenamefont {Rigol}}]{dalessio_quantum_2016}%
  \BibitemOpen
  \bibfield  {author} {\bibinfo {author} {\bibfnamefont {L.}~\bibnamefont
  {D'Alessio}}, \bibinfo {author} {\bibfnamefont {Y.}~\bibnamefont {Kafri}},
  \bibinfo {author} {\bibfnamefont {A.}~\bibnamefont {Polkovnikov}},\ and\
  \bibinfo {author} {\bibfnamefont {M.}~\bibnamefont {Rigol}},\ }\href
  {https://doi.org/10.1080/00018732.2016.1198134} {\bibfield  {journal}
  {\bibinfo  {journal} {Adv. Phys.}\ }\textbf {\bibinfo {volume} {65}},\
  \bibinfo {pages} {239} (\bibinfo {year} {2016})}\BibitemShut {NoStop}%
\bibitem [{\citenamefont {Deutsch}(1991)}]{deutsch_quantum_1991}%
  \BibitemOpen
  \bibfield  {author} {\bibinfo {author} {\bibfnamefont {J.~M.}\ \bibnamefont
  {Deutsch}},\ }\href {https://doi.org/10.1103/PhysRevA.43.2046} {\bibfield
  {journal} {\bibinfo  {journal} {Phys. Rev. A}\ }\textbf {\bibinfo {volume}
  {43}},\ \bibinfo {pages} {2046} (\bibinfo {year} {1991})}\BibitemShut
  {NoStop}%
\bibitem [{\citenamefont {Srednicki}(1994)}]{srednicki_chaos_1994}%
  \BibitemOpen
  \bibfield  {author} {\bibinfo {author} {\bibfnamefont {M.}~\bibnamefont
  {Srednicki}},\ }\href {https://doi.org/10.1103/PhysRevE.50.888} {\bibfield
  {journal} {\bibinfo  {journal} {Phys. Rev. E}\ }\textbf {\bibinfo {volume}
  {50}},\ \bibinfo {pages} {888} (\bibinfo {year} {1994})}\BibitemShut
  {NoStop}%
\bibitem [{\citenamefont {Rigol}\ \emph {et~al.}(2008)\citenamefont {Rigol},
  \citenamefont {Dunjko},\ and\ \citenamefont
  {Olshanii}}]{rigol_thermalization_2008}%
  \BibitemOpen
  \bibfield  {author} {\bibinfo {author} {\bibfnamefont {M.}~\bibnamefont
  {Rigol}}, \bibinfo {author} {\bibfnamefont {V.}~\bibnamefont {Dunjko}},\ and\
  \bibinfo {author} {\bibfnamefont {M.}~\bibnamefont {Olshanii}},\ }\href
  {https://doi.org/10.1038/nature06838} {\bibfield  {journal} {\bibinfo
  {journal} {Nature}\ }\textbf {\bibinfo {volume} {452}},\ \bibinfo {pages}
  {854} (\bibinfo {year} {2008})}\BibitemShut {NoStop}%
\bibitem [{\citenamefont {Lux}\ \emph {et~al.}(2014)\citenamefont {Lux},
  \citenamefont {Müller}, \citenamefont {Mitra},\ and\ \citenamefont
  {Rosch}}]{lux_hydrodynamic_2014}%
  \BibitemOpen
  \bibfield  {author} {\bibinfo {author} {\bibfnamefont {J.}~\bibnamefont
  {Lux}}, \bibinfo {author} {\bibfnamefont {J.}~\bibnamefont {Müller}},
  \bibinfo {author} {\bibfnamefont {A.}~\bibnamefont {Mitra}},\ and\ \bibinfo
  {author} {\bibfnamefont {A.}~\bibnamefont {Rosch}},\ }\href
  {https://doi.org/10.1103/PhysRevA.89.053608} {\bibfield  {journal} {\bibinfo
  {journal} {Phys. Rev. A}\ }\textbf {\bibinfo {volume} {89}},\ \bibinfo
  {pages} {053608} (\bibinfo {year} {2014})}\BibitemShut {NoStop}%
\bibitem [{\citenamefont {Spohn}(1991)}]{spohn_large_1991}%
  \BibitemOpen
  \bibfield  {author} {\bibinfo {author} {\bibfnamefont {H.}~\bibnamefont
  {Spohn}},\ }\href {http://link.springer.com/10.1007/978-3-642-84371-6}
  {{\selectlanguage {english}\emph {\bibinfo {title} {Large {Scale} {Dynamics}
  of {Interacting} {Particles}}}}}\ (\bibinfo  {publisher} {Springer},\
  \bibinfo {year} {1991})\BibitemShut {NoStop}%
\bibitem [{\citenamefont {Kaufman}\ \emph {et~al.}(2016)\citenamefont
  {Kaufman}, \citenamefont {Tai}, \citenamefont {Lukin}, \citenamefont
  {Rispoli}, \citenamefont {Schittko}, \citenamefont {Preiss},\ and\
  \citenamefont {Greiner}}]{kaufman_quantum_2016}%
  \BibitemOpen
  \bibfield  {author} {\bibinfo {author} {\bibfnamefont {A.~M.}\ \bibnamefont
  {Kaufman}}, \bibinfo {author} {\bibfnamefont {M.~E.}\ \bibnamefont {Tai}},
  \bibinfo {author} {\bibfnamefont {A.}~\bibnamefont {Lukin}}, \bibinfo
  {author} {\bibfnamefont {M.}~\bibnamefont {Rispoli}}, \bibinfo {author}
  {\bibfnamefont {R.}~\bibnamefont {Schittko}}, \bibinfo {author}
  {\bibfnamefont {P.~M.}\ \bibnamefont {Preiss}},\ and\ \bibinfo {author}
  {\bibfnamefont {M.}~\bibnamefont {Greiner}},\ }\href
  {https://doi.org/10.1126/science.aaf6725} {\bibfield  {journal} {\bibinfo
  {journal} {Science}\ }\textbf {\bibinfo {volume} {353}},\ \bibinfo {pages}
  {794} (\bibinfo {year} {2016})}\BibitemShut {NoStop}%
\bibitem [{\citenamefont {Rispoli}\ \emph {et~al.}(2019)\citenamefont
  {Rispoli}, \citenamefont {Lukin}, \citenamefont {Schittko}, \citenamefont
  {Kim}, \citenamefont {Tai}, \citenamefont {Léonard},\ and\ \citenamefont
  {Greiner}}]{rispoli_quantum_2019}%
  \BibitemOpen
  \bibfield  {author} {\bibinfo {author} {\bibfnamefont {M.}~\bibnamefont
  {Rispoli}}, \bibinfo {author} {\bibfnamefont {A.}~\bibnamefont {Lukin}},
  \bibinfo {author} {\bibfnamefont {R.}~\bibnamefont {Schittko}}, \bibinfo
  {author} {\bibfnamefont {S.}~\bibnamefont {Kim}}, \bibinfo {author}
  {\bibfnamefont {M.~E.}\ \bibnamefont {Tai}}, \bibinfo {author} {\bibfnamefont
  {J.}~\bibnamefont {Léonard}},\ and\ \bibinfo {author} {\bibfnamefont
  {M.}~\bibnamefont {Greiner}},\ }\href
  {https://doi.org/10.1038/s41586-019-1527-2} {\bibfield  {journal} {\bibinfo
  {journal} {Nature}\ }\textbf {\bibinfo {volume} {573}},\ \bibinfo {pages}
  {385} (\bibinfo {year} {2019})}\BibitemShut {NoStop}%
\bibitem [{\citenamefont {von Keyserlingk}\ \emph {et~al.}(2022)\citenamefont
  {von Keyserlingk}, \citenamefont {Pollmann},\ and\ \citenamefont
  {Rakovszky}}]{von_keyserlingk_operator_2022}%
  \BibitemOpen
  \bibfield  {author} {\bibinfo {author} {\bibfnamefont {C.}~\bibnamefont {von
  Keyserlingk}}, \bibinfo {author} {\bibfnamefont {F.}~\bibnamefont
  {Pollmann}},\ and\ \bibinfo {author} {\bibfnamefont {T.}~\bibnamefont
  {Rakovszky}},\ }\href {https://doi.org/10.1103/PhysRevB.105.245101}
  {\bibfield  {journal} {\bibinfo  {journal} {Phys. Rev. B}\ }\textbf {\bibinfo
  {volume} {105}},\ \bibinfo {pages} {245101} (\bibinfo {year}
  {2022})}\BibitemShut {NoStop}%
\bibitem [{\citenamefont {Bertini}\ \emph {et~al.}(2015)\citenamefont
  {Bertini}, \citenamefont {De~Sole}, \citenamefont {Gabrielli}, \citenamefont
  {Jona-Lasinio},\ and\ \citenamefont {Landim}}]{bertini_macroscopic_2015}%
  \BibitemOpen
  \bibfield  {author} {\bibinfo {author} {\bibfnamefont {L.}~\bibnamefont
  {Bertini}}, \bibinfo {author} {\bibfnamefont {A.}~\bibnamefont {De~Sole}},
  \bibinfo {author} {\bibfnamefont {D.}~\bibnamefont {Gabrielli}}, \bibinfo
  {author} {\bibfnamefont {G.}~\bibnamefont {Jona-Lasinio}},\ and\ \bibinfo
  {author} {\bibfnamefont {C.}~\bibnamefont {Landim}},\ }\href
  {https://doi.org/10.1103/RevModPhys.87.593} {\bibfield  {journal} {\bibinfo
  {journal} {Rev. Mod. Phys.}\ }\textbf {\bibinfo {volume} {87}},\ \bibinfo
  {pages} {593} (\bibinfo {year} {2015})}\BibitemShut {NoStop}%
\bibitem [{\citenamefont {McCulloch}\ \emph {et~al.}()\citenamefont
  {McCulloch}, \citenamefont {De~Nardis}, \citenamefont {Gopalakrishnan},\ and\
  \citenamefont {Vasseur}}]{mcculloch_full_2023}%
  \BibitemOpen
  \bibfield  {author} {\bibinfo {author} {\bibfnamefont {E.}~\bibnamefont
  {McCulloch}}, \bibinfo {author} {\bibfnamefont {J.}~\bibnamefont
  {De~Nardis}}, \bibinfo {author} {\bibfnamefont {S.}~\bibnamefont
  {Gopalakrishnan}},\ and\ \bibinfo {author} {\bibfnamefont {R.}~\bibnamefont
  {Vasseur}},\ }\href {http://arxiv.org/abs/2302.01355} {\bibinfo  {journal}
  {arXiv:2302.01355}\ }\BibitemShut {NoStop}%
\bibitem [{\citenamefont {Doyon}\ \emph {et~al.}(2022)\citenamefont {Doyon},
  \citenamefont {Perfetto}, \citenamefont {Sasamoto},\ and\ \citenamefont
  {Yoshimura}}]{doyon_ballistic_2022}%
  \BibitemOpen
\bibfield  {journal} {  }\bibfield  {author} {\bibinfo {author} {\bibfnamefont
  {B.}~\bibnamefont {Doyon}}, \bibinfo {author} {\bibfnamefont
  {G.}~\bibnamefont {Perfetto}}, \bibinfo {author} {\bibfnamefont
  {T.}~\bibnamefont {Sasamoto}},\ and\ \bibinfo {author} {\bibfnamefont
  {T.}~\bibnamefont {Yoshimura}},\ }\href
  {https://doi.org/10.48550/arXiv.2206.14167} {\bibinfo {title} {Ballistic
  macroscopic fluctuation theory}} (\bibinfo {year} {2022}),\ \bibinfo {note}
  {arXiv:2206.14167}\BibitemShut {NoStop}%
\bibitem [{\citenamefont {Bakr}\ \emph {et~al.}(2009)\citenamefont {Bakr},
  \citenamefont {Gillen}, \citenamefont {Peng}, \citenamefont {Fölling},\ and\
  \citenamefont {Greiner}}]{bakr_quantum_2009}%
  \BibitemOpen
  \bibfield  {author} {\bibinfo {author} {\bibfnamefont {W.~S.}\ \bibnamefont
  {Bakr}}, \bibinfo {author} {\bibfnamefont {J.~I.}\ \bibnamefont {Gillen}},
  \bibinfo {author} {\bibfnamefont {A.}~\bibnamefont {Peng}}, \bibinfo {author}
  {\bibfnamefont {S.}~\bibnamefont {Fölling}},\ and\ \bibinfo {author}
  {\bibfnamefont {M.}~\bibnamefont {Greiner}},\ }\href
  {https://doi.org/10.1038/nature08482} {\bibfield  {journal} {\bibinfo
  {journal} {Nature}\ }\textbf {\bibinfo {volume} {462}},\ \bibinfo {pages}
  {74} (\bibinfo {year} {2009})}\BibitemShut {NoStop}%
\bibitem [{\citenamefont {Sherson}\ \emph {et~al.}(2010)\citenamefont
  {Sherson}, \citenamefont {Weitenberg}, \citenamefont {Endres}, \citenamefont
  {Cheneau}, \citenamefont {Bloch},\ and\ \citenamefont
  {Kuhr}}]{sherson_single-atom-resolved_2010}%
  \BibitemOpen
  \bibfield  {author} {\bibinfo {author} {\bibfnamefont {J.~F.}\ \bibnamefont
  {Sherson}}, \bibinfo {author} {\bibfnamefont {C.}~\bibnamefont {Weitenberg}},
  \bibinfo {author} {\bibfnamefont {M.}~\bibnamefont {Endres}}, \bibinfo
  {author} {\bibfnamefont {M.}~\bibnamefont {Cheneau}}, \bibinfo {author}
  {\bibfnamefont {I.}~\bibnamefont {Bloch}},\ and\ \bibinfo {author}
  {\bibfnamefont {S.}~\bibnamefont {Kuhr}},\ }\href
  {https://doi.org/10.1038/nature09378} {\bibfield  {journal} {\bibinfo
  {journal} {Nature}\ }\textbf {\bibinfo {volume} {467}},\ \bibinfo {pages}
  {68} (\bibinfo {year} {2010})}\BibitemShut {NoStop}%
\bibitem [{\citenamefont {Gross}\ and\ \citenamefont
  {Bakr}(2021)}]{gross_quantum_2021}%
  \BibitemOpen
  \bibfield  {author} {\bibinfo {author} {\bibfnamefont {C.}~\bibnamefont
  {Gross}}\ and\ \bibinfo {author} {\bibfnamefont {W.~S.}\ \bibnamefont
  {Bakr}},\ }\href {https://doi.org/10.1038/s41567-021-01370-5} {\bibfield
  {journal} {\bibinfo  {journal} {Nat. Phys.}\ }\textbf {\bibinfo {volume}
  {17}},\ \bibinfo {pages} {1316} (\bibinfo {year} {2021})}\BibitemShut
  {NoStop}%
\bibitem [{\citenamefont {Lukin}\ \emph {et~al.}(2019)\citenamefont {Lukin},
  \citenamefont {Rispoli}, \citenamefont {Schittko}, \citenamefont {Tai},
  \citenamefont {Kaufman}, \citenamefont {Choi}, \citenamefont {Khemani},
  \citenamefont {Léonard},\ and\ \citenamefont
  {Greiner}}]{lukin_probing_2019}%
  \BibitemOpen
  \bibfield  {author} {\bibinfo {author} {\bibfnamefont {A.}~\bibnamefont
  {Lukin}}, \bibinfo {author} {\bibfnamefont {M.}~\bibnamefont {Rispoli}},
  \bibinfo {author} {\bibfnamefont {R.}~\bibnamefont {Schittko}}, \bibinfo
  {author} {\bibfnamefont {M.~E.}\ \bibnamefont {Tai}}, \bibinfo {author}
  {\bibfnamefont {A.~M.}\ \bibnamefont {Kaufman}}, \bibinfo {author}
  {\bibfnamefont {S.}~\bibnamefont {Choi}}, \bibinfo {author} {\bibfnamefont
  {V.}~\bibnamefont {Khemani}}, \bibinfo {author} {\bibfnamefont
  {J.}~\bibnamefont {Léonard}},\ and\ \bibinfo {author} {\bibfnamefont
  {M.}~\bibnamefont {Greiner}},\ }\href
  {https://doi.org/10.1126/science.aau0818} {\bibfield  {journal} {\bibinfo
  {journal} {Science}\ }\textbf {\bibinfo {volume} {364}},\ \bibinfo {pages}
  {256} (\bibinfo {year} {2019})}\BibitemShut {NoStop}%
\bibitem [{\citenamefont {Klostermann}\ \emph {et~al.}(2022)\citenamefont
  {Klostermann}, \citenamefont {Cabrera}, \citenamefont {von Raven},
  \citenamefont {Wienand}, \citenamefont {Schweizer}, \citenamefont {Bloch},\
  and\ \citenamefont {Aidelsburger}}]{klostermann_fast_2022}%
  \BibitemOpen
  \bibfield  {author} {\bibinfo {author} {\bibfnamefont {T.}~\bibnamefont
  {Klostermann}}, \bibinfo {author} {\bibfnamefont {C.~R.}\ \bibnamefont
  {Cabrera}}, \bibinfo {author} {\bibfnamefont {H.}~\bibnamefont {von Raven}},
  \bibinfo {author} {\bibfnamefont {J.~F.}\ \bibnamefont {Wienand}}, \bibinfo
  {author} {\bibfnamefont {C.}~\bibnamefont {Schweizer}}, \bibinfo {author}
  {\bibfnamefont {I.}~\bibnamefont {Bloch}},\ and\ \bibinfo {author}
  {\bibfnamefont {M.}~\bibnamefont {Aidelsburger}},\ }\href
  {https://doi.org/10.1103/PhysRevA.105.043319} {\bibfield  {journal} {\bibinfo
   {journal} {Phys. Rev. A}\ }\textbf {\bibinfo {volume} {105}},\ \bibinfo
  {pages} {043319} (\bibinfo {year} {2022})}\BibitemShut {NoStop}%
\bibitem [{\citenamefont {Impertro}\ \emph {et~al.}()\citenamefont {Impertro},
  \citenamefont {Wienand}, \citenamefont {Häfele}, \citenamefont {von Raven},
  \citenamefont {Hubele}, \citenamefont {Klostermann}, \citenamefont {Cabrera},
  \citenamefont {Bloch},\ and\ \citenamefont
  {Aidelsburger}}]{impertro_unsupervised_2022}%
  \BibitemOpen
  \bibfield  {author} {\bibinfo {author} {\bibfnamefont {A.}~\bibnamefont
  {Impertro}}, \bibinfo {author} {\bibfnamefont {J.~F.}\ \bibnamefont
  {Wienand}}, \bibinfo {author} {\bibfnamefont {S.}~\bibnamefont {Häfele}},
  \bibinfo {author} {\bibfnamefont {H.}~\bibnamefont {von Raven}}, \bibinfo
  {author} {\bibfnamefont {S.}~\bibnamefont {Hubele}}, \bibinfo {author}
  {\bibfnamefont {T.}~\bibnamefont {Klostermann}}, \bibinfo {author}
  {\bibfnamefont {C.~R.}\ \bibnamefont {Cabrera}}, \bibinfo {author}
  {\bibfnamefont {I.}~\bibnamefont {Bloch}},\ and\ \bibinfo {author}
  {\bibfnamefont {M.}~\bibnamefont {Aidelsburger}},\ }\href@noop {} {\ }\Eprint
  {https://arxiv.org/abs/2212.11974} {arXiv:2212.11974} \BibitemShut {NoStop}%
\bibitem [{\citenamefont {Kinoshita}\ \emph {et~al.}(2006)\citenamefont
  {Kinoshita}, \citenamefont {Wenger},\ and\ \citenamefont
  {Weiss}}]{kinoshita_quantum_2006}%
  \BibitemOpen
  \bibfield  {author} {\bibinfo {author} {\bibfnamefont {T.}~\bibnamefont
  {Kinoshita}}, \bibinfo {author} {\bibfnamefont {T.}~\bibnamefont {Wenger}},\
  and\ \bibinfo {author} {\bibfnamefont {D.~S.}\ \bibnamefont {Weiss}},\ }\href
  {https://doi.org/10.1038/nature04693} {\bibfield  {journal} {\bibinfo
  {journal} {Nature}\ }\textbf {\bibinfo {volume} {440}},\ \bibinfo {pages}
  {900} (\bibinfo {year} {2006})}\BibitemShut {NoStop}%
\bibitem [{\citenamefont {Trotzky}\ \emph {et~al.}(2012)\citenamefont
  {Trotzky}, \citenamefont {Chen}, \citenamefont {Flesch}, \citenamefont
  {McCulloch}, \citenamefont {Schollwöck}, \citenamefont {Eisert},\ and\
  \citenamefont {Bloch}}]{trotzky_probing_2012}%
  \BibitemOpen
  \bibfield  {author} {\bibinfo {author} {\bibfnamefont {S.}~\bibnamefont
  {Trotzky}}, \bibinfo {author} {\bibfnamefont {Y.-A.}\ \bibnamefont {Chen}},
  \bibinfo {author} {\bibfnamefont {A.}~\bibnamefont {Flesch}}, \bibinfo
  {author} {\bibfnamefont {I.~P.}\ \bibnamefont {McCulloch}}, \bibinfo {author}
  {\bibfnamefont {U.}~\bibnamefont {Schollwöck}}, \bibinfo {author}
  {\bibfnamefont {J.}~\bibnamefont {Eisert}},\ and\ \bibinfo {author}
  {\bibfnamefont {I.}~\bibnamefont {Bloch}},\ }\href
  {https://doi.org/10.1038/nphys2232} {\bibfield  {journal} {\bibinfo
  {journal} {Nat. Phys.}\ }\textbf {\bibinfo {volume} {8}},\ \bibinfo {pages}
  {325} (\bibinfo {year} {2012})}\BibitemShut {NoStop}%
\bibitem [{\citenamefont {Schreiber}\ \emph {et~al.}(2015)\citenamefont
  {Schreiber}, \citenamefont {Hodgman}, \citenamefont {Bordia}, \citenamefont
  {Lüschen}, \citenamefont {Fischer}, \citenamefont {Vosk}, \citenamefont
  {Altman}, \citenamefont {Schneider},\ and\ \citenamefont
  {Bloch}}]{schreiber_observation_2015}%
  \BibitemOpen
  \bibfield  {author} {\bibinfo {author} {\bibfnamefont {M.}~\bibnamefont
  {Schreiber}}, \bibinfo {author} {\bibfnamefont {S.~S.}\ \bibnamefont
  {Hodgman}}, \bibinfo {author} {\bibfnamefont {P.}~\bibnamefont {Bordia}},
  \bibinfo {author} {\bibfnamefont {H.~P.}\ \bibnamefont {Lüschen}}, \bibinfo
  {author} {\bibfnamefont {M.~H.}\ \bibnamefont {Fischer}}, \bibinfo {author}
  {\bibfnamefont {R.}~\bibnamefont {Vosk}}, \bibinfo {author} {\bibfnamefont
  {E.}~\bibnamefont {Altman}}, \bibinfo {author} {\bibfnamefont
  {U.}~\bibnamefont {Schneider}},\ and\ \bibinfo {author} {\bibfnamefont
  {I.}~\bibnamefont {Bloch}},\ }\href {https://doi.org/10.1126/science.aaa7432}
  {\bibfield  {journal} {\bibinfo  {journal} {Science}\ }\textbf {\bibinfo
  {volume} {349}},\ \bibinfo {pages} {842} (\bibinfo {year}
  {2015})}\BibitemShut {NoStop}%
\bibitem [{\citenamefont {Levitov}\ \emph {et~al.}(1996)\citenamefont
  {Levitov}, \citenamefont {Lee},\ and\ \citenamefont
  {Lesovik}}]{levitov_electron_1996}%
  \BibitemOpen
  \bibfield  {author} {\bibinfo {author} {\bibfnamefont {L.~S.}\ \bibnamefont
  {Levitov}}, \bibinfo {author} {\bibfnamefont {H.}~\bibnamefont {Lee}},\ and\
  \bibinfo {author} {\bibfnamefont {G.~B.}\ \bibnamefont {Lesovik}},\ }\href
  {https://doi.org/10.1063/1.531672} {\bibfield  {journal} {\bibinfo  {journal}
  {J. Math. Phys.}\ }\textbf {\bibinfo {volume} {37}},\ \bibinfo {pages} {4845}
  (\bibinfo {year} {1996})}\BibitemShut {NoStop}%
\bibitem [{\citenamefont {Ranabhat}\ and\ \citenamefont
  {Collura}(2022)}]{ranabhat_dynamics_2022}%
  \BibitemOpen
  \bibfield  {author} {\bibinfo {author} {\bibfnamefont {N.}~\bibnamefont
  {Ranabhat}}\ and\ \bibinfo {author} {\bibfnamefont {M.}~\bibnamefont
  {Collura}},\ }\href {https://doi.org/10.21468/SciPostPhys.12.4.126}
  {\bibfield  {journal} {\bibinfo  {journal} {SciPost Phys.}\ }\textbf
  {\bibinfo {volume} {12}},\ \bibinfo {pages} {126} (\bibinfo {year}
  {2022})}\BibitemShut {NoStop}%
\bibitem [{\citenamefont {Humeniuk}\ and\ \citenamefont
  {Büchler}(2017)}]{humeniuk_full_2017}%
  \BibitemOpen
  \bibfield  {author} {\bibinfo {author} {\bibfnamefont {S.}~\bibnamefont
  {Humeniuk}}\ and\ \bibinfo {author} {\bibfnamefont {H.~P.}\ \bibnamefont
  {Büchler}},\ }\href {https://doi.org/10.1103/PhysRevLett.119.236401}
  {\bibfield  {journal} {\bibinfo  {journal} {Phys. Rev. Lett.}\ }\textbf
  {\bibinfo {volume} {119}},\ \bibinfo {pages} {236401} (\bibinfo {year}
  {2017})}\BibitemShut {NoStop}%
\bibitem [{\citenamefont {Honeychurch}\ and\ \citenamefont
  {Kosov}(2020)}]{honeychurch_full_2020}%
  \BibitemOpen
  \bibfield  {author} {\bibinfo {author} {\bibfnamefont {T.~D.}\ \bibnamefont
  {Honeychurch}}\ and\ \bibinfo {author} {\bibfnamefont {D.~S.}\ \bibnamefont
  {Kosov}},\ }\href {https://doi.org/10.1103/PhysRevB.102.195409} {\bibfield
  {journal} {\bibinfo  {journal} {Phys. Rev. B}\ }\textbf {\bibinfo {volume}
  {102}},\ \bibinfo {pages} {195409} (\bibinfo {year} {2020})}\BibitemShut
  {NoStop}%
\bibitem [{\citenamefont {Malossi}\ \emph {et~al.}(2014)\citenamefont
  {Malossi}, \citenamefont {Valado}, \citenamefont {Scotto}, \citenamefont
  {Huillery}, \citenamefont {Pillet}, \citenamefont {Ciampini}, \citenamefont
  {Arimondo},\ and\ \citenamefont {Morsch}}]{malossi_full_2014}%
  \BibitemOpen
  \bibfield  {author} {\bibinfo {author} {\bibfnamefont {N.}~\bibnamefont
  {Malossi}}, \bibinfo {author} {\bibfnamefont {M.}~\bibnamefont {Valado}},
  \bibinfo {author} {\bibfnamefont {S.}~\bibnamefont {Scotto}}, \bibinfo
  {author} {\bibfnamefont {P.}~\bibnamefont {Huillery}}, \bibinfo {author}
  {\bibfnamefont {P.}~\bibnamefont {Pillet}}, \bibinfo {author} {\bibfnamefont
  {D.}~\bibnamefont {Ciampini}}, \bibinfo {author} {\bibfnamefont
  {E.}~\bibnamefont {Arimondo}},\ and\ \bibinfo {author} {\bibfnamefont
  {O.}~\bibnamefont {Morsch}},\ }\href
  {https://doi.org/10.1103/PhysRevLett.113.023006} {\bibfield  {journal}
  {\bibinfo  {journal} {Phys. Rev. Lett.}\ }\textbf {\bibinfo {volume} {113}},\
  \bibinfo {pages} {023006} (\bibinfo {year} {2014})}\BibitemShut {NoStop}%
\bibitem [{\citenamefont {Calabrese}\ \emph {et~al.}(2020)\citenamefont
  {Calabrese}, \citenamefont {Collura}, \citenamefont {Giulio},\ and\
  \citenamefont {Murciano}}]{calabrese_full_2020}%
  \BibitemOpen
  \bibfield  {author} {\bibinfo {author} {\bibfnamefont {P.}~\bibnamefont
  {Calabrese}}, \bibinfo {author} {\bibfnamefont {M.}~\bibnamefont {Collura}},
  \bibinfo {author} {\bibfnamefont {G.~D.}\ \bibnamefont {Giulio}},\ and\
  \bibinfo {author} {\bibfnamefont {S.}~\bibnamefont {Murciano}},\ }\href
  {https://doi.org/10.1209/0295-5075/129/60007} {\bibfield  {journal} {\bibinfo
   {journal} {Europhys. Lett.}\ }\textbf {\bibinfo {volume} {129}},\ \bibinfo
  {pages} {60007} (\bibinfo {year} {2020})}\BibitemShut {NoStop}%
\bibitem [{\citenamefont {Stéphan}\ and\ \citenamefont
  {Pollmann}(2017)}]{stephan_full_2017}%
  \BibitemOpen
  \bibfield  {author} {\bibinfo {author} {\bibfnamefont {J.-M.}\ \bibnamefont
  {Stéphan}}\ and\ \bibinfo {author} {\bibfnamefont {F.}~\bibnamefont
  {Pollmann}},\ }\href {https://doi.org/10.1103/PhysRevB.95.035119} {\bibfield
  {journal} {\bibinfo  {journal} {Phys. Rev. B}\ }\textbf {\bibinfo {volume}
  {95}},\ \bibinfo {pages} {035119} (\bibinfo {year} {2017})}\BibitemShut
  {NoStop}%
\bibitem [{\citenamefont {Collura}\ \emph {et~al.}(2017)\citenamefont
  {Collura}, \citenamefont {Essler},\ and\ \citenamefont
  {Groha}}]{collura_full_2017}%
  \BibitemOpen
  \bibfield  {author} {\bibinfo {author} {\bibfnamefont {M.}~\bibnamefont
  {Collura}}, \bibinfo {author} {\bibfnamefont {F.~H.~L.}\ \bibnamefont
  {Essler}},\ and\ \bibinfo {author} {\bibfnamefont {S.}~\bibnamefont
  {Groha}},\ }\href {https://doi.org/10.1088/1751-8121/aa87dd} {\bibfield
  {journal} {\bibinfo  {journal} {J. Phys. A: Math. Theor.}\ }\textbf {\bibinfo
  {volume} {50}},\ \bibinfo {pages} {414002} (\bibinfo {year}
  {2017})}\BibitemShut {NoStop}%
\bibitem [{\citenamefont {Nazarov}\ and\ \citenamefont
  {Kindermann}(2003)}]{nazarov_full_2003}%
  \BibitemOpen
  \bibfield  {author} {\bibinfo {author} {\bibfnamefont {Y.~V.}\ \bibnamefont
  {Nazarov}}\ and\ \bibinfo {author} {\bibfnamefont {M.}~\bibnamefont
  {Kindermann}},\ }\href {https://doi.org/10.1140/epjb/e2003-00293-1}
  {\bibfield  {journal} {\bibinfo  {journal} {Eur. Phys. B}\ }\textbf {\bibinfo
  {volume} {35}},\ \bibinfo {pages} {413} (\bibinfo {year} {2003})}\BibitemShut
  {NoStop}%
\bibitem [{\citenamefont {Devillard}\ \emph {et~al.}(2020)\citenamefont
  {Devillard}, \citenamefont {Chevallier}, \citenamefont {Vignolo},\ and\
  \citenamefont {Albert}}]{devillard_full_2020}%
  \BibitemOpen
  \bibfield  {author} {\bibinfo {author} {\bibfnamefont {P.}~\bibnamefont
  {Devillard}}, \bibinfo {author} {\bibfnamefont {D.}~\bibnamefont
  {Chevallier}}, \bibinfo {author} {\bibfnamefont {P.}~\bibnamefont
  {Vignolo}},\ and\ \bibinfo {author} {\bibfnamefont {M.}~\bibnamefont
  {Albert}},\ }\href {https://doi.org/10.1103/PhysRevA.101.063604} {\bibfield
  {journal} {\bibinfo  {journal} {Phys. Rev. A}\ }\textbf {\bibinfo {volume}
  {101}},\ \bibinfo {pages} {063604} (\bibinfo {year} {2020})}\BibitemShut
  {NoStop}%
\bibitem [{\citenamefont {Lovas}\ \emph {et~al.}(2017)\citenamefont {Lovas},
  \citenamefont {Dóra}, \citenamefont {Demler},\ and\ \citenamefont
  {Zaránd}}]{lovas_full_2017}%
  \BibitemOpen
  \bibfield  {author} {\bibinfo {author} {\bibfnamefont {I.}~\bibnamefont
  {Lovas}}, \bibinfo {author} {\bibfnamefont {B.}~\bibnamefont {Dóra}},
  \bibinfo {author} {\bibfnamefont {E.}~\bibnamefont {Demler}},\ and\ \bibinfo
  {author} {\bibfnamefont {G.}~\bibnamefont {Zaránd}},\ }\href
  {https://doi.org/10.1103/PhysRevA.95.053621} {\bibfield  {journal} {\bibinfo
  {journal} {Phys. Rev. A}\ }\textbf {\bibinfo {volume} {95}},\ \bibinfo
  {pages} {053621} (\bibinfo {year} {2017})}\BibitemShut {NoStop}%
\bibitem [{\citenamefont {{Gopalakrishnan}}\ \emph {et~al.}()\citenamefont
  {{Gopalakrishnan}}, \citenamefont {{Morningstar}}, \citenamefont
  {{Vasseur}},\ and\ \citenamefont {{Khemani}}}]{gopalakrishnan_theory_2022}%
  \BibitemOpen
  \bibfield  {author} {\bibinfo {author} {\bibfnamefont {S.}~\bibnamefont
  {{Gopalakrishnan}}}, \bibinfo {author} {\bibfnamefont {A.}~\bibnamefont
  {{Morningstar}}}, \bibinfo {author} {\bibfnamefont {R.}~\bibnamefont
  {{Vasseur}}},\ and\ \bibinfo {author} {\bibfnamefont {V.}~\bibnamefont
  {{Khemani}}},\ }\href@noop {} {\ }\Eprint {https://arxiv.org/abs/2203.09526}
  {arXiv:2203.09526} \BibitemShut {NoStop}%
\bibitem [{\citenamefont {Zheng}\ \emph {et~al.}()\citenamefont {Zheng},
  \citenamefont {Zhang}, \citenamefont {Shen}, \citenamefont {Luo},
  \citenamefont {Liu}, \citenamefont {He}, \citenamefont {Zhang}, \citenamefont
  {Lin}, \citenamefont {Wang}, \citenamefont {Zhu}, \citenamefont {Chen},
  \citenamefont {Lu}, \citenamefont {Thanasilp}, \citenamefont {Angelakis},
  \citenamefont {Yuan},\ and\ \citenamefont {Pan}}]{zheng_efficiently_2022}%
  \BibitemOpen
  \bibfield  {author} {\bibinfo {author} {\bibfnamefont {Y.-G.}\ \bibnamefont
  {Zheng}}, \bibinfo {author} {\bibfnamefont {W.-Y.}\ \bibnamefont {Zhang}},
  \bibinfo {author} {\bibfnamefont {Y.-C.}\ \bibnamefont {Shen}}, \bibinfo
  {author} {\bibfnamefont {A.}~\bibnamefont {Luo}}, \bibinfo {author}
  {\bibfnamefont {Y.}~\bibnamefont {Liu}}, \bibinfo {author} {\bibfnamefont
  {M.-G.}\ \bibnamefont {He}}, \bibinfo {author} {\bibfnamefont {H.-R.}\
  \bibnamefont {Zhang}}, \bibinfo {author} {\bibfnamefont {W.}~\bibnamefont
  {Lin}}, \bibinfo {author} {\bibfnamefont {H.-Y.}\ \bibnamefont {Wang}},
  \bibinfo {author} {\bibfnamefont {Z.-H.}\ \bibnamefont {Zhu}}, \bibinfo
  {author} {\bibfnamefont {M.-C.}\ \bibnamefont {Chen}}, \bibinfo {author}
  {\bibfnamefont {C.-Y.}\ \bibnamefont {Lu}}, \bibinfo {author} {\bibfnamefont
  {S.}~\bibnamefont {Thanasilp}}, \bibinfo {author} {\bibfnamefont {D.~G.}\
  \bibnamefont {Angelakis}}, \bibinfo {author} {\bibfnamefont {Z.-S.}\
  \bibnamefont {Yuan}},\ and\ \bibinfo {author} {\bibfnamefont {J.-W.}\
  \bibnamefont {Pan}},\ }\href@noop {} {\ }\Eprint
  {https://arxiv.org/abs/2210.08556} {arXiv:2210.08556} \BibitemShut {NoStop}%
\bibitem [{\citenamefont {Dall}\ \emph {et~al.}(2013)\citenamefont {Dall},
  \citenamefont {Manning}, \citenamefont {Hodgman}, \citenamefont {RuGway},
  \citenamefont {Kheruntsyan},\ and\ \citenamefont
  {Truscott}}]{dall_ideal_2013}%
  \BibitemOpen
  \bibfield  {author} {\bibinfo {author} {\bibfnamefont {R.~G.}\ \bibnamefont
  {Dall}}, \bibinfo {author} {\bibfnamefont {A.~G.}\ \bibnamefont {Manning}},
  \bibinfo {author} {\bibfnamefont {S.~S.}\ \bibnamefont {Hodgman}}, \bibinfo
  {author} {\bibfnamefont {W.}~\bibnamefont {RuGway}}, \bibinfo {author}
  {\bibfnamefont {K.~V.}\ \bibnamefont {Kheruntsyan}},\ and\ \bibinfo {author}
  {\bibfnamefont {A.~G.}\ \bibnamefont {Truscott}},\ }\href
  {https://doi.org/10.1038/nphys2632} {\bibfield  {journal} {\bibinfo
  {journal} {Nat. Phys.}\ }\textbf {\bibinfo {volume} {9}},\ \bibinfo {pages}
  {341} (\bibinfo {year} {2013})}\BibitemShut {NoStop}%
\bibitem [{\citenamefont {Wei}\ \emph {et~al.}(2022)\citenamefont {Wei},
  \citenamefont {Rubio-Abadal}, \citenamefont {Ye}, \citenamefont {Machado},
  \citenamefont {Kemp}, \citenamefont {Srakaew}, \citenamefont {Hollerith},
  \citenamefont {Rui}, \citenamefont {Gopalakrishnan}, \citenamefont {Yao},
  \citenamefont {Bloch},\ and\ \citenamefont {Zeiher}}]{wei_quantum_2022}%
  \BibitemOpen
  \bibfield  {author} {\bibinfo {author} {\bibfnamefont {D.}~\bibnamefont
  {Wei}}, \bibinfo {author} {\bibfnamefont {A.}~\bibnamefont {Rubio-Abadal}},
  \bibinfo {author} {\bibfnamefont {B.}~\bibnamefont {Ye}}, \bibinfo {author}
  {\bibfnamefont {F.}~\bibnamefont {Machado}}, \bibinfo {author} {\bibfnamefont
  {J.}~\bibnamefont {Kemp}}, \bibinfo {author} {\bibfnamefont {K.}~\bibnamefont
  {Srakaew}}, \bibinfo {author} {\bibfnamefont {S.}~\bibnamefont {Hollerith}},
  \bibinfo {author} {\bibfnamefont {J.}~\bibnamefont {Rui}}, \bibinfo {author}
  {\bibfnamefont {S.}~\bibnamefont {Gopalakrishnan}}, \bibinfo {author}
  {\bibfnamefont {N.~Y.}\ \bibnamefont {Yao}}, \bibinfo {author} {\bibfnamefont
  {I.}~\bibnamefont {Bloch}},\ and\ \bibinfo {author} {\bibfnamefont
  {J.}~\bibnamefont {Zeiher}},\ }\href
  {https://doi.org/10.1126/science.abk2397} {\bibfield  {journal} {\bibinfo
  {journal} {Science}\ }\textbf {\bibinfo {volume} {376}},\ \bibinfo {pages}
  {716} (\bibinfo {year} {2022})}\BibitemShut {NoStop}%
\bibitem [{\citenamefont {Herc\'e}\ \emph {et~al.}(2023)\citenamefont
  {Herc\'e}, \citenamefont {Bureik}, \citenamefont {T\'enart}, \citenamefont
  {Aspect}, \citenamefont {Dareau},\ and\ \citenamefont
  {Cl\'ement}}]{herce_full_2022}%
  \BibitemOpen
  \bibfield  {author} {\bibinfo {author} {\bibfnamefont {G.}~\bibnamefont
  {Herc\'e}}, \bibinfo {author} {\bibfnamefont {J.-P.}\ \bibnamefont {Bureik}},
  \bibinfo {author} {\bibfnamefont {A.}~\bibnamefont {T\'enart}}, \bibinfo
  {author} {\bibfnamefont {A.}~\bibnamefont {Aspect}}, \bibinfo {author}
  {\bibfnamefont {A.}~\bibnamefont {Dareau}},\ and\ \bibinfo {author}
  {\bibfnamefont {D.}~\bibnamefont {Cl\'ement}},\ }\href
  {https://doi.org/10.1103/PhysRevResearch.5.L012037} {\bibfield  {journal}
  {\bibinfo  {journal} {Phys. Rev. Res.}\ }\textbf {\bibinfo {volume} {5}},\
  \bibinfo {pages} {L012037} (\bibinfo {year} {2023})}\BibitemShut {NoStop}%
\bibitem [{\citenamefont {Rakovszky}\ \emph {et~al.}(2022)\citenamefont
  {Rakovszky}, \citenamefont {von Keyserlingk},\ and\ \citenamefont
  {Pollmann}}]{DAOE2020}%
  \BibitemOpen
  \bibfield  {author} {\bibinfo {author} {\bibfnamefont {T.}~\bibnamefont
  {Rakovszky}}, \bibinfo {author} {\bibfnamefont {C.~W.}\ \bibnamefont {von
  Keyserlingk}},\ and\ \bibinfo {author} {\bibfnamefont {F.}~\bibnamefont
  {Pollmann}},\ }\href {https://doi.org/10.1103/PhysRevB.105.075131} {\bibfield
   {journal} {\bibinfo  {journal} {Phys. Rev. B}\ }\textbf {\bibinfo {volume}
  {105}},\ \bibinfo {pages} {075131} (\bibinfo {year} {2022})}\BibitemShut
  {NoStop}%
\bibitem [{\citenamefont {Lieb}\ and\ \citenamefont
  {Robinson}(1972)}]{lieb_finite_1972}%
  \BibitemOpen
  \bibfield  {author} {\bibinfo {author} {\bibfnamefont {E.~H.}\ \bibnamefont
  {Lieb}}\ and\ \bibinfo {author} {\bibfnamefont {D.~W.}\ \bibnamefont
  {Robinson}},\ }\href {https://doi.org/10.1007/BF01645779} {\bibfield
  {journal} {\bibinfo  {journal} {Commun. Math. Phys.}\ }\textbf {\bibinfo
  {volume} {28}},\ \bibinfo {pages} {251} (\bibinfo {year} {1972})}\BibitemShut
  {NoStop}%
\bibitem [{sup()}]{supp}%
  \BibitemOpen
  \href@noop {} {}\bibinfo {note} {See online supplemental materials for
  details.}\BibitemShut {Stop}%
\bibitem [{\citenamefont {Crépin}\ \emph {et~al.}(2011)\citenamefont
  {Crépin}, \citenamefont {Laflorencie}, \citenamefont {Roux},\ and\
  \citenamefont {Simon}}]{crepin_phase_2011}%
  \BibitemOpen
  \bibfield  {author} {\bibinfo {author} {\bibfnamefont {F.}~\bibnamefont
  {Crépin}}, \bibinfo {author} {\bibfnamefont {N.}~\bibnamefont
  {Laflorencie}}, \bibinfo {author} {\bibfnamefont {G.}~\bibnamefont {Roux}},\
  and\ \bibinfo {author} {\bibfnamefont {P.}~\bibnamefont {Simon}},\ }\href
  {https://doi.org/10.1103/PhysRevB.84.054517} {\bibfield  {journal} {\bibinfo
  {journal} {Phys. Rev. B}\ }\textbf {\bibinfo {volume} {84}},\ \bibinfo
  {pages} {054517} (\bibinfo {year} {2011})}\BibitemShut {NoStop}%
\bibitem [{\citenamefont {Donohue}\ and\ \citenamefont
  {Giamarchi}(2001)}]{donohue_mott-superfluid_2001}%
  \BibitemOpen
  \bibfield  {author} {\bibinfo {author} {\bibfnamefont {P.}~\bibnamefont
  {Donohue}}\ and\ \bibinfo {author} {\bibfnamefont {T.}~\bibnamefont
  {Giamarchi}},\ }\href {https://doi.org/10.1103/PhysRevB.63.180508} {\bibfield
   {journal} {\bibinfo  {journal} {Phys. Rev. B}\ }\textbf {\bibinfo {volume}
  {63}},\ \bibinfo {pages} {180508} (\bibinfo {year} {2001})}\BibitemShut
  {NoStop}%
\bibitem [{\citenamefont {Lüschen}\ \emph {et~al.}(2017)\citenamefont
  {Lüschen}, \citenamefont {Bordia}, \citenamefont {Scherg}, \citenamefont
  {Alet}, \citenamefont {Altman}, \citenamefont {Schneider},\ and\
  \citenamefont {Bloch}}]{luschen_observation_2017}%
  \BibitemOpen
  \bibfield  {author} {\bibinfo {author} {\bibfnamefont {H.~P.}\ \bibnamefont
  {Lüschen}}, \bibinfo {author} {\bibfnamefont {P.}~\bibnamefont {Bordia}},
  \bibinfo {author} {\bibfnamefont {S.}~\bibnamefont {Scherg}}, \bibinfo
  {author} {\bibfnamefont {F.}~\bibnamefont {Alet}}, \bibinfo {author}
  {\bibfnamefont {E.}~\bibnamefont {Altman}}, \bibinfo {author} {\bibfnamefont
  {U.}~\bibnamefont {Schneider}},\ and\ \bibinfo {author} {\bibfnamefont
  {I.}~\bibnamefont {Bloch}},\ }\href
  {https://doi.org/10.1103/PhysRevLett.119.260401} {\bibfield  {journal}
  {\bibinfo  {journal} {Phys. Rev. Lett.}\ }\textbf {\bibinfo {volume} {119}},\
  \bibinfo {pages} {260401} (\bibinfo {year} {2017})}\BibitemShut {NoStop}%
\bibitem [{\citenamefont {Rubio-Abadal}\ \emph {et~al.}(2019)\citenamefont
  {Rubio-Abadal}, \citenamefont {Choi}, \citenamefont {Zeiher}, \citenamefont
  {Hollerith}, \citenamefont {Rui}, \citenamefont {Bloch},\ and\ \citenamefont
  {Gross}}]{rubio-abadal_many-body_2019}%
  \BibitemOpen
  \bibfield  {author} {\bibinfo {author} {\bibfnamefont {A.}~\bibnamefont
  {Rubio-Abadal}}, \bibinfo {author} {\bibfnamefont {J.-Y.}\ \bibnamefont
  {Choi}}, \bibinfo {author} {\bibfnamefont {J.}~\bibnamefont {Zeiher}},
  \bibinfo {author} {\bibfnamefont {S.}~\bibnamefont {Hollerith}}, \bibinfo
  {author} {\bibfnamefont {J.}~\bibnamefont {Rui}}, \bibinfo {author}
  {\bibfnamefont {I.}~\bibnamefont {Bloch}},\ and\ \bibinfo {author}
  {\bibfnamefont {C.}~\bibnamefont {Gross}},\ }\href
  {https://doi.org/10.1103/PhysRevX.9.041014} {\bibfield  {journal} {\bibinfo
  {journal} {Phys. Rev. X}\ }\textbf {\bibinfo {volume} {9}},\ \bibinfo {pages}
  {041014} (\bibinfo {year} {2019})}\BibitemShut {NoStop}%
\bibitem [{\citenamefont {Kohlert}\ \emph {et~al.}(2019)\citenamefont
  {Kohlert}, \citenamefont {Scherg}, \citenamefont {Li}, \citenamefont
  {Lüschen}, \citenamefont {Das~Sarma}, \citenamefont {Bloch},\ and\
  \citenamefont {Aidelsburger}}]{kohlert_observation_2019}%
  \BibitemOpen
  \bibfield  {author} {\bibinfo {author} {\bibfnamefont {T.}~\bibnamefont
  {Kohlert}}, \bibinfo {author} {\bibfnamefont {S.}~\bibnamefont {Scherg}},
  \bibinfo {author} {\bibfnamefont {X.}~\bibnamefont {Li}}, \bibinfo {author}
  {\bibfnamefont {H.~P.}\ \bibnamefont {Lüschen}}, \bibinfo {author}
  {\bibfnamefont {S.}~\bibnamefont {Das~Sarma}}, \bibinfo {author}
  {\bibfnamefont {I.}~\bibnamefont {Bloch}},\ and\ \bibinfo {author}
  {\bibfnamefont {M.}~\bibnamefont {Aidelsburger}},\ }\href
  {https://doi.org/10.1103/PhysRevLett.122.170403} {\bibfield  {journal}
  {\bibinfo  {journal} {Phys. Rev. Lett.}\ }\textbf {\bibinfo {volume} {122}},\
  \bibinfo {pages} {170403} (\bibinfo {year} {2019})}\BibitemShut {NoStop}%
\bibitem [{\citenamefont {Cramer}\ \emph {et~al.}(2008)\citenamefont {Cramer},
  \citenamefont {Flesch}, \citenamefont {McCulloch}, \citenamefont
  {Schollwöck},\ and\ \citenamefont {Eisert}}]{cramer_exploring_2008}%
  \BibitemOpen
  \bibfield  {author} {\bibinfo {author} {\bibfnamefont {M.}~\bibnamefont
  {Cramer}}, \bibinfo {author} {\bibfnamefont {A.}~\bibnamefont {Flesch}},
  \bibinfo {author} {\bibfnamefont {I.~P.}\ \bibnamefont {McCulloch}}, \bibinfo
  {author} {\bibfnamefont {U.}~\bibnamefont {Schollwöck}},\ and\ \bibinfo
  {author} {\bibfnamefont {J.}~\bibnamefont {Eisert}},\ }\href
  {https://doi.org/10.1103/PhysRevLett.101.063001} {\bibfield  {journal}
  {\bibinfo  {journal} {Phys. Rev. Lett.}\ }\textbf {\bibinfo {volume} {101}},\
  \bibinfo {pages} {063001} (\bibinfo {year} {2008})}\BibitemShut {NoStop}%
\bibitem [{\citenamefont {Mallick}\ \emph {et~al.}(2022)\citenamefont
  {Mallick}, \citenamefont {Moriya},\ and\ \citenamefont
  {Sasamoto}}]{mallick_exact_2022}%
  \BibitemOpen
  \bibfield  {author} {\bibinfo {author} {\bibfnamefont {K.}~\bibnamefont
  {Mallick}}, \bibinfo {author} {\bibfnamefont {H.}~\bibnamefont {Moriya}},\
  and\ \bibinfo {author} {\bibfnamefont {T.}~\bibnamefont {Sasamoto}},\ }\href
  {https://doi.org/10.1103/PhysRevLett.129.040601} {\bibfield  {journal}
  {\bibinfo  {journal} {Phys. Rev. Lett.}\ }\textbf {\bibinfo {volume} {129}},\
  \bibinfo {pages} {040601} (\bibinfo {year} {2022})}\BibitemShut {NoStop}%
\bibitem [{\citenamefont {{Steinigeweg}}\ \emph {et~al.}(2014)\citenamefont
  {{Steinigeweg}}, \citenamefont {{Heidrich-Meisner}}, \citenamefont
  {{Gemmer}}, \citenamefont {{Michielsen}},\ and\ \citenamefont {{De
  Raedt}}}]{steinigeweg_2014}%
  \BibitemOpen
  \bibfield  {author} {\bibinfo {author} {\bibfnamefont {R.}~\bibnamefont
  {{Steinigeweg}}}, \bibinfo {author} {\bibfnamefont {F.}~\bibnamefont
  {{Heidrich-Meisner}}}, \bibinfo {author} {\bibfnamefont {J.}~\bibnamefont
  {{Gemmer}}}, \bibinfo {author} {\bibfnamefont {K.}~\bibnamefont
  {{Michielsen}}},\ and\ \bibinfo {author} {\bibfnamefont {H.}~\bibnamefont
  {{De Raedt}}},\ }\href {https://doi.org/10.1103/PhysRevB.90.094417}
  {\bibfield  {journal} {\bibinfo  {journal} {\prb}\ }\textbf {\bibinfo
  {volume} {90}},\ \bibinfo {eid} {094417} (\bibinfo {year}
  {2014})}\BibitemShut {NoStop}%
\bibitem [{\citenamefont {Cheneau}\ \emph {et~al.}(2012)\citenamefont
  {Cheneau}, \citenamefont {Barmettler}, \citenamefont {Poletti}, \citenamefont
  {Endres}, \citenamefont {Schauß}, \citenamefont {Fukuhara}, \citenamefont
  {Gross}, \citenamefont {Bloch}, \citenamefont {Kollath},\ and\ \citenamefont
  {Kuhr}}]{cheneau_light-cone-like_2012}%
  \BibitemOpen
  \bibfield  {author} {\bibinfo {author} {\bibfnamefont {M.}~\bibnamefont
  {Cheneau}}, \bibinfo {author} {\bibfnamefont {P.}~\bibnamefont {Barmettler}},
  \bibinfo {author} {\bibfnamefont {D.}~\bibnamefont {Poletti}}, \bibinfo
  {author} {\bibfnamefont {M.}~\bibnamefont {Endres}}, \bibinfo {author}
  {\bibfnamefont {P.}~\bibnamefont {Schauß}}, \bibinfo {author} {\bibfnamefont
  {T.}~\bibnamefont {Fukuhara}}, \bibinfo {author} {\bibfnamefont
  {C.}~\bibnamefont {Gross}}, \bibinfo {author} {\bibfnamefont
  {I.}~\bibnamefont {Bloch}}, \bibinfo {author} {\bibfnamefont
  {C.}~\bibnamefont {Kollath}},\ and\ \bibinfo {author} {\bibfnamefont
  {S.}~\bibnamefont {Kuhr}},\ }\href {https://doi.org/10.1038/nature10748}
  {\bibfield  {journal} {\bibinfo  {journal} {Nature}\ }\textbf {\bibinfo
  {volume} {481}},\ \bibinfo {pages} {484} (\bibinfo {year}
  {2012})}\BibitemShut {NoStop}%
\bibitem [{\citenamefont {Takasu}\ \emph {et~al.}(2020)\citenamefont {Takasu},
  \citenamefont {Yagami}, \citenamefont {Asaka}, \citenamefont {Fukushima},
  \citenamefont {Nagao}, \citenamefont {Goto}, \citenamefont {Danshita},\ and\
  \citenamefont {Takahashi}}]{takasu_energy_2020}%
  \BibitemOpen
  \bibfield  {author} {\bibinfo {author} {\bibfnamefont {Y.}~\bibnamefont
  {Takasu}}, \bibinfo {author} {\bibfnamefont {T.}~\bibnamefont {Yagami}},
  \bibinfo {author} {\bibfnamefont {H.}~\bibnamefont {Asaka}}, \bibinfo
  {author} {\bibfnamefont {Y.}~\bibnamefont {Fukushima}}, \bibinfo {author}
  {\bibfnamefont {K.}~\bibnamefont {Nagao}}, \bibinfo {author} {\bibfnamefont
  {S.}~\bibnamefont {Goto}}, \bibinfo {author} {\bibfnamefont {I.}~\bibnamefont
  {Danshita}},\ and\ \bibinfo {author} {\bibfnamefont {Y.}~\bibnamefont
  {Takahashi}},\ }\href {https://doi.org/10.1126/sciadv.aba9255} {\bibfield
  {journal} {\bibinfo  {journal} {Sci. Adv.}\ }\textbf {\bibinfo {volume}
  {6}},\ \bibinfo {pages} {eaba9255} (\bibinfo {year} {2020})}\BibitemShut
  {NoStop}%
\bibitem [{\citenamefont {Tajik}\ \emph {et~al.}(2023)\citenamefont {Tajik},
  \citenamefont {Gluza}, \citenamefont {Sebe}, \citenamefont {Schüttelkopf},
  \citenamefont {Cataldini}, \citenamefont {Sabino}, \citenamefont {Møller},
  \citenamefont {Ji}, \citenamefont {Erne}, \citenamefont {Guarnieri},
  \citenamefont {Sotiriadis}, \citenamefont {Eisert},\ and\ \citenamefont
  {Schmiedmayer}}]{tajik_experimental_2022}%
  \BibitemOpen
  \bibfield  {author} {\bibinfo {author} {\bibfnamefont {M.}~\bibnamefont
  {Tajik}}, \bibinfo {author} {\bibfnamefont {M.}~\bibnamefont {Gluza}},
  \bibinfo {author} {\bibfnamefont {N.}~\bibnamefont {Sebe}}, \bibinfo {author}
  {\bibfnamefont {P.}~\bibnamefont {Schüttelkopf}}, \bibinfo {author}
  {\bibfnamefont {F.}~\bibnamefont {Cataldini}}, \bibinfo {author}
  {\bibfnamefont {J.}~\bibnamefont {Sabino}}, \bibinfo {author} {\bibfnamefont
  {F.}~\bibnamefont {Møller}}, \bibinfo {author} {\bibfnamefont {S.-C.}\
  \bibnamefont {Ji}}, \bibinfo {author} {\bibfnamefont {S.}~\bibnamefont
  {Erne}}, \bibinfo {author} {\bibfnamefont {G.}~\bibnamefont {Guarnieri}},
  \bibinfo {author} {\bibfnamefont {S.}~\bibnamefont {Sotiriadis}}, \bibinfo
  {author} {\bibfnamefont {J.}~\bibnamefont {Eisert}},\ and\ \bibinfo {author}
  {\bibfnamefont {J.}~\bibnamefont {Schmiedmayer}},\ }\href
  {https://doi.org/10.1073/pnas.2301287120} {\bibfield  {journal} {\bibinfo
  {journal} {Proc. Natl. Acad. Sci.}\ }\textbf {\bibinfo {volume} {120}},\
  \bibinfo {pages} {e2301287120} (\bibinfo {year} {2023})}\BibitemShut
  {NoStop}%
\bibitem [{\citenamefont {Bernard}(2021)}]{Bernard_QMFT}%
  \BibitemOpen
  \bibfield  {author} {\bibinfo {author} {\bibfnamefont {D.}~\bibnamefont
  {Bernard}},\ }\href {https://doi.org/10.1088/1751-8121/ac2597} {\bibfield
  {journal} {\bibinfo  {journal} {J. Phys. A Math. Theor.}\ }\textbf {\bibinfo
  {volume} {54}},\ \bibinfo {pages} {433001} (\bibinfo {year}
  {2021})}\BibitemShut {NoStop}%
\bibitem [{\citenamefont {Mark}\ \emph {et~al.}()\citenamefont {Mark},
  \citenamefont {Choi}, \citenamefont {Shaw}, \citenamefont {Endres},\ and\
  \citenamefont {Choi}}]{mark2022benchmarking}%
  \BibitemOpen
  \bibfield  {author} {\bibinfo {author} {\bibfnamefont {D.~K.}\ \bibnamefont
  {Mark}}, \bibinfo {author} {\bibfnamefont {J.}~\bibnamefont {Choi}}, \bibinfo
  {author} {\bibfnamefont {A.~L.}\ \bibnamefont {Shaw}}, \bibinfo {author}
  {\bibfnamefont {M.}~\bibnamefont {Endres}},\ and\ \bibinfo {author}
  {\bibfnamefont {S.}~\bibnamefont {Choi}},\ }\href@noop {} {\ }\Eprint
  {https://arxiv.org/abs/2205.12211} {arXiv:2205.12211} \BibitemShut {NoStop}%
\bibitem [{\citenamefont {Choi}\ \emph {et~al.}(2023)\citenamefont {Choi},
  \citenamefont {Shaw}, \citenamefont {Madjarov}, \citenamefont {Xie},
  \citenamefont {Finkelstein}, \citenamefont {Covey}, \citenamefont {Cotler},
  \citenamefont {Mark}, \citenamefont {Huang}, \citenamefont {Kale},
  \citenamefont {Pichler}, \citenamefont {Brand{\~{a}}o}, \citenamefont
  {Choi},\ and\ \citenamefont {Endres}}]{Choi2023}%
  \BibitemOpen
  \bibfield  {author} {\bibinfo {author} {\bibfnamefont {J.}~\bibnamefont
  {Choi}}, \bibinfo {author} {\bibfnamefont {A.~L.}\ \bibnamefont {Shaw}},
  \bibinfo {author} {\bibfnamefont {I.~S.}\ \bibnamefont {Madjarov}}, \bibinfo
  {author} {\bibfnamefont {X.}~\bibnamefont {Xie}}, \bibinfo {author}
  {\bibfnamefont {R.}~\bibnamefont {Finkelstein}}, \bibinfo {author}
  {\bibfnamefont {J.~P.}\ \bibnamefont {Covey}}, \bibinfo {author}
  {\bibfnamefont {J.~S.}\ \bibnamefont {Cotler}}, \bibinfo {author}
  {\bibfnamefont {D.~K.}\ \bibnamefont {Mark}}, \bibinfo {author}
  {\bibfnamefont {H.-Y.}\ \bibnamefont {Huang}}, \bibinfo {author}
  {\bibfnamefont {A.}~\bibnamefont {Kale}}, \bibinfo {author} {\bibfnamefont
  {H.}~\bibnamefont {Pichler}}, \bibinfo {author} {\bibfnamefont {F.~G. S.~L.}\
  \bibnamefont {Brand{\~{a}}o}}, \bibinfo {author} {\bibfnamefont
  {S.}~\bibnamefont {Choi}},\ and\ \bibinfo {author} {\bibfnamefont
  {M.}~\bibnamefont {Endres}},\ }\href
  {https://doi.org/10.1038/s41586-022-05442-1} {\bibfield  {journal} {\bibinfo
  {journal} {Nature}\ }\textbf {\bibinfo {volume} {613}},\ \bibinfo {pages}
  {468} (\bibinfo {year} {2023})}\BibitemShut {NoStop}%
\bibitem [{\citenamefont {Le}\ \emph {et~al.}()\citenamefont {Le},
  \citenamefont {Zhang}, \citenamefont {Gopalakrishnan}, \citenamefont
  {Rigol},\ and\ \citenamefont {Weiss}}]{le_direct_2022}%
  \BibitemOpen
  \bibfield  {author} {\bibinfo {author} {\bibfnamefont {Y.}~\bibnamefont
  {Le}}, \bibinfo {author} {\bibfnamefont {Y.}~\bibnamefont {Zhang}}, \bibinfo
  {author} {\bibfnamefont {S.}~\bibnamefont {Gopalakrishnan}}, \bibinfo
  {author} {\bibfnamefont {M.}~\bibnamefont {Rigol}},\ and\ \bibinfo {author}
  {\bibfnamefont {D.~S.}\ \bibnamefont {Weiss}},\ }\href@noop {} {\ }\Eprint
  {https://arxiv.org/abs/2210.07318} {arXiv:2210.07318} \BibitemShut {NoStop}%
\bibitem [{\citenamefont {Choi}\ \emph {et~al.}(2016)\citenamefont {Choi},
  \citenamefont {Hild}, \citenamefont {Zeiher}, \citenamefont {Schauß},
  \citenamefont {Rubio-Abadal}, \citenamefont {Yefsah}, \citenamefont
  {Khemani}, \citenamefont {Huse}, \citenamefont {Bloch},\ and\ \citenamefont
  {Gross}}]{choi_exploring_2016}%
  \BibitemOpen
  \bibfield  {author} {\bibinfo {author} {\bibfnamefont {J.-Y.}\ \bibnamefont
  {Choi}}, \bibinfo {author} {\bibfnamefont {S.}~\bibnamefont {Hild}}, \bibinfo
  {author} {\bibfnamefont {J.}~\bibnamefont {Zeiher}}, \bibinfo {author}
  {\bibfnamefont {P.}~\bibnamefont {Schauß}}, \bibinfo {author} {\bibfnamefont
  {A.}~\bibnamefont {Rubio-Abadal}}, \bibinfo {author} {\bibfnamefont
  {T.}~\bibnamefont {Yefsah}}, \bibinfo {author} {\bibfnamefont
  {V.}~\bibnamefont {Khemani}}, \bibinfo {author} {\bibfnamefont {D.~A.}\
  \bibnamefont {Huse}}, \bibinfo {author} {\bibfnamefont {I.}~\bibnamefont
  {Bloch}},\ and\ \bibinfo {author} {\bibfnamefont {C.}~\bibnamefont {Gross}},\
  }\href@noop {} {\bibfield  {journal} {\bibinfo  {journal} {Science}\ }\textbf
  {\bibinfo {volume} {352}},\ \bibinfo {pages} {1547} (\bibinfo {year}
  {2016})}\BibitemShut {NoStop}%
\bibitem [{\citenamefont {Castro-Alvaredo}\ \emph {et~al.}(2016)\citenamefont
  {Castro-Alvaredo}, \citenamefont {Doyon},\ and\ \citenamefont
  {Yoshimura}}]{castro-alvaredo_emergent_2016}%
  \BibitemOpen
  \bibfield  {author} {\bibinfo {author} {\bibfnamefont {O.~A.}\ \bibnamefont
  {Castro-Alvaredo}}, \bibinfo {author} {\bibfnamefont {B.}~\bibnamefont
  {Doyon}},\ and\ \bibinfo {author} {\bibfnamefont {T.}~\bibnamefont
  {Yoshimura}},\ }\href {https://doi.org/10.1103/PhysRevX.6.041065} {\bibfield
  {journal} {\bibinfo  {journal} {Phys. Rev. X}\ }\textbf {\bibinfo {volume}
  {6}},\ \bibinfo {pages} {041065} (\bibinfo {year} {2016})}\BibitemShut
  {NoStop}%
\bibitem [{\citenamefont {Gring}\ \emph {et~al.}(2012)\citenamefont {Gring},
  \citenamefont {Kuhnert}, \citenamefont {Langen}, \citenamefont {Kitagawa},
  \citenamefont {Rauer}, \citenamefont {Schreitl}, \citenamefont {Mazets},
  \citenamefont {Smith}, \citenamefont {Demler},\ and\ \citenamefont
  {Schmiedmayer}}]{gring_relaxation_2012}%
  \BibitemOpen
  \bibfield  {author} {\bibinfo {author} {\bibfnamefont {M.}~\bibnamefont
  {Gring}}, \bibinfo {author} {\bibfnamefont {M.}~\bibnamefont {Kuhnert}},
  \bibinfo {author} {\bibfnamefont {T.}~\bibnamefont {Langen}}, \bibinfo
  {author} {\bibfnamefont {T.}~\bibnamefont {Kitagawa}}, \bibinfo {author}
  {\bibfnamefont {B.}~\bibnamefont {Rauer}}, \bibinfo {author} {\bibfnamefont
  {M.}~\bibnamefont {Schreitl}}, \bibinfo {author} {\bibfnamefont
  {I.}~\bibnamefont {Mazets}}, \bibinfo {author} {\bibfnamefont {D.~A.}\
  \bibnamefont {Smith}}, \bibinfo {author} {\bibfnamefont {E.}~\bibnamefont
  {Demler}},\ and\ \bibinfo {author} {\bibfnamefont {J.}~\bibnamefont
  {Schmiedmayer}},\ }\href {https://doi.org/10.1126/science.1224953} {\bibfield
   {journal} {\bibinfo  {journal} {Science}\ }\textbf {\bibinfo {volume}
  {337}},\ \bibinfo {pages} {1318} (\bibinfo {year} {2012})}\BibitemShut
  {NoStop}%
\bibitem [{\citenamefont {Ueda}(2020)}]{ueda_quantum_2020}%
  \BibitemOpen
  \bibfield  {author} {\bibinfo {author} {\bibfnamefont {M.}~\bibnamefont
  {Ueda}},\ }\href {https://doi.org/10.1038/s42254-020-0237-x} {\bibfield
  {journal} {\bibinfo  {journal} {Nat. Rev. Phys.}\ }\textbf {\bibinfo {volume}
  {2}},\ \bibinfo {pages} {669} (\bibinfo {year} {2020})}\BibitemShut {NoStop}%
\bibitem [{\citenamefont {Léonard}\ \emph {et~al.}(2023)\citenamefont
  {Léonard}, \citenamefont {Kim}, \citenamefont {Rispoli}, \citenamefont
  {Lukin}, \citenamefont {Schittko}, \citenamefont {Kwan}, \citenamefont
  {Demler}, \citenamefont {Sels},\ and\ \citenamefont
  {Greiner}}]{leonard_probing_2023}%
  \BibitemOpen
  \bibfield  {author} {\bibinfo {author} {\bibfnamefont {J.}~\bibnamefont
  {Léonard}}, \bibinfo {author} {\bibfnamefont {S.}~\bibnamefont {Kim}},
  \bibinfo {author} {\bibfnamefont {M.}~\bibnamefont {Rispoli}}, \bibinfo
  {author} {\bibfnamefont {A.}~\bibnamefont {Lukin}}, \bibinfo {author}
  {\bibfnamefont {R.}~\bibnamefont {Schittko}}, \bibinfo {author}
  {\bibfnamefont {J.}~\bibnamefont {Kwan}}, \bibinfo {author} {\bibfnamefont
  {E.}~\bibnamefont {Demler}}, \bibinfo {author} {\bibfnamefont
  {D.}~\bibnamefont {Sels}},\ and\ \bibinfo {author} {\bibfnamefont
  {M.}~\bibnamefont {Greiner}},\ }\href
  {https://doi.org/10.1038/s41567-022-01887-3} {\bibfield  {journal} {\bibinfo
  {journal} {Nat. Phys.}\ }\textbf {\bibinfo {volume} {19}},\ \bibinfo {pages}
  {481} (\bibinfo {year} {2023})}\BibitemShut {NoStop}%
\bibitem [{\citenamefont {Bordia}\ \emph {et~al.}(2017)\citenamefont {Bordia},
  \citenamefont {Lüschen}, \citenamefont {Scherg}, \citenamefont
  {Gopalakrishnan}, \citenamefont {Knap}, \citenamefont {Schneider},\ and\
  \citenamefont {Bloch}}]{bordia_probing_2017}%
  \BibitemOpen
  \bibfield  {author} {\bibinfo {author} {\bibfnamefont {P.}~\bibnamefont
  {Bordia}}, \bibinfo {author} {\bibfnamefont {H.}~\bibnamefont {Lüschen}},
  \bibinfo {author} {\bibfnamefont {S.}~\bibnamefont {Scherg}}, \bibinfo
  {author} {\bibfnamefont {S.}~\bibnamefont {Gopalakrishnan}}, \bibinfo
  {author} {\bibfnamefont {M.}~\bibnamefont {Knap}}, \bibinfo {author}
  {\bibfnamefont {U.}~\bibnamefont {Schneider}},\ and\ \bibinfo {author}
  {\bibfnamefont {I.}~\bibnamefont {Bloch}},\ }\href
  {https://doi.org/10.1103/PhysRevX.7.041047} {\bibfield  {journal} {\bibinfo
  {journal} {Phys. Rev. X}\ }\textbf {\bibinfo {volume} {7}},\ \bibinfo {pages}
  {041047} (\bibinfo {year} {2017})}\BibitemShut {NoStop}%
\bibitem [{\citenamefont {Moudgalya}\ \emph {et~al.}(2022)\citenamefont
  {Moudgalya}, \citenamefont {Bernevig},\ and\ \citenamefont
  {Regnault}}]{moudgalya_quantum_2022}%
  \BibitemOpen
  \bibfield  {author} {\bibinfo {author} {\bibfnamefont {S.}~\bibnamefont
  {Moudgalya}}, \bibinfo {author} {\bibfnamefont {B.~A.}\ \bibnamefont
  {Bernevig}},\ and\ \bibinfo {author} {\bibfnamefont {N.}~\bibnamefont
  {Regnault}},\ }\href {https://doi.org/10.1088/1361-6633/ac73a0} {\bibfield
  {journal} {\bibinfo  {journal} {Rep. Prog. Phys.}\ }\textbf {\bibinfo
  {volume} {85}},\ \bibinfo {pages} {086501} (\bibinfo {year}
  {2022})}\BibitemShut {NoStop}%
\bibitem [{\citenamefont {Turner}\ \emph {et~al.}(2018)\citenamefont {Turner},
  \citenamefont {Michailidis}, \citenamefont {Abanin}, \citenamefont {Serbyn},\
  and\ \citenamefont {Papic}}]{turner_quantum_2018}%
  \BibitemOpen
  \bibfield  {author} {\bibinfo {author} {\bibfnamefont {C.~J.}\ \bibnamefont
  {Turner}}, \bibinfo {author} {\bibfnamefont {A.~A.}\ \bibnamefont
  {Michailidis}}, \bibinfo {author} {\bibfnamefont {D.~A.}\ \bibnamefont
  {Abanin}}, \bibinfo {author} {\bibfnamefont {M.}~\bibnamefont {Serbyn}},\
  and\ \bibinfo {author} {\bibfnamefont {Z.}~\bibnamefont {Papic}},\ }\href
  {https://doi.org/10.1038/s41567-018-0137-5} {\bibfield  {journal} {\bibinfo
  {journal} {Nat. Phys.}\ }\textbf {\bibinfo {volume} {14}},\ \bibinfo {pages}
  {745} (\bibinfo {year} {2018})}\BibitemShut {NoStop}%
\bibitem [{\citenamefont {Scherg}\ \emph {et~al.}(2021)\citenamefont {Scherg},
  \citenamefont {Kohlert}, \citenamefont {Sala}, \citenamefont {Pollmann},
  \citenamefont {Hebbe~Madhusudhana}, \citenamefont {Bloch},\ and\
  \citenamefont {Aidelsburger}}]{scherg_observing_2021}%
  \BibitemOpen
  \bibfield  {author} {\bibinfo {author} {\bibfnamefont {S.}~\bibnamefont
  {Scherg}}, \bibinfo {author} {\bibfnamefont {T.}~\bibnamefont {Kohlert}},
  \bibinfo {author} {\bibfnamefont {P.}~\bibnamefont {Sala}}, \bibinfo {author}
  {\bibfnamefont {F.}~\bibnamefont {Pollmann}}, \bibinfo {author}
  {\bibfnamefont {B.}~\bibnamefont {Hebbe~Madhusudhana}}, \bibinfo {author}
  {\bibfnamefont {I.}~\bibnamefont {Bloch}},\ and\ \bibinfo {author}
  {\bibfnamefont {M.}~\bibnamefont {Aidelsburger}},\ }\href
  {https://doi.org/10.1038/s41467-021-24726-0} {\bibfield  {journal} {\bibinfo
  {journal} {Nat. Comm.}\ }\textbf {\bibinfo {volume} {12}},\ \bibinfo {pages}
  {4490} (\bibinfo {year} {2021})}\BibitemShut {NoStop}%
\bibitem [{\citenamefont {Kohlert}\ \emph {et~al.}(2023)\citenamefont
  {Kohlert}, \citenamefont {Scherg}, \citenamefont {Sala}, \citenamefont
  {Pollmann}, \citenamefont {Hebbe~Madhusudhana}, \citenamefont {Bloch},\ and\
  \citenamefont {Aidelsburger}}]{kohlert_exploring_2023}%
  \BibitemOpen
  \bibfield  {author} {\bibinfo {author} {\bibfnamefont {T.}~\bibnamefont
  {Kohlert}}, \bibinfo {author} {\bibfnamefont {S.}~\bibnamefont {Scherg}},
  \bibinfo {author} {\bibfnamefont {P.}~\bibnamefont {Sala}}, \bibinfo {author}
  {\bibfnamefont {F.}~\bibnamefont {Pollmann}}, \bibinfo {author}
  {\bibfnamefont {B.}~\bibnamefont {Hebbe~Madhusudhana}}, \bibinfo {author}
  {\bibfnamefont {I.}~\bibnamefont {Bloch}},\ and\ \bibinfo {author}
  {\bibfnamefont {M.}~\bibnamefont {Aidelsburger}},\ }\href
  {https://doi.org/10.1103/PhysRevLett.130.010201} {\bibfield  {journal}
  {\bibinfo  {journal} {Phys. Rev. Lett.}\ }\textbf {\bibinfo {volume} {130}},\
  \bibinfo {pages} {010201} (\bibinfo {year} {2023})}\BibitemShut {NoStop}%
\bibitem [{\citenamefont {Su}\ \emph {et~al.}(2023)\citenamefont {Su},
  \citenamefont {Sun}, \citenamefont {Hudomal}, \citenamefont {Desaules},
  \citenamefont {Zhou}, \citenamefont {Yang}, \citenamefont {Halimeh},
  \citenamefont {Yuan}, \citenamefont {Papić},\ and\ \citenamefont
  {Pan}}]{su_observation_2023}%
  \BibitemOpen
  \bibfield  {author} {\bibinfo {author} {\bibfnamefont {G.-X.}\ \bibnamefont
  {Su}}, \bibinfo {author} {\bibfnamefont {H.}~\bibnamefont {Sun}}, \bibinfo
  {author} {\bibfnamefont {A.}~\bibnamefont {Hudomal}}, \bibinfo {author}
  {\bibfnamefont {J.-Y.}\ \bibnamefont {Desaules}}, \bibinfo {author}
  {\bibfnamefont {Z.-Y.}\ \bibnamefont {Zhou}}, \bibinfo {author}
  {\bibfnamefont {B.}~\bibnamefont {Yang}}, \bibinfo {author} {\bibfnamefont
  {J.~C.}\ \bibnamefont {Halimeh}}, \bibinfo {author} {\bibfnamefont {Z.-S.}\
  \bibnamefont {Yuan}}, \bibinfo {author} {\bibfnamefont {Z.}~\bibnamefont
  {Papić}},\ and\ \bibinfo {author} {\bibfnamefont {J.-W.}\ \bibnamefont
  {Pan}},\ }\href {https://doi.org/10.1103/PhysRevResearch.5.023010} {\bibfield
   {journal} {\bibinfo  {journal} {Phys. Rev. Research}\ }\textbf {\bibinfo
  {volume} {5}},\ \bibinfo {pages} {023010} (\bibinfo {year}
  {2023})}\BibitemShut {NoStop}%
\end{thebibliography}%


\begin{thebibliography}{13}%
\makeatletter
\providecommand \@ifxundefined [1]{%
 \@ifx{#1\undefined}
}%
\providecommand \@ifnum [1]{%
 \ifnum #1\expandafter \@firstoftwo
 \else \expandafter \@secondoftwo
 \fi
}%
\providecommand \@ifx [1]{%
 \ifx #1\expandafter \@firstoftwo
 \else \expandafter \@secondoftwo
 \fi
}%
\providecommand \natexlab [1]{#1}%
\providecommand \enquote  [1]{``#1''}%
\providecommand \bibnamefont  [1]{#1}%
\providecommand \bibfnamefont [1]{#1}%
\providecommand \citenamefont [1]{#1}%
\providecommand \href@noop [0]{\@secondoftwo}%
\providecommand \href [0]{\begingroup \@sanitize@url \@href}%
\providecommand \@href[1]{\@@startlink{#1}\@@href}%
\providecommand \@@href[1]{\endgroup#1\@@endlink}%
\providecommand \@sanitize@url [0]{\catcode `\\12\catcode `\$12\catcode
  `\&12\catcode `\#12\catcode `\^12\catcode `\_12\catcode `\%12\relax}%
\providecommand \@@startlink[1]{}%
\providecommand \@@endlink[0]{}%
\providecommand \url  [0]{\begingroup\@sanitize@url \@url }%
\providecommand \@url [1]{\endgroup\@href {#1}{\urlprefix }}%
\providecommand \urlprefix  [0]{URL }%
\providecommand \Eprint [0]{\href }%
\providecommand \doibase [0]{https://doi.org/}%
\providecommand \selectlanguage [0]{\@gobble}%
\providecommand \bibinfo  [0]{\@secondoftwo}%
\providecommand \bibfield  [0]{\@secondoftwo}%
\providecommand \translation [1]{[#1]}%
\providecommand \BibitemOpen [0]{}%
\providecommand \bibitemStop [0]{}%
\providecommand \bibitemNoStop [0]{.\EOS\space}%
\providecommand \EOS [0]{\spacefactor3000\relax}%
\providecommand \BibitemShut  [1]{\csname bibitem#1\endcsname}%
\let\auto@bib@innerbib\@empty
\bibitem [{\citenamefont {Klostermann}\ \emph {et~al.}(2022)\citenamefont
  {Klostermann}, \citenamefont {Cabrera}, \citenamefont {von Raven},
  \citenamefont {Wienand}, \citenamefont {Schweizer}, \citenamefont {Bloch},\
  and\ \citenamefont {Aidelsburger}}]{klostermann_fast_2022}%
  \BibitemOpen
  \bibfield  {author} {\bibinfo {author} {\bibfnamefont {T.}~\bibnamefont
  {Klostermann}}, \bibinfo {author} {\bibfnamefont {C.~R.}\ \bibnamefont
  {Cabrera}}, \bibinfo {author} {\bibfnamefont {H.}~\bibnamefont {von Raven}},
  \bibinfo {author} {\bibfnamefont {J.~F.}\ \bibnamefont {Wienand}}, \bibinfo
  {author} {\bibfnamefont {C.}~\bibnamefont {Schweizer}}, \bibinfo {author}
  {\bibfnamefont {I.}~\bibnamefont {Bloch}},\ and\ \bibinfo {author}
  {\bibfnamefont {M.}~\bibnamefont {Aidelsburger}},\ }\href
  {https://doi.org/10.1103/PhysRevA.105.043319} {\bibfield  {journal} {\bibinfo
   {journal} {Phys. Rev. A}\ }\textbf {\bibinfo {volume} {105}},\ \bibinfo
  {pages} {043319} (\bibinfo {year} {2022})}\BibitemShut {NoStop}%
\bibitem [{\citenamefont {Impertro}\ \emph {et~al.}()\citenamefont {Impertro},
  \citenamefont {Wienand}, \citenamefont {Häfele}, \citenamefont {von Raven},
  \citenamefont {Hubele}, \citenamefont {Klostermann}, \citenamefont {Cabrera},
  \citenamefont {Bloch},\ and\ \citenamefont
  {Aidelsburger}}]{impertro_unsupervised_2022}%
  \BibitemOpen
  \bibfield  {author} {\bibinfo {author} {\bibfnamefont {A.}~\bibnamefont
  {Impertro}}, \bibinfo {author} {\bibfnamefont {J.~F.}\ \bibnamefont
  {Wienand}}, \bibinfo {author} {\bibfnamefont {S.}~\bibnamefont {Häfele}},
  \bibinfo {author} {\bibfnamefont {H.}~\bibnamefont {von Raven}}, \bibinfo
  {author} {\bibfnamefont {S.}~\bibnamefont {Hubele}}, \bibinfo {author}
  {\bibfnamefont {T.}~\bibnamefont {Klostermann}}, \bibinfo {author}
  {\bibfnamefont {C.~R.}\ \bibnamefont {Cabrera}}, \bibinfo {author}
  {\bibfnamefont {I.}~\bibnamefont {Bloch}},\ and\ \bibinfo {author}
  {\bibfnamefont {M.}~\bibnamefont {Aidelsburger}},\ }\href@noop {} {\ }\Eprint
  {https://arxiv.org/abs/2212.11974} {arXiv:2212.11974} \BibitemShut {NoStop}%
\bibitem [{\citenamefont {Fölling}\ \emph {et~al.}(2007)\citenamefont
  {Fölling}, \citenamefont {Trotzky}, \citenamefont {Cheinet}, \citenamefont
  {Feld}, \citenamefont {Saers}, \citenamefont {Widera}, \citenamefont
  {Müller},\ and\ \citenamefont {Bloch}}]{folling_direct_2007}%
  \BibitemOpen
  \bibfield  {author} {\bibinfo {author} {\bibfnamefont {S.}~\bibnamefont
  {Fölling}}, \bibinfo {author} {\bibfnamefont {S.}~\bibnamefont {Trotzky}},
  \bibinfo {author} {\bibfnamefont {P.}~\bibnamefont {Cheinet}}, \bibinfo
  {author} {\bibfnamefont {M.}~\bibnamefont {Feld}}, \bibinfo {author}
  {\bibfnamefont {R.}~\bibnamefont {Saers}}, \bibinfo {author} {\bibfnamefont
  {A.}~\bibnamefont {Widera}}, \bibinfo {author} {\bibfnamefont
  {T.}~\bibnamefont {Müller}},\ and\ \bibinfo {author} {\bibfnamefont
  {I.}~\bibnamefont {Bloch}},\ }\href {https://doi.org/10.1038/nature06112}
  {\bibfield  {journal} {\bibinfo  {journal} {Nature}\ }\textbf {\bibinfo
  {volume} {448}},\ \bibinfo {pages} {1029} (\bibinfo {year}
  {2007})}\BibitemShut {NoStop}%
\bibitem [{\citenamefont {Trotzky}\ \emph {et~al.}(2012)\citenamefont
  {Trotzky}, \citenamefont {Chen}, \citenamefont {Flesch}, \citenamefont
  {McCulloch}, \citenamefont {Schollwöck}, \citenamefont {Eisert},\ and\
  \citenamefont {Bloch}}]{trotzky_probing_2012}%
  \BibitemOpen
  \bibfield  {author} {\bibinfo {author} {\bibfnamefont {S.}~\bibnamefont
  {Trotzky}}, \bibinfo {author} {\bibfnamefont {Y.-A.}\ \bibnamefont {Chen}},
  \bibinfo {author} {\bibfnamefont {A.}~\bibnamefont {Flesch}}, \bibinfo
  {author} {\bibfnamefont {I.~P.}\ \bibnamefont {McCulloch}}, \bibinfo {author}
  {\bibfnamefont {U.}~\bibnamefont {Schollwöck}}, \bibinfo {author}
  {\bibfnamefont {J.}~\bibnamefont {Eisert}},\ and\ \bibinfo {author}
  {\bibfnamefont {I.}~\bibnamefont {Bloch}},\ }\href
  {https://doi.org/10.1038/nphys2232} {\bibfield  {journal} {\bibinfo
  {journal} {Nat. Phys.}\ }\textbf {\bibinfo {volume} {8}},\ \bibinfo {pages}
  {325} (\bibinfo {year} {2012})}\BibitemShut {NoStop}%
\bibitem [{\citenamefont {Cheuk}\ \emph {et~al.}(2015)\citenamefont {Cheuk},
  \citenamefont {Nichols}, \citenamefont {Okan}, \citenamefont {Gersdorf},
  \citenamefont {Ramasesh}, \citenamefont {Bakr}, \citenamefont {Lompe},\ and\
  \citenamefont {Zwierlein}}]{cheuk_quantum-gas_2015}%
  \BibitemOpen
  \bibfield  {author} {\bibinfo {author} {\bibfnamefont {L.~W.}\ \bibnamefont
  {Cheuk}}, \bibinfo {author} {\bibfnamefont {M.~A.}\ \bibnamefont {Nichols}},
  \bibinfo {author} {\bibfnamefont {M.}~\bibnamefont {Okan}}, \bibinfo {author}
  {\bibfnamefont {T.}~\bibnamefont {Gersdorf}}, \bibinfo {author}
  {\bibfnamefont {V.~V.}\ \bibnamefont {Ramasesh}}, \bibinfo {author}
  {\bibfnamefont {W.~S.}\ \bibnamefont {Bakr}}, \bibinfo {author}
  {\bibfnamefont {T.}~\bibnamefont {Lompe}},\ and\ \bibinfo {author}
  {\bibfnamefont {M.~W.}\ \bibnamefont {Zwierlein}},\ }\href
  {https://doi.org/10.1103/PhysRevLett.114.193001} {\bibfield  {journal}
  {\bibinfo  {journal} {Phys. Rev. Lett.}\ }\textbf {\bibinfo {volume} {114}},\
  \bibinfo {pages} {193001} (\bibinfo {year} {2015})}\BibitemShut {NoStop}%
\bibitem [{\citenamefont {Smith}(2019)}]{smith_disorder-free_2019}%
  \BibitemOpen
  \bibfield  {author} {\bibinfo {author} {\bibfnamefont {A.}~\bibnamefont
  {Smith}},\ }\href@noop {} {\emph {\bibinfo {title} {Disorder-{Free}
  {Localization}}}},\ \bibinfo {edition} {1st}\ ed.\ (\bibinfo  {publisher}
  {Springer},\ \bibinfo {year} {2019})\BibitemShut {NoStop}%
\bibitem [{\citenamefont {Karamlou}\ \emph {et~al.}(2022)\citenamefont
  {Karamlou}, \citenamefont {Braumüller}, \citenamefont {Yanay}, \citenamefont
  {Di~Paolo}, \citenamefont {Harrington}, \citenamefont {Kannan}, \citenamefont
  {Kim}, \citenamefont {Kjaergaard}, \citenamefont {Melville}, \citenamefont
  {Muschinske}, \citenamefont {Niedzielski}, \citenamefont {Vepsäläinen},
  \citenamefont {Winik}, \citenamefont {Yoder}, \citenamefont {Schwartz},
  \citenamefont {Tahan}, \citenamefont {Orlando}, \citenamefont {Gustavsson},\
  and\ \citenamefont {Oliver}}]{karamlou_quantum_2022}%
  \BibitemOpen
  \bibfield  {author} {\bibinfo {author} {\bibfnamefont {A.~H.}\ \bibnamefont
  {Karamlou}}, \bibinfo {author} {\bibfnamefont {J.}~\bibnamefont
  {Braumüller}}, \bibinfo {author} {\bibfnamefont {Y.}~\bibnamefont {Yanay}},
  \bibinfo {author} {\bibfnamefont {A.}~\bibnamefont {Di~Paolo}}, \bibinfo
  {author} {\bibfnamefont {P.~M.}\ \bibnamefont {Harrington}}, \bibinfo
  {author} {\bibfnamefont {B.}~\bibnamefont {Kannan}}, \bibinfo {author}
  {\bibfnamefont {D.}~\bibnamefont {Kim}}, \bibinfo {author} {\bibfnamefont
  {M.}~\bibnamefont {Kjaergaard}}, \bibinfo {author} {\bibfnamefont
  {A.}~\bibnamefont {Melville}}, \bibinfo {author} {\bibfnamefont
  {S.}~\bibnamefont {Muschinske}}, \bibinfo {author} {\bibfnamefont {B.~M.}\
  \bibnamefont {Niedzielski}}, \bibinfo {author} {\bibfnamefont
  {A.}~\bibnamefont {Vepsäläinen}}, \bibinfo {author} {\bibfnamefont
  {R.}~\bibnamefont {Winik}}, \bibinfo {author} {\bibfnamefont {J.~L.}\
  \bibnamefont {Yoder}}, \bibinfo {author} {\bibfnamefont {M.}~\bibnamefont
  {Schwartz}}, \bibinfo {author} {\bibfnamefont {C.}~\bibnamefont {Tahan}},
  \bibinfo {author} {\bibfnamefont {T.~P.}\ \bibnamefont {Orlando}}, \bibinfo
  {author} {\bibfnamefont {S.}~\bibnamefont {Gustavsson}},\ and\ \bibinfo
  {author} {\bibfnamefont {W.~D.}\ \bibnamefont {Oliver}},\ }\href
  {https://doi.org/10.1038/s41534-022-00528-0} {\bibfield  {journal} {\bibinfo
  {journal} {NPJ Quantum Inf.}\ }\textbf {\bibinfo {volume} {8}},\ \bibinfo
  {pages} {35} (\bibinfo {year} {2022})}\BibitemShut {NoStop}%
\bibitem [{\citenamefont {Hauschild}\ and\ \citenamefont
  {Pollmann}(2018)}]{hauschild_efficient_2018}%
  \BibitemOpen
  \bibfield  {author} {\bibinfo {author} {\bibfnamefont {J.}~\bibnamefont
  {Hauschild}}\ and\ \bibinfo {author} {\bibfnamefont {F.}~\bibnamefont
  {Pollmann}},\ }\href {https://scipost.org/10.21468/SciPostPhysLectNotes.5}
  {\bibfield  {journal} {\bibinfo  {journal} {SciPost Phys. Lect. Notes}\ }
  (\bibinfo {year} {2018})}\BibitemShut {NoStop}%
\bibitem [{\citenamefont {von Keyserlingk}\ \emph {et~al.}(2022)\citenamefont
  {von Keyserlingk}, \citenamefont {Pollmann},\ and\ \citenamefont
  {Rakovszky}}]{von_keyserlingk_operator_2022}%
  \BibitemOpen
  \bibfield  {author} {\bibinfo {author} {\bibfnamefont {C.}~\bibnamefont {von
  Keyserlingk}}, \bibinfo {author} {\bibfnamefont {F.}~\bibnamefont
  {Pollmann}},\ and\ \bibinfo {author} {\bibfnamefont {T.}~\bibnamefont
  {Rakovszky}},\ }\href {https://doi.org/10.1103/PhysRevB.105.245101}
  {\bibfield  {journal} {\bibinfo  {journal} {Phys. Rev. B}\ }\textbf {\bibinfo
  {volume} {105}},\ \bibinfo {pages} {245101} (\bibinfo {year}
  {2022})}\BibitemShut {NoStop}%
\bibitem [{\citenamefont {{Steinigeweg}}\ \emph {et~al.}(2014)\citenamefont
  {{Steinigeweg}}, \citenamefont {{Heidrich-Meisner}}, \citenamefont
  {{Gemmer}}, \citenamefont {{Michielsen}},\ and\ \citenamefont {{De
  Raedt}}}]{steinigeweg_2014}%
  \BibitemOpen
  \bibfield  {author} {\bibinfo {author} {\bibfnamefont {R.}~\bibnamefont
  {{Steinigeweg}}}, \bibinfo {author} {\bibfnamefont {F.}~\bibnamefont
  {{Heidrich-Meisner}}}, \bibinfo {author} {\bibfnamefont {J.}~\bibnamefont
  {{Gemmer}}}, \bibinfo {author} {\bibfnamefont {K.}~\bibnamefont
  {{Michielsen}}},\ and\ \bibinfo {author} {\bibfnamefont {H.}~\bibnamefont
  {{De Raedt}}},\ }\href {https://doi.org/10.1103/PhysRevB.90.094417}
  {\bibfield  {journal} {\bibinfo  {journal} {\prb}\ }\textbf {\bibinfo
  {volume} {90}},\ \bibinfo {eid} {094417} (\bibinfo {year}
  {2014})}\BibitemShut {NoStop}%
\bibitem [{\citenamefont {McCulloch}\ \emph {et~al.}()\citenamefont
  {McCulloch}, \citenamefont {De~Nardis}, \citenamefont {Gopalakrishnan},\ and\
  \citenamefont {Vasseur}}]{mcculloch_full_2023}%
  \BibitemOpen
  \bibfield  {author} {\bibinfo {author} {\bibfnamefont {E.}~\bibnamefont
  {McCulloch}}, \bibinfo {author} {\bibfnamefont {J.}~\bibnamefont
  {De~Nardis}}, \bibinfo {author} {\bibfnamefont {S.}~\bibnamefont
  {Gopalakrishnan}},\ and\ \bibinfo {author} {\bibfnamefont {R.}~\bibnamefont
  {Vasseur}},\ }\href {http://arxiv.org/abs/2302.01355} {\bibinfo  {journal}
  {arXiv:2302.01355}\ }\BibitemShut {NoStop}%
\bibitem [{\citenamefont {Derrida}\ and\ \citenamefont
  {Gerschenfeld}(2009)}]{derrida_current_2009}%
  \BibitemOpen
\bibfield  {journal} {  }\bibfield  {author} {\bibinfo {author} {\bibfnamefont
  {B.}~\bibnamefont {Derrida}}\ and\ \bibinfo {author} {\bibfnamefont
  {A.}~\bibnamefont {Gerschenfeld}},\ }\href
  {https://doi.org/10.1007/s10955-009-9772-7} {\bibfield  {journal} {\bibinfo
  {journal} {J. Stat. Phys.}\ }\textbf {\bibinfo {volume} {136}},\ \bibinfo
  {pages} {1} (\bibinfo {year} {2009})}\BibitemShut {NoStop}%
\bibitem [{\citenamefont {Mallick}\ \emph {et~al.}(2022)\citenamefont
  {Mallick}, \citenamefont {Moriya},\ and\ \citenamefont
  {Sasamoto}}]{mallick_exact_2022}%
  \BibitemOpen
  \bibfield  {author} {\bibinfo {author} {\bibfnamefont {K.}~\bibnamefont
  {Mallick}}, \bibinfo {author} {\bibfnamefont {H.}~\bibnamefont {Moriya}},\
  and\ \bibinfo {author} {\bibfnamefont {T.}~\bibnamefont {Sasamoto}},\ }\href
  {https://doi.org/10.1103/PhysRevLett.129.040601} {\bibfield  {journal}
  {\bibinfo  {journal} {Phys. Rev. Lett.}\ }\textbf {\bibinfo {volume} {129}},\
  \bibinfo {pages} {040601} (\bibinfo {year} {2022})}\BibitemShut {NoStop}%
\end{thebibliography}%


\begin{thebibliography}{0}%
\makeatletter
\providecommand \@ifxundefined [1]{%
 \@ifx{#1\undefined}
}%
\providecommand \@ifnum [1]{%
 \ifnum #1\expandafter \@firstoftwo
 \else \expandafter \@secondoftwo
 \fi
}%
\providecommand \@ifx [1]{%
 \ifx #1\expandafter \@firstoftwo
 \else \expandafter \@secondoftwo
 \fi
}%
\providecommand \natexlab [1]{#1}%
\providecommand \enquote  [1]{``#1''}%
\providecommand \bibnamefont  [1]{#1}%
\providecommand \bibfnamefont [1]{#1}%
\providecommand \citenamefont [1]{#1}%
\providecommand \href@noop [0]{\@secondoftwo}%
\providecommand \href [0]{\begingroup \@sanitize@url \@href}%
\providecommand \@href[1]{\@@startlink{#1}\@@href}%
\providecommand \@@href[1]{\endgroup#1\@@endlink}%
\providecommand \@sanitize@url [0]{\catcode `\\12\catcode `\$12\catcode
  `\&12\catcode `\#12\catcode `\^12\catcode `\_12\catcode `\%12\relax}%
\providecommand \@@startlink[1]{}%
\providecommand \@@endlink[0]{}%
\providecommand \url  [0]{\begingroup\@sanitize@url \@url }%
\providecommand \@url [1]{\endgroup\@href {#1}{\urlprefix }}%
\providecommand \urlprefix  [0]{URL }%
\providecommand \Eprint [0]{\href }%
\providecommand \doibase [0]{https://doi.org/}%
\providecommand \selectlanguage [0]{\@gobble}%
\providecommand \bibinfo  [0]{\@secondoftwo}%
\providecommand \bibfield  [0]{\@secondoftwo}%
\providecommand \translation [1]{[#1]}%
\providecommand \BibitemOpen [0]{}%
\providecommand \bibitemStop [0]{}%
\providecommand \bibitemNoStop [0]{.\EOS\space}%
\providecommand \EOS [0]{\spacefactor3000\relax}%
\providecommand \BibitemShut  [1]{\csname bibitem#1\endcsname}%
\let\auto@bib@innerbib\@empty
\end{thebibliography}%
\end{document}